\documentstyle[12pt,epsf]{article}
\def\cmm2{{\,\rm cm^{-2}}}
\def\cm2{{\,{\rm cm}^2}}
\def\cmm3{{\,{\rm cm}^{-3}}}
\def\gcmm3{{\,{\rm g\,cm^{-3}}}}

\def\la{\mathrel{\mathpalette\fun <}}
\def\ga{\mathrel{\mathpalette\fun >}}
\def\fun#1#2{\lower3.6pt\vbox{\baselineskip0pt\lineskip.9pt
  \ialign{$\mathsurround=0pt#1\hfil##\hfil$\crcr#2\crcr\sim\crcr}}}
\begin{document}
\pagestyle{empty}
\begin{center}
\rightline{}
\vspace{.5in}
\rightline{FERMILAB--Pub--96/066-A}
\rightline{submitted to {\sl Physical Review D}}

\vspace{.3in}
{\Large \bf On the Propagation of\\
\medskip
Extragalactic High Energy Cosmic and $\gamma$-Rays}\\

\vspace{.3in}
Sangjin Lee\\
\vspace{.2in}
{\sl Department of Physics\\
Enrico Fermi Institute, The University of Chicago, Chicago, IL~~60637-1433}\\

\vspace{0.1in}
{\sl NASA/Fermilab Astrophysics Center\\
Fermi National Accelerator Laboratory, Batavia, IL~~60510-0500}\\
\end{center}

\vspace{.5in}

\centerline{\bf Abstract}
\medskip
\noindent{The observation of air showers from elementary particles
with energies exceeding $10^{20}{\,{\rm eV}}$ poses a puzzle to
the physics and astrophysics of cosmic rays which is still
unresolved.} Explaining the origin and nature of these particles
is a challenge. In order to constrain production mechanisms and
sites, one has to account for the processing of particle spectra
by interactions with radiation backgrounds and magnetic fields
on the way to the observer.
In this paper, I report on an extensive study on the
propagation of extragalactic nucleons, $\gamma$-rays, and electrons
in the energy range between $10^8{\,{\rm eV}}$ and $10^{23}{\,{\rm eV}}$.
I have devised an efficient numerical
method to solve the transport equations for cosmic ray spectral
evolution. The universal radiation background spectrum in the energy range
between $\simeq 10^{-9}{\,{\rm eV}}$ and
$\simeq 1{\,{\rm eV}}$ is considered in the numerical code, including the
diffuse radio background, the cosmic microwave background, and
the infrared/optical background, as well as a possible extragalactic
magnetic field. I apply the code to compute the particle
spectra predicted by various models of ultrahigh energy cosmic
ray origin.
A comparison with the observed fluxes, especially the
diffuse $\gamma$-ray background in several energy ranges, allows one to
constrain certain classes of models. I conclude that scenarios
which attribute the highest energy cosmic rays to Grand
Unification Scale physics or to cosmological Gamma Ray Bursts
are viable at the present time.


\newpage
\pagestyle{plain}
\setcounter{page}{1}
\newpage

\section{Introduction}
Shortly after the cosmic microwave background (CMB) was
discovered~\cite{Penzias} it became
clear that this universal radiation field has profound
implications for the astrophysics of ultrahigh energy cosmic
rays (UHE CR) of energies above $10^{18}{\,{\rm eV}}$. For nucleons the
most profound effect is photoproduction of pions on the CMB.
Known as the Greisen-Zatsepin-Kuz'min (GZK)
``cutoff''~\cite{GZK,Stecker1}, this effect leads to a steep drop in
their energy attenuation length by
about a factor 100 at around $6\times10^{19}{\,{\rm eV}}$ which
corresponds to the threshold for this process.
The nucleon attenuation length
above this threshold is about $10{\,{\rm Mpc}}$. Heavy nuclei
with energies above about $10^{19}{\,{\rm eV}}$ are
photodisintegrated in the field of the CMB within a few
${\,{\rm Mpc}}$~\cite{psb}. One of the major unresolved questions in
cosmic ray physics is the existence or non-existence of a cutoff
in the UHE CR spectrum at a few $10^{19}\,$eV which, in the case of
extragalactic sources, could be attributed to these effects.

Therefore, there has been renewed interest in UHE CR research
since events with energies exceeding $10^{20}{\,{\rm eV}}$ have been
detected.
The Haverah Park experiment~\cite{Watson} reported several
events with energies near or slightly above $10^{20}{\,{\rm eV}}$. The
Fly's Eye experiment~\cite{FE1,FE2} detected the world's
highest energy CR event to date, with an energy
$\simeq3\times10^{20}{\,{\rm eV}}$. Near the arrival direction of
this event the Yakutsk experiment~\cite{Efiego} recorded
another event of energy $\simeq1.1\times10^{20}{\,{\rm eV}}$. More
recently, the AGASA experiment~\cite{AG1,AG2} has also reported an
event with energy $1.7-2.6\times10^{20}{\,{\rm eV}}$. It is currently
unclear whether these events indicate a spectrum continuing
beyond $10^{20}{\,{\rm eV}}$ without any cutoff or the existence of
a cutoff followed by a recovery in the form of a
``gap'' in the spectrum~\cite{slsb}.

There has been much speculation about the nature and
origin of these highest energy cosmic rays
(HECRs)~\cite{Hillas,Sorrell,Sommers,Elbert,SSB}.
Concerning the production mechanism
one can distinguish between two broad classes of models:
Within acceleration models, charged primaries, namely protons and
heavy nuclei are accelerated to very high
energies~\cite{Blandford,Gaisser} in a ``bottom-up'' manner.
Preferred sites are large-scale
astrophysical shocks which occur for instance in
radio galaxies~\cite{Biermann}. Even there it seems barely
possible to accelerate CRs to the required
energies~\cite{Hillas,Cesarsky,Norman}. Recently it has also been
suggested that acceleration of
UHE CRs could be associated with cosmological gamma-ray bursts
(GRBs)~\cite{Wax1,Vietri,Wax2}.
In the second class of so called ``top-down'' models, charged
and neutral primaries are produced at UHEs in the first place,
typically by quantum mechanical decay of supermassive elementary
X particles related to grand unified theories (GUTs). Sources of
such particles at present could be topological defects
(TDs) left over from early universe phase transitions caused by
the spontaneous breaking of symmetries underlying these
GUTs~\cite{Hill,HSW,Bh0,Bh1,BR,bhs,BS,Sigl1}. The injection
spectra in top-down models tend to be considerably harder
(flatter) than in acceleration models.

The particle identity of the UHE CRs is not known either. The Fly's Eye
analysis~\cite{FE1}
suggested a transition from a spectrum dominated by heavy nuclei
to a predominantly light
composition, i.e. nucleons or even $\gamma$-rays, above a few
times $10^{19}{\,{\rm eV}}$. However, this has not
been confirmed by the AGASA experiment~\cite{AG3}.
Although there have been claims that the shower profile of the
highest energy Fly's Eye event may be inconsistent with a primary
photon~\cite{hvsv} or even with a proton
primary~\cite{Gaisser:pc}, the situation is not settled because
of many uncertainties which can affect the shower development in the
atmosphere. 

Other options discussed for the nature of the HECRs include heavy
nuclei and even
neutrinos~\cite{hvsv}. Heavy nuclei have their own merits because they can
be deflected considerably by the galactic magnetic field which
relaxes the source direction requirements \cite{FE1,AG1}.
In addition, for shock acceleration, heavy nuclei can be
accelerated to higher terminal energies because of their higher
charge. However, one should note that the range for heavy nuclei is limited to
a few Mpc as mentioned above. Neutrinos, on the other hand, do not
lose much energy over
cosmological distances~\cite{Weiler,Yoshida}, but by the same token the
probability for interacting in the atmosphere is
small. Attributing the HECRs to neutrinos would therefore
require a neutrino flux at UHEs which is much higher than the
observed CR flux at the same energies. This poses severe
constraints on the possible sources for these neutrinos~\cite{sl}.
In addition, neutrinos would be expected to give rise to
predominantly deeply penetrating showers in the atmosphere.

The production spectrum of UHE CRs is modified during their
propagation. There are many studies on nucleon propagation in
the literature using analytical~\cite{bg,akv,Rachen,Geddes} as
well as numerical approaches~\cite{Elbert,hs,yt,ac}, and the
propagation of heavy nuclei has also been
considered~\cite{Elbert}. This was mainly motivated by
the conventional acceleration models which usually predict UHE
CR fluxes to be dominated by these particles.
However, secondary $\gamma$-rays and
neutrinos can also be produced, for example as decay products of
pions created by interactions with various radiation backgrounds
at the source or during propagation~\cite{yt}. Under certain
circumstances their flux can become comparable with the primary
flux~\cite{Wolfendale1}. Furthermore, within TD models $\gamma$-rays
are expected to dominate to begin with~\cite{abs}.
A study on $\gamma$-ray propagation in this context has been performed
recently~\cite{pj} using a quantitative treatment on the cascade initiated by
UHE photons. In my opinion, however, it suffers from 
several unrealistic assumptions with respect to the injection
scenarios considered. I improve on their treatment of the propagation of
$\gamma$-rays.
Apart from that I find three reasons to explore UHE $\gamma$-ray
propagation in more detail in this paper:

First, due to the absence of threshold effects
similar to photopion production which causes the GZK ``cutoff'' for
nucleons, the $\gamma$-ray
spectrum is not expected to have a break around $10^{20}{\,{\rm eV}}$.
Furthermore, $\gamma$-rays can generate electromagnetic (EM) cascades
while propagating rather than being absorbed right away. UHE
electrons produced by pair production upscatter background photons and
transfer most of the energy back to photons. This effect
considerably increases
the effective energy attenuation length of the ``cascade''
photons~\cite{los,ls}. At a few times $10^{20}{\,{\rm eV}}$ this
attenuation length
may be even greater than that for protons which drops
precipitously at the threshold for photopion production.
Extragalactic $\gamma$-rays could therefore have some potential to
produce a recovery beyond the GZK ``cutoff''.

Second, in contrast to
the case of nucleons, the propagation of $\gamma$-rays is presently
fraught by certain ambiguities which are mainly due to
uncertainties in the intensity of
the universal radio background and the strength and spectrum of
the extragalactic magnetic field (EGMF). We hope that
an application of the general framework presented here under
different assumptions for such parameters could in turn
provide some insights into their actual values once the
UHE $\gamma$-ray flux is known to some accuracy. This would be in some
analogy to the method of using TeV $\gamma$-ray observations to
constrain or detect the universal infrared/optical
background~\cite{Stecker3}. In previous work~\cite{los}
it is shown that, depending on its strength, the large-scale EGMF
could produce a feature in the $\gamma$-ray spectrum
which might be observable in the future.

Finally, the study of high energy cosmic and $\gamma$-ray propagation
can place stringent constraints on the nature and origin of UHE
CRs. Such constraints can be obtained by computing the
propagation modified spectra especially of lower energy $\gamma$-rays
expected within 
a certain scenario and comparing the predictions with the
observed fluxes~\cite{Wolfendale1,hpsv,SJSB}.
At UHEs there are some
experimental prospects to distinguish $\gamma$-rays from other
primaries in the future, possibly even on an
event by event basis~\cite{Cronin2}. This would allow comparing
not only the total fluxes of UHE nucleons, heavy nuclei, and
$\gamma$-rays, but also their composition with model predictions.

This motivated the present comprehensive study of propagation of
nucleons and $\gamma$-rays and its application
to models which attribute UHE CRs to top-down mechanisms within
GUT-scale physics or
associate them with cosmological GRBs.
I explore the energy range of $10^8 < E <
10^{23} {\,{\rm eV}}$. The low end is chosen such that
we can draw constraints by comparing the propagated
spectra with existing measurements of the diffuse $\gamma$-ray
background around
$100{\,{\rm MeV}}$~\cite{Digel,Fichtel,Osborne}. The high end is chosen
beyond the highest CR
energies ever observed enabling us to study top-down models. I include
not only the CMB but also the diffuse radio background which plays a
big role at the highest energies and the infrared/optical (IR/O) background
which influences 
the flux at somewhat lower energies. I also include the EGMF as a free
parameter. The propagation of nucleons is also studied with
special emphasis on the
production of secondary $\gamma$-rays, electrons, and neutrinos.

The rest of the paper is organized as follows:
In Section 2, I present the general ingredients of
calculating the propagation of extragalactic $\gamma$-rays and
nucleons. I discuss the role and nature of the low energy
photon background and the EGMF, and explain in detail the
implicit method used in solving the transport equations
numerically. Section 3 is devoted to the treatment of the
relevant interactions of $\gamma$-rays and nucleons. I compare
our analysis with other work in Section 4.
Section 5 discusses the generic forms of the injection spectra and the
source distribution for typical top-down models and the GRB
scenario. Results and constraints from the spectra predicted at
Earth are presented in detail. In Section 6, I summarize the
findings and discuss future prospects.

\section{Formalism}

\subsection{Radiation backgrounds}

UHE CRs undergo reactions with the universal diffuse radiation
backgrounds permeating the universe~\cite{Ressell}. The most
relevant among them are the CMB, and the radio and IR/O
backgrounds.

Photon primaries of energy $E$ can be absorbed by pair
production with a background photon of energy $\epsilon$ if ($c=\hbar=1$
throughout)
\begin{equation}
E\geq E_{\rm th}\equiv\frac{m_e^2}{\epsilon} = 2.611 \times 10^{11}
\left(\frac{\epsilon}{1{\,{\rm eV}}}\right)^{-1} {\,{\rm eV}}\,.\label{Ethr}
\end{equation}
Therefore, for a typical CMB photon ($\epsilon \sim 10^{-3} {\,{\rm eV}}$) the
threshold energy $E_{\rm th}$ is $\sim3\times 10^{14}{\,{\rm eV}}$,
whereas for a typical radio photon
($\epsilon \la 10^{-8} {\,{\rm eV}}$) the threshold is
$\ga 3 \times 10^{19}
{\,{\rm eV}}$, thus affecting UHE $\gamma$-rays. Furthermore,
since the pair production cross section peaks near the threshold,
pair production on the radio background dominates over pair
production on the CMB in that energy range although the number
density of radio background photons is much smaller than that of
CMB photons.
On the other hand, the IR/O background affects the lower energy photons for the
same reason. The threshold for pair production on the IR/O background lies at
about $10^{12} - 10^{13}{\,{\rm eV}}$. Similarly, the contribution of the
IR/O background to the
total photopion production rate by protons is not negligible in the lower
energy range ($E \la 10^{18} {\,{\rm eV}}$).

All these backgrounds evolve with time (i.e. redshift) by cooling with the
expansion of the universe. However, on top of it, the radio
background and the IR/O background evolve due to the evolution of the respective
sources. The evolution of the radio background is tied to the evolution of the
radio sources such as radio galaxies, and the evolution of the IR/O
background to that of normal galaxies. Treating the evolution of these
backgrounds carefully is important if we are to go back to the redshift where
there existed not many of these sources ($z \ga 5 - 6$). The
flux of an isotropic radiation background component produced by an ensemble of
sources is 
given by the following relation:
\begin{equation}
j(\epsilon,z)=\int_{z}^{z_i} \left(\frac{1+z}{1+z'} \right)^3
\Phi\left(\epsilon \frac{1+z'}{1+z},z' \right) \frac{1+z'}{1+z} \frac{c}{H_0}
(1+z')^{-5/2} dz',\label{evol}
\end{equation}
where $j(\epsilon,z)$ is the radiation flux (in units of number
per area per time per
solid angle per energy) at redshift $z$, $z_i$ is the initial redshift when
the sources begin to appear, $H_0$ is the Hubble constant, and
$\Phi(\epsilon,z)$ is the production spectrum of the relevant background
(in units of number of background photons per volume per
time per solid angle per energy) at redshift $z$. Throughout
this paper we assume $H_0=75\,{\rm km}\,{\rm sec}^{-1}\,{\rm
Mpc}^{-1}$ and a critical density universe (i.e.~$\Omega_0=1$) for simplicity,
but I keep $H_0$ in the formulae to show the dependence. If we assume that
\begin{equation}
\Phi(\epsilon,z)=\Phi_0(\epsilon)(1+z)^3 N_c(z)\,,\label{backinj}
\end{equation}
where $\Phi_0 (\epsilon)$ is the typical intensity spectrum of an individual
source and $N_c(z)$ is the comoving density of the sources as a function of
the redshift, Eq.~(\ref{evol}) may be rewritten as
\begin{equation}
j(\epsilon,z)=\frac{c}{H_0} (1+z)^2 \int_z^{z_i} (1+z')^{-3/2} \Phi_0
\left(\epsilon\,\frac{1+z'}{1+z}\right) N_c (z') dz'.\label{evol2}
\end{equation}
The background photon number density is then obtained from the relation
$n(\epsilon,z)=4\pi/c \cdot 
j(\epsilon,z)$. Note that it is important to self-consistently derive the
background in this way rather than following the often-used approach of
assuming a current background photon distribution, $n_0 (\epsilon)$, and then
extrapolating it back to higher redshifts via the formula $n(\epsilon,z) =
(1+z)^3 n_0[(1+z) \epsilon]$. While easy to implement, this approach is only
valid for a truly primordial background formed at extremely high redshifts
(e.g. the CMB) and can lead to misleading results if one is not careful. 

The role of the IR/O background in determining the level of 
the cascade radiation background below $\sim 1{\,{\rm TeV}}$ 
which is the diffuse $\gamma$-ray background due to cascades by CRs
was examined
in more detail in
Ref.~\cite{coaha}. Even for rather extreme assumptions for the IR/O
background, the cascade background level typically does not vary by more than a
factor of a few. Accordingly, for this paper I have chosen to adopt a simple,
``middle of the road'' model for the formation of the IR/O background (a
discussion of the various possibilities can be found in
Refs.~\cite{stecketal,mmpri}). I assumed that the dominant contribution to the
IR/O background comes from ordinary galaxies which formed early in the
universe, at $z_{i} \simeq 5$. The typical galaxy was assumed to have a spectrum
like that of the 5 Gyr disk galaxy spectrum shown in Fig.~4 of
Ref.~\cite{mazzei}, which has a component peaking at $\simeq
1\,\mu{\rm m}$ in
wavelength due to direct emission from stars and a second component peaking at
$\simeq 100\,\mu{\rm m}$ due to reprocessing of the starlight by
interstellar dust. The
combined number and luminosity evolution of the galaxies was taken to go as
$(1+z)^7$, i.e. most of the background was produced in a strong, initial burst
of star formation in the galaxies, and the intensity of their emission was
adjusted to give an optical background density today of $n_{\rm opt}\simeq 2
\times 10^{-3}\,{\rm cm}^{-3}$.

For the present diffuse
extragalactic radio background spectrum I use the estimate given in
Ref.~\cite{clark} (see also Refs.~\cite{Ressell,Longair}). This spectrum can
be parametrized by a power 
law with a lower frequency cutoff for which I use $f_c=2{\,{\rm MHz}}$. One
can also estimate the contribution to this background in the
power law regime caused by radio galaxies. I did that
by inserting the injection flux $\Phi(\epsilon,z)$
resulting from the radio luminosity function given by Eq.~(7) of
Ref.~\cite{Dunlop} into Eq.~(\ref{evol}). The intensities
resulting at $z=0$ are within a factor $\simeq2$ of
the estimate given in Ref.~\cite{clark}. I adopt the functional
redshift dependence for the power law regime following from this
calculation and normalize it to the present intensities given in
Ref.~\cite{clark}. In addition, I assume a redshift-independent
lower frequency cutoff at $f_c=2{\,{\rm MHz}}$.

The combined radiation spectrum at
$z=0$ used in this paper is presented in Fig.~1. In
Fig.~2
I plot $N_c(z)$, i.e. the effective comoving densities of radio
and IR/O sources whose luminosities are normalized at $z=0$, as
functions of redshift.

\subsection{Extragalactic Magnetic Field}

The long range EGMF affects the propagation of CR
particles via synchrotron radiation and deflection (or even
diffusion).

Synchrotron radiation is much more straightforward to consider than
deflection. The synchrotron loss rate for a charged particle
with mass $m$, energy $E$, and charge
$qe$ ($e$ is the electron charge) subject to a magnetic field of
strength $B$ is given by \cite{jackson}
\begin{equation}
\frac{dE}{dt} = - \frac{4}{3}\sigma_T \frac{B^2}{8\pi} \left(\frac{qm_e}{m}
\right)^4 \left( \frac{E}{m_e} \right)^2.
\end{equation}
where $\sigma_T$ is the Thomson cross section, and $m_e$ is the electron mass.
Here, the average over random magnetic field orientations was taken.
Synchrotron loss influences the electronic component of the
cascade most strongly in the UHE regime~\cite{los}.
On the other hand, at a given energy the synchrotron loss rate
for protons is much smaller than
that for electrons because the loss rate is proportional to $m^{-4}$.
Thus, for protons synchrotron loss is completely negligible for the energies,
magnetic field values, and distances I consider in this paper.

The relevant synchrotron power spectrum radiated by the
electrons is given by~\cite{jackson}
\begin{equation}
\frac{dP}{dE_{\gamma}} = \frac{\sqrt{3}}{2\pi} \frac{e^3 B}{m_e}
\;G(E_\gamma/E_c)\,,\label{Psynch}
\end{equation}
where
\begin{equation}
G(x) \equiv x \int_x^{\infty} \sqrt{1-(x/\xi)^2}
K_{5/3} (\xi) d\xi\,,
\end{equation}
and the critical energy $E_c$ is defined as
\begin{equation}
E_c \equiv \frac{3eB}{2m_e} \left( \frac{E_e}{m_e} \right)^2
\simeq2.2\times10^{14}\left({E_e\over10^{21}{\,{\rm eV}}}\right)^2\,
\left({B\over10^{-9}{\rm G}}\right){\,{\rm eV}}\,.\label{Esyn}
\end{equation}
The power spectrum peaks at $E_\gamma\simeq0.23E_c$. The number
spectrum, which is obtained by dividing Eq.~(\ref{Psynch}) by the
photon energy $E_\gamma$, is a monotonically decreasing function of energy.

Deflection is another important factor when dealing with the propagation
problem in 
general \cite{los}. The straight line propagation (SLP) approximation which
treats the motion of CR particles in one dimension fails if
the effect of 
the deflection becomes large. The gyroradius of a charged particle with
charge $qe$ and momentum $p$ (energy $E$) is given by
\begin{equation}
R_g=\frac{p}{qeB_\perp} \simeq\frac{E}{qeB_\perp}
\simeq1.1\times10^3\;\frac{1}{q}\left({E\over10^{21}{\,{\rm eV}}}\right)
\left({B_\perp\over10^{-9}{\rm G}}\right)^{-1}{\,{\rm Mpc}}\,,
\end{equation}
where $B_\perp$ is the field component perpendicular to the
particle's motion. Note that the EGMF deflects
protons and electrons by the same amount at a given energy once they are
relativistic.  
If the gyroradius of a charged particle is considerably longer than the source
distance, the effect of the deflection is practically negligible. On the other
hand, if the gyroradius is comparable or shorter than the source distance, the
deflection may not be neglected and one now has to keep track of the
transversal motion, which makes the problem much more complicated.
However, if the sources are distributed homogenously and
isotropically throughout the universe, then the influence of the deflection
on the shape of the spectrum becomes small.
Although this is a purely mathematical model, it is a good
approximation for many realistic situations. On the other hand, if one
considers the CR flux from a single source, deflection becomes important.

In the case of an EM cascade the propagating particle basically
alternates between a photon
and an electron, and only electrons are affected by the EGMF. Therefore, the
effective gyroradius of a cascade photon $R^{\gamma}_g$ can be expressed as
\begin{equation}
R_g^{\gamma} \simeq R_g^{e^-} \left( 1+\frac{L_{\gamma}}{L_e} \right)
\end{equation}
where $R^{e^-}_g$ is the electron gyroradius, $L_{\gamma}$ is
the photon interaction length,
and $L_e$ is the energy loss length for the electron.
If the effective gyroradius is considerably shorter than the source distance,
the real 
spectrum would be very different from what one obtains by using the SLP
assumption, and below 
the energy where the gyroradius is comparable to the source
distance the flux is expected to be heavily suppressed. This
point has been ignored in most of the work on CR propagation~\cite{hs,yt,pj}.
Fig.~3 illustrates the gyroradii and the synchrotron 
loss rates of electrons for various strengths of the EGMF.
In this paper, the strength of the EGMF is assumed as a free parameter
between $10^{-12}$ G and $10^{-9}$ G as in Ref.~\cite{los}.

\subsection{Transport Equations}

I adopt a transport equation scheme to solve the propagation problem. Since
we have an EM cascade ensuing, it is often inadequate to use the simple
continuous energy loss (CEL) approximation which neglects non-leading
particles. In addition, since
particle numbers grow fast with time, using a full-blown Monte
Carlo calculation will require
excessive computing time. In this problem, using the transport equation approach
is very economical in terms of computing time as well as sufficiently
accurate. Previous work done by Protheroe \& Johnson \cite{pj} uses a
mixture of transport equations and Monte Carlo techniques. 

A sample transport equation for
electrons which includes pair production (PP) and inverse
Compton scattering (ICS) can be written as follows:
\begin{eqnarray}
\frac{d}{dt} N_e (E_e,t) & = & - N_e (E_e,t) \int
d\epsilon\,n(\epsilon) \int d\mu \frac{1-\beta_e \mu}{2} \sigma_{\rm
ICS} (E_e,\epsilon,\mu) + \label{step1}\\
& & \int dE^\prime_{e}N_e(E^\prime_{e},t) \int d\epsilon\,n(\epsilon) \int d\mu
\frac{1-\beta^\prime_{e} \mu}{2} 
\frac{d\sigma_{\rm ICS}}{dE_e} (E_e;E^\prime_{e},\epsilon,\mu) + \nonumber \\ 
& & \int dE_{\gamma}N_{\gamma} (E_{\gamma},t) \int d\epsilon\,n(\epsilon)
\int d\mu \frac{1-\mu}{2} \frac{d\sigma_{\rm PP}}{dE_e}
(E_e;E_{\gamma},\epsilon,\mu) + Q(E_e,t),\nonumber
\end{eqnarray}
where $N_e (E_e,t)$ is the (differential) number density of electrons at
energy $E_e$ at time $t$, $n(\epsilon)$ is the number density of background
photons at energy 
$\epsilon$, $Q(E_e,t)$ is an external source term for electrons at energy
$E_e$ and time $t$, 
$\mu$ is the interaction angle between the CR electron and the 
background photon ($\mu=-1$ for a head-on collision), and $\beta_e$ is the
velocity of the CR electron. The terms describe the loss of
electrons due to ICS, the influx of electrons scattered into the energy range
due to ICS, the influx of electrons produced due to PP by
photons, and the external injection. The factor
$(1-\beta_e \mu)/2$ is the flux factor. I define the angle averaged cross
sections $R(E,\epsilon)$ and $P(E^\prime;E,\epsilon)$ as
\begin{equation}
R(E,\epsilon) \equiv \int d\mu \frac{1-\beta \mu}{2} \sigma(E,\epsilon,\mu),
\end{equation}
and
\begin{equation}
P(E^\prime;E,\epsilon) \equiv \int d\mu \frac{1-\beta\mu}{2}
\frac{d\sigma}{dE^\prime}(E^\prime;E,\epsilon,\mu).
\end{equation}
Then Eq.~(\ref{step1}) is rewritten as
\begin{eqnarray}
\frac{d}{dt} N_e (E_e,t) & = &
-N_e (E_e,t)
\int d\epsilon\,n(\epsilon) R_{\rm ICS} (E_e,\epsilon) + 
\int dE^\prime_{e}N_e(E^\prime_{e},t)
\int d\epsilon\,n(\epsilon) P_{e,\rm ICS}
(E_e;E^\prime_{e},\epsilon) \nonumber \\
 & & + \int dE_{\gamma}N_{\gamma} (E_{\gamma},t)
\int d\epsilon\,n(\epsilon) P_{e,\rm PP}
(E_e;E_{\gamma},\epsilon)+Q(E_e,t).\label{step2}
\end{eqnarray}

In order to solve this differential equation numerically, we first bin the
energies of the CR electrons, CR photons, and background photons. We divide
each decade of energy into 20 equidistant logarithmic bins and call the
central value of 
the $i$-th bin $E_i$ and boundary values $E_{i-1/2}$ and $E_{i+1/2}$. And we
replace the continuous integrals by finite sums, 
and integrate Eq.~(\ref{step2}) over one CR energy bin. Then we get
\begin{eqnarray}
\frac{d}{dt} N_e^k & = &
- N_e^k \sum_j \Delta\epsilon_j n(\epsilon_j) R_{\rm ICS}^{kj}
+ \sum_i \sum_j N_e^i
\Delta\epsilon_j n(\epsilon_j) P_{e,\rm ICS}^{ijk} + \nonumber \\
& & \sum_i \sum_j N_{\gamma}^i
\Delta\epsilon_j n(\epsilon_j) P_{e,\rm PP}^{ijk} + Q^k,\label{diff}
\end{eqnarray}
where $N^i \equiv \int_{E_{i-1/2}}^{E_{i+1/2}} dE\,N(E,t)$, 
$R^{kj} \equiv
R(E_k,\epsilon_j)$, $P^{ijk} \equiv \int_{E_{k-1/2}}^{E_{k+1/2}}
dE\,P(E;E_i,\epsilon_j)$, $Q^i \equiv \int_{E_{i-1/2}}^{E_{i+1/2}}
dE\,Q(E,t)$, and $\Delta\epsilon_j \equiv \epsilon_{i+1/2}-\epsilon_{i-1/2}$.

I adopt a first order implicit scheme to solve this difference equation
(\ref{diff}); i.e. 
\begin{equation}
{N_e^k}^\prime = \frac{\frac{1}{\Delta t} N_e^k + \sum_{i \neq k} 
\sum_j {N_e^i}^\prime \Delta\epsilon_j n(\epsilon_j) P_{e,\rm ICS}^{ijk} +
\sum_{i,j} {N_{\gamma}^i}^\prime 
\Delta\epsilon_j n(\epsilon_j) P_{e,\rm PP}^{ijk} + Q^k} {\frac{1}{\Delta t} + 
\sum_j \Delta\epsilon_j n(\epsilon_j) (R_{\rm ICS}^{kj}-P_{e,\rm
ICS}^{kjk})},\label{step3}
\end{equation}
where ${N_e^k}^\prime$ is a solution advanced by a timestep
$\Delta t$ from $N_e^k$.
Eq.~(\ref{step3}) can be understood as follows: the second term in the
denominator corresponds to the {\it net\/} loss of the particles from
bin $k$. The second term in the numerator corresponds to the net {\it
scattered\/}
particle influx into bin $k$ due to scattering from other bins
which conserves the particle species. The third and the last terms in
the numerator describe the influx of the particles either due to production
by different particles or due to external injection. Various different
interactions can be included in this main equation according to the general
scheme laid out above, including certain energy losses which can
be treated as continuous such as synchrotron radiation for which
I use a simple first order upwind scheme.

The implicit method has the advantage that the solution converges for arbitrary
size of the timestep we take. Therefore, we are allowed to use a bigger
timestep than is allowed by an explicit Euler scheme. However, to ensure the
desired accuracy, we need to optimize the stepsize for a given problem by
trial and error. See Ref.~\cite{ck} for a more detailed discussion of this
implicit method.

It is important to monitor
conservation of particle numbers and total energy in order to
obtain reliable results. For example, for ICS
the coefficients should satisfy
\begin{equation}
R^{kj} = \sum_i P_e^{ijk} = \sum_i P_{\gamma}^{ijk}~~~~({\rm
number})\label{num}
\end{equation}
and
\begin{equation}
(E_k+\epsilon_j) R^{kj} = \sum_i E_i P_e^{ijk} + \sum_i E_i
P_{\gamma}^{ijk}~~~~({\rm energy}).\label{energy}
\end{equation}
It is sometimes necessary to adjust the coefficients in order to
obey these relations. This stabilizes the calculation against
growing errors due to discretization of variables.

In Eq.~(\ref{step3}) the coefficient matrices are multiplied
with and summed over the background photon spectrum vector
and/or the CR spectrum vector at a given redshift in order to
advance the solution. This procedure has the advantage
that one can deal with an arbitrary evolution of the radiation
backgrounds in time, which is important in this problem.
If one would integrate the background spectrum
into the coefficients beforehand, it would become
extremely difficult and time-consuming to handle an arbitrary
background evolution because one would have to recalculate the
coefficients at each redshift.

In Ref.~\cite{pj}, it was assumed that all radiation backgrounds exhibit a
trivial evolution by redshifting. This allowed them to adopt a matrix doubling
method~\cite{ps} for the propagation calculations. However, in case of a more
realistic background evolution, matrix doubling is almost
impossible. In contrast, our approach is always guaranteed to work efficiently
and is sufficiently accurate in the more general case.

Eq.~(\ref{step3}) is then solved iteratively by inserting the initial values
for ${N^i}^\prime$'s on the right hand side and re-inserting the
new values until a
convergence is achieved. Since there are four main
particle species (nucleons, photons, electrons, and neutrinos),
one should converge all spectra
simultaneously. However, it is economical and equally valid to
converge 
each particle spectrum separately while holding the others fixed,
and repeat this whole procedure until all spectra converge.

In addition, we account for redshifting by performing the
operation $N_a(E,z)\rightarrow[1+\Delta
z/(1+z)]^{-2}N_a\left(E[1+\Delta z/(1+z)],z\right)$ for each particle
species $a$ after a step $\Delta z$ in redshift. Here I
match $\Delta z$ conveniently with the logarithmic energy
bin size, $\log_{10}[1+\Delta
z/(1+z)]=\log_{10}(E_i/E_{i-1})=1/20$, which corresponds to the
transformation $N^i_a(z)\rightarrow[1+\Delta
z/(1+z)]^{-3}N^{i+1}_a(z)$.

The numerical code is a combination and extension of the codes developed
by Coppi \& K\"{o}nigl~\cite{ck}, and by Lee \& Sigl~\cite{ls}.

\section{Interactions of Relativistic Nucleons and $\gamma$-Rays}

In this section I discuss various relevant interactions and their cross
sections from which the coefficients $R$ and $P$ used in
the transport equations are calculated.

\subsection{``Cascade'' Photons}

Pair production (PP) and inverse Compton scattering (ICS) are the two main
processes that drive the EM cascade. First, let us define the inelasticity
which is the fraction of the energy that is transferred from the scattering
particle to the scattered (or produced). It is given by
\begin{equation}
\eta(s) \equiv 1-\frac{1}{\sigma_{\rm tot}(s)} \int d\epsilon^\prime
\epsilon^\prime\frac{d\sigma}{d\epsilon^\prime}(\epsilon^\prime,s),
\end{equation}
where $s$ is the squared center of mass (CM) energy and
$\epsilon^\prime$ is the energy of the recoiling (leading) particle in
units of the initial particle energy.

In the extreme Klein-Nishina limit where $s \gg m_e^2$, either
the electron or the positron produced in a pair production event
typically carries almost all of the initial
total energy. The produced electron (positron) then undergoes ICS,
and the 
inelasticity for ICS in this high energy limit is
more than 90 \%. Therefore, $e^- (e^+)$ loses most of its energy and the
background photon is upscattered with almost all of the initial energy of the
UHE photon. This cycle of the ``cascade'' photon is responsible for slowing
down the energy attenuation of the leading particle. In some
previous work it was 
incorrectly claimed that the UHE photons lose energy very fast based on the
fact that the {\it mean free path\/} of the UHE photon is fairly short, but this
sequence of PP and ICS makes the actual energy attenuation much
slower. In addition, the contribution of non-leading particles
to the flux which are neglected in the CEL approximation can be
substantial for cascades which are not fully developed. This will be
important in some of the
applications considered in this paper.

If the EGMF is present, however, the above scenario changes somewhat. In the
energy range where the synchrotron loss rate for the electrons is greater than
the ICS rate, the
development of the EM cascade is heavily suppressed. Its
penetration depth is basically reduced to the photon mean free
path in this energy regime. A more
detailed discussion is found in Ref.~\cite{los}.

\subsubsection{Pair Production}

The total cross section for PP ($\gamma \gamma_b \rightarrow e^- e^+$) is
well-known, and is given by 
\begin{equation}
\sigma_{\rm PP}=\sigma_T \cdot \frac{3}{16} (1-\beta^2) \left[ (3-\beta^4) \ln
\frac{1+\beta}{1-\beta} -2\beta (2-\beta^2) \right]
\end{equation}
where $\beta \equiv(1-4m_e^2/s)^{1/2}$ is the velocity of the outgoing
electron in the CM frame.
In order to calculate the differential cross section, I adopt the simplifying
approximation where the dependence of the cross section on the
azimuthal angle of the outgoing particle in the CR frame is
ignored. Since the terms that depend on the azimuthal angle is smaller by more
than $10^{-11}$ than the leading order terms, this approximation is very
accurate for practical purposes. 
The differential cross section for a photon of energy $E_\gamma$
to produce an electron of energy $E^\prime_e$ is then given by~\cite{zdz,cb}:
\begin{eqnarray}
\frac{d\sigma_{\rm PP}}{dE^\prime_e}=\sigma_T\cdot\frac{3}{4}\frac{m_e^2}{s}
\frac{1}{E_\gamma}
&\Biggl[&\frac{E^\prime_e}{E_\gamma-E^\prime_e} +
\frac{E_\gamma-E^\prime_e}{E^\prime_e}
+ E_\gamma (1-\beta^2) \left(
\frac{1}{E^\prime_e} + \frac{1}{E_\gamma-E^\prime_e} \right)\\
&&- \frac{E_\gamma^2 (1-\beta^2)^2}{4} \left(
\frac{1}{E^\prime_e} + \frac{1}{E_\gamma-E^\prime_e} \right)^2
\Biggr]\,, \nonumber
\end{eqnarray}
where the range is restricted to $(1-\beta)/2\leq
E^\prime_e/E_\gamma\leq(1+\beta)/2$.
The differential cross section with respect to the positron energy is
identical due to symmetry.

\subsubsection{Double Pair Production}

Double pair production (DPP; $\gamma \gamma_b \rightarrow e^-e^+e^-e^+$) is
a higher order QED process that affects the UHE photons. It is known
that the DPP total cross section is a sharply rising function of $s$ at the
threshold and approaches the asymptotic value quickly at $\sigma (\infty)
\simeq 6.45~\mu$b \cite{brown}. For interactions with the microwave
background, the DPP rate begins to dominate over the PP rate above $\sim
10^{21}{\,{\rm eV}}$. If we take the contribution of the radio
background into account, this energy goes up somewhat.

The differential cross sections of DPP may be obtained through second order
QED calculations, but it is extremely involved, and I could not find a
suitable reference in which the differential cross section is calculated. In
addition, since it is still a small sized effect, I think that introducing a
reasonable assumption about the differential cross section is adequate for our
purpose. Therefore, I use the assumption where one pair of the two carries
all the initial energy and two particles in the pair share the energy
equally. 
I presume that this assumption does not change the calculations in a
significant way.

In Fig.~4(a) we plot $\sigma(s)$ and
$\sigma(s)\eta(s)$ of which
the latter is proportional to the fractional energy loss rate of
the leading particle, for PP and DPP.

\subsubsection{Inverse Compton Scattering}

The total cross section for ICS ($e \gamma_b \rightarrow e \gamma$) is given
by the well-known Klein-Nishina formula: 
\begin{equation}
\sigma_{\rm ICS} = \sigma_T \cdot \frac{3}{8} \frac{m_e^2}{s\beta} \left[
\frac{2}{\beta (1+\beta)} (2+2 \beta-\beta^2 - 2 \beta^3) - \frac{1}{\beta^2}
(2-3\beta^2-\beta^3) \ln \frac{1+\beta}{1-\beta} \right]
\end{equation}
where $\beta \equiv (s-m_e^2)/(s+m_e^2)$ is the velocity of the outgoing
electron in the center of mass frame.
Most part of the energy range of interest is in the extreme
Klein-Nishina regime, but nonetheless I use the exact formula.

The differential cross section for an electron of energy $E_e$
to produce an electron of energy $E^\prime_e$ is then given
by~\cite{zdz,cb}:
\begin{equation}
\frac{d\sigma_{\rm ICS}}{dE^\prime_e} = \sigma_T \cdot
\frac{3}{8} \frac{m_e^2}{s} \frac{1}{E_e}
\frac{1+\beta}{\beta} \left[ \frac{E^\prime_e}{E_e} +
\frac{E_e}{E^\prime_e} +
\frac{2(1-\beta)}{\beta} \left(1-\frac{E_e}{E^\prime_e} \right) +
\frac{(1-\beta)^2}{\beta^2} \left( 1-\frac{E_e}{E^\prime_e} \right)^2
\right].\label{ics} 
\end{equation}
where the range is restricted to $(1-\beta)/(1+\beta)\leq
E^\prime_e/E_e\leq1$.
The differential cross section with respect to the energy
$E^\prime_\gamma$ of the outgoing photon is obtained
by substituting $E_e-E^\prime_{\gamma}$ for $E^\prime_e$ in
Eq.~(\ref{ics}).

\subsubsection{Triplet Pair Production}

Triplet pair production (TPP; $e \gamma_b \rightarrow e e^-
e^+$) is a rather significant contribution to the interactions
of UHE electrons. This process is discussed in detail in
Refs.~\cite{ck,mast1,haug,bors}.
Although the total cross section for TPP on CMB photons becomes
comparable to the ICS cross section already at $\sim 10^{17}
{\,{\rm eV}}$, the actual energy attenuation is not important until much higher
energies because the inelasticity is very small ($\la
10^{-3}$). Nonetheless,
it is fairly efficient in channelling the energy content to lower energies,
and may not be ignored.

I use the formulation given by \cite{haug} in calculating the total cross
section, and the detailed expressions are
given in Appendix A. The total cross section of TPP increases asymptotically
logarithmically with $s$:
\begin{equation}
\sigma_{\rm TPP} = \sigma_T \frac{3\alpha}{8\pi} \left[ \frac{28}{9} \ln
\frac{s}{m_e^2} - \frac{218}{27} \right]~~~~(s \gg m_e^2),\label{tpp}
\end{equation}
where $\alpha$ is the fine structure constant.

While it is possible to calculate the differential cross sections numerically
using the expressions given in Ref.~\cite{mast1,haug,bors}, it is extremely
time-consuming because it involves multi-dimensional integrations of very
complicated functions. Furthermore, some of the variables introduced
there become very large or very small, and hence create
problems with the finite computing precision.
The detailed behavior of the TPP cross sections near
threshold is unimportant since TPP is dominated by ICS in this
energy regime. Thus, it will suffice to use a
simple and efficient approximation that works very well for the region away
from the threshold.
First, I make note of the fact that the differential cross
section with respect to the energy of one of the particles of
the produced pair tends to $d\sigma/dE^\prime \propto
{E^\prime}^{-7/4}$ for $s\gg m^2_e$~\cite{mast1}. Furthermore,
in the same regime the inelasticity for TPP can be well
approximated by~\cite{mast1}
\begin{equation}
\eta(s) \simeq 1.768 (s/m_e^2)^{-3/4}.
\end{equation}
I then make the assumption that the differential rate $P(E'; E,\epsilon)$ for
the produced particle with energy $E'$ and for the incoming electron with
energy $E$ and the incoming background photon with energy $\epsilon$ is given
as a power law with spectral index $\delta$:
\begin{equation}
P(E'; E,\epsilon) = R(E,\epsilon) C(E,\epsilon) E'^{-\delta},
\end{equation}
where $C(E,\epsilon)$ is a normalization factor.
Then using the requirement that the integrated differential rates must be the
same as the total rate and energy conservation [i.e. the analogues of
Eqs.~(\ref{num}) and
(\ref{energy})], one can uniquely determine the coefficient $C(E,\epsilon)$ and
$\delta$. The spectral index $\delta$ approaches $7/4$ for large $s$.

For the recoiling electron, on the other hand, I may assume
continuous energy loss whose rate is given by
\begin{equation}
\frac{dE}{dt}\simeq- E \int d\epsilon\,n(\epsilon) R(E,\epsilon) \eta(s).
\end{equation}

The importance of TPP again depends on the presence and the strength of the
EGMF. If the EGMF is stronger than about $10^{-12}$ G, then TPP energy loss is
dominated by synchrotron cooling, and it is no longer very important. Since
various arguments and indirect measurements of the EGMF~\cite{magnet}
suggest that EGMF is at least $10^{-12}$ G, TPP may not play a big role in the
propagation of UHE photons. However, in the absence of the EGMF, the
contribution of TPP to the energy attenuation of electrons and photons is
comparable to or even greater than ICS above $\sim10^{22}{\,{\rm eV}}$
and thus may not be ignored.

In Fig.~4(b) we plot $\sigma(s)$ and
$\sigma(s)\eta(s)$ of which
the latter is proportional to the fractional energy loss rate of
the ``leading particle'', for ICS and TPP.
Fig.~5 shows all the rates at redshift $z=0$ that affect the
photons and electrons in the energy range we consider.

\subsubsection{Other Processes}

Other interactions that are neglected in this paper are all processes
involving the productions of one or more $e^-e^+$ pairs substituted by muon,
tau lepton, and pion pairs, double Compton scattering
($e \gamma_b \rightarrow e\gamma\gamma$), $\gamma\gamma$
scattering ($\gamma\gamma_b\rightarrow\gamma\gamma$),
Bethe-Heitler pair production ($\gamma X\rightarrow Xe^+e^-$,
where $X$ can be an atom, an ion, or a free electron), the
process $\gamma\gamma_b\rightarrow e^+e^-\gamma$, and pair
production on a magnetic
field ($\gamma B \rightarrow e^- e^+$). The total cross section of single muon
pair production ($\gamma \gamma_b \rightarrow \mu^- \mu^+$), for
example, is smaller than 
electron pair production by about a factor of 10.
Energy loss rates for TPP involving heavier
pairs are suppressed by a factor $\simeq(m/m_e)^{1/2}$ in the
limit of large $s$. Similarly, double pair
production involving heavier pairs is also negligible
\cite{brown}. The double Compton scattering cross section is of
order $\alpha^3$ and must be treated together with the radiative
corrections to ordinary Compton scattering of the same
order. The corrections to the lowest order ICS cross
section by processes involving $m_\gamma$ additional photons in the final
state, $e\gamma_b\rightarrow e+(m_\gamma+1)\gamma$,
$m_\gamma\geq1$, turn out
to be less than 10\% in the energy range under
consideration~\cite{Gould}. A similar remark applies to
corrections to the lowest order PP cross section by
the processes $\gamma\gamma_b\rightarrow e^+e^-+m_\gamma \gamma$,
$m_\gamma \geq 1$.
Photon-photon scattering can only play a role for $z\ga100$ and
energies below the redshift dependent pair production threshold
Eq.~(\ref{Ethr})~\cite{szdz,zdzs}. A similar remark applies to
Bethe-Heitler pair production on atoms, ions and free
electrons~\cite{zdzs}.
Pair production on a magnetic field of order $10^{-6}{\,{\rm G}}$ which
is typical for the field of Our Galaxy, is only relevant for
$E \ga 10^{24} {\,{\rm eV}}$. This critical energy is even higher for the
EGMF and this process is thus negligible in the analysis.

\subsection{Nucleons}

There are three major processes that affect the propagation of protons and
neutrons: Electron-positron pair production by protons (PPP; $p
\gamma_b \rightarrow pe^-e^+$), photopion production ($N
\gamma_b \rightarrow N(n\pi),n\ge1$), and neutron $\beta$-decay
($n\rightarrow pe^-\bar\nu_e$).

\subsubsection{Pair Production by Protons}

PPP provides the main energy attenuation for protons with energies below the
GZK cutoff~\cite{blum}. The energy threshold for this process is
\begin{equation}
E_{\rm th} = \frac{m_e (m_N + m_e)}{\epsilon} \simeq 4.8 \times 10^{14}
\left(\frac{\epsilon}{{\,{\rm eV}}} \right)^{-1} {\,{\rm eV}}.
\end{equation}
Thus, for a microwave background photon ($\epsilon \sim 10^{-3} {\,{\rm
eV}}$), PPP 
ensues at a proton energy $E\sim 5 \times 10^{17} {\,{\rm eV}}$. Below this
energy, 
the protons cool essentially only by redshifting with the expansion.
 
The PPP total cross section behaves very similarly to that for
triplet pair production because PPP is almost identical to TPP~\cite{mok}, and
the expression for the total cross section away from the threshold may be
given by Eq.~(\ref{tpp}). However, while TPP near its threshold is
dominated by other processes,
the exact behavior of PPP rates near the threshold are very important
because PPP dominates the proton energy loss in that
energy range. I use the parametric fits given in
Ref.~\cite{czs} for the cross section and inelasticity. Then
I use the same approach in calculating the differential rates
as we did for TPP. It can be shown that these rates are well
approximated by a power law. On the other hand,
the proton spectrum evolution due to PPP is well described by
CEL because the inelasticities are smaller than $10^{-3}$ at all relevant
energies. Production of heavier pairs like $\mu^+\mu^-$ is
suppressed similarly to the case of TPP. The energy ranges for
the produced pairs and the
recoiling proton are given in Appendix A.

\subsubsection{Photopion Production}

Photopion production provides the main energy attenuation for
nucleons above $E \simeq 10^{19}{\,{\rm eV}}$.
The energy threshold for this process is
\begin{equation}
E_{\rm th} = \frac{m_Nm_\pi+m^2_\pi/2}{\epsilon} \simeq 6.8 \times 10^{16}
\left(\frac{\epsilon}{{\,{\rm eV}}} \right)^{-1} {\,{\rm eV}}.
\end{equation}
Thus, for a microwave background photon ($\epsilon \sim 10^{-3}
{\,{\rm eV}}$), photopion production ensues at a nucleon energy $E\sim7
\times 10^{19}{\,{\rm eV}}$.
Since publications on numerical studies of nucleon and
$\gamma$-ray propagation usually do not contain detailed
information on the implementation especially of multiple pion
production, I present our approach here in some detail. First,
I define a few suitable kinematic variables which depend only on
the incoming particles.
If $E_\gamma^{\rm lab}$ is
the photon energy in the laboratory frame (LF) where the nucleon
is at rest, $m_N$ is the nucleon mass, and $s$ is the squared
center of mass (CM) energy, then the following relations hold:
\begin{eqnarray}
  \kappa&\equiv& \frac{E_{\gamma}^{\rm lab}}{m_N} = \frac{\epsilon
E_N}{m_N^2} (1-\mu)\,,\nonumber\\
  s&=&m_N^2(1+2\kappa)\,.\label{kin}
\end{eqnarray}
Since laboratory measurements of cross sections are usually given
in terms of $E_\gamma^{\rm lab}$, I will conveniently express
everything in terms of $\kappa$ in the following.

Concerning single pion production I consider the following
reactions:
\begin{eqnarray}
  \gamma+N&\rightarrow&\pi^0+N\,:\,{d\sigma_1\over d\Omega^*}=
  \sum_{j=0}^Ka_{1j}(\kappa)\left(x^*\right)^j\,,\label{cr}\\
  \gamma+p&\rightarrow&\pi^++n\,:\,{d\sigma_2\over d\Omega^*}=
  \left(1-\beta^*x^*\right)^{-1/2}
  \sum_{j=0}^Ka_{2j}(\kappa)\left(x^*\right)^j\,,\label{ce1}\\
  \gamma+n&\rightarrow&\pi^-+p\,:\,{d\sigma_3\over d\Omega^*}=
  \left(1-\beta^*x^*\right)^{-1/2}
  \sum_{j=0}^Ka_{3j}(\kappa)\left(x^*\right)^j\,.\label{ce2}
\end{eqnarray}
The differential cross sections for these processes are
expressed here in terms of $\kappa$ and the CM quantities $\Omega^*$,
$\beta^*$, and $x^*$ which denote solid angle, pion velocity,
and the cosine of the scattering angle, respectively. The
functions $a_{ij}(\kappa)$, $i=1,2,3$ and $j=1,\cdots,N$ are
fitted to laboratory cross section data and I use fits up to
order $K=3$~\cite{Landolt}. The expressions in
Eqs.~(\ref{cr})-(\ref{ce2}) can be easily rewritten in terms
of the energies of the outgoing nucleons and pions in the
cosmic ray frame (CRF) which I denote by $E'_a$
for $a=p,n,\pi^+,\pi^-,\pi^0$. The relevant formulae are
be given in Appendix B.

Note that in Eq.~(\ref{cr}) I have assumed identical cross
sections for the two charge retention processes involving
protons and neutrons. This is a very good approximation
(see, e.g. Ref.~\cite{Landolt}). Reactions~(\ref{ce1}) and
(\ref{ce2}) constitute the charge exchange reactions for
single pion production.

We now turn to multiple pion production. Let us first
consider the channel $\gamma+p\rightarrow\pi^-+X$ where
$X$ stands for anything. This channel has been discussed
in detail in
Ref.~\cite{Moffeit}. There, the Feynman $x$-variable
$x=p_\|^*/p_{\rm max}^*$ was introduced, which is the
fraction of the pion parallel momentum $p_\|^*$ in the
CMF to its maximal value
\begin{equation}
  p_{\rm max}^*=m_N \frac{D}{(1+2\kappa)^{1/2}}
\,,\label{pmax}
\end{equation}
where $D \equiv [(\kappa-\varepsilon^2/2)^2-\varepsilon^2]^{1/2}$, and
$\varepsilon=m_\pi/m_N$ is the ratio of the pion and
nucleon mass. Denoting the transverse momentum with
$p_\bot$ and the $\pi^-$ energy in the CMF by $E_{\pi^-}^*$,
the differential cross section for $\pi^-$ production
was written in terms of a structure function
$f(x,p_\bot^2,s)$:
\begin{equation}
  d^2\sigma_{\gamma p\rightarrow\pi^-X}=\pi\,
  {p_{\rm max}^*\over E_{\pi^-}^*}\,f(x,p_\bot^2,s)\,
  dx\,dp_\bot^2\,.\label{dsig1}
\end{equation}
Performing Lorentz transformations into the CRF where the
proton and $\pi^-$ energies are $E_p$ and
$E'_{\pi^-}$ (see Appendix B), this can
be written as
\begin{eqnarray}
  {d\sigma_{\gamma p\rightarrow\pi^-X}\over dE'_{\pi^-}}
  &=&{2\pi m_Np_{\rm max}^*\over E_p}
  \left(1+2\kappa\right)^{1/2}\int_{x_{\rm min}}^{x_{\rm max}}dx
  \label{dsig2}\\
  &&\times f\left[x,-m_\pi^2+
  \left({E'_{\pi^-}m_N\over E_N}\right)^2(1+2\kappa)
  +2p_{\rm max}^*\left({E'_{\pi^-}m_N\over E_N}\right)
  (1+2\kappa)^{1/2}x,s\right]\,,\nonumber
\end{eqnarray}
where $x_{\rm min}\geq-1$ is chosen such that $p_\bot^2\geq0$
which is the second argument of $f$, and $x_{\rm max}$ such that
$p_\bot^2/(p_{\rm max}^*)^2 + x^2 \le 1$.

For $s\rightarrow\infty$ the structure function $f$ is
independent of $s$~\cite{Moffeit,Feynman}. I will therefore
neglect any $s$-dependence altogether. Furthermore, I
take into account these processes only above some threshold
which is sufficiently high such that the contribution of
single pion production is negligible; I take
$\kappa\geq2(\varepsilon+\varepsilon^2)$.

Finally, for our purposes I assume that the remaining
dependence of $f(x,p_\bot^2)$ factorizes into an $x$-dependent
part and an exponential dependence on $p_\bot^2$:
\begin{equation}
  f(x,p_\bot^2,s)\simeq{1\over\Lambda^2}\
  \exp[-p_\bot^2/\Lambda^2]\,f(x)\,.\label{fact}
\end{equation}
Here, $\Lambda\simeq{\rm GeV}/6.4$ is roughly of the order
of the QCD scale and $f(x)$ can be fitted to the data
presented in Ref.~\cite{Moffeit}.

Within these approximations we have finally
\begin{eqnarray}
  {d\sigma_4\over dE'_{\pi^-}}&\equiv&
  {d\sigma_{\gamma p\rightarrow\pi^-X}\over dE'_{\pi^-}}
  ={2\pi\over E_p}\left({m_N\over\Lambda}\right)^2
D
  \int_{x_{\rm min}}^{x_{\rm max}} dxf(x)\label{dsig3}\\
  &&\times\exp\left\{
  -\left({m_N\over\Lambda}\right)^2\left[-\varepsilon^2+
  \left({E'_{\pi^-}\over E_p}\right)^2(1+2\kappa)
  +2D
  \left({E'_{\pi^-}\over E_p}\right)x\right]\right\}
  \,,\nonumber
\end{eqnarray}
where $x_{\rm min}$ and $x_{\rm max}$ are given by
\begin{eqnarray}
x_{\rm min}&=&\max\left[-1,\frac{\varepsilon^2-(E'_{\pi^-}/E_p)^2(1+2\kappa)}
{2D(E'_{\pi^-}/E_p)}\right]\,,\nonumber\\
x_{\rm max}&=&
-\frac{E'_{\pi}}{E_p} \frac{1+2\kappa}{D}
+ \left(1+
\varepsilon^2 \frac{1+2\kappa}{D^2} \right)^{1/2}
\,.\label{minmax}
\end{eqnarray}

At this point it is important to realize that
$f(x)=f_c(x)+f_\rho(x)$ can be divided into a contribution
$f_c(x)$ from the ``central" pions and a contribution
$f_\rho(x)$ from production of multiple $\rho_0$ mesons
(sometimes also called the leading pion contribution)
which subsequently decay into equal distributions of
$\pi^+$ and $\pi^-$. Therefore, $f_\rho(x)$ exclusively
contributes to the production of $\pi^+$ and $\pi^-$
and corresponds to a charge retention process where
the nature of the nucleon is unchanged. In contrast,
$f_c(x)$ describes a process resulting in approximately
equal distributions of $\pi^0$ and $\pi^-$ with the
probability for change of nucleon isospin being about
$2/5$ (from simple quark counting).

From these assumptions it follows immediately that
$d\sigma_5/dE'_{\pi^0}\equiv d\sigma_{\gamma p\rightarrow
\pi^0X}/dE'_{\pi^0}$ is obtained by substituting
$f_c(x)$ for $f(x)$ in Eq.~(\ref{dsig3}).

In addition, I assume that inclusive and leading pion
production takes place with the approximately constant
cross sections $\sigma_{\rm tot}\simeq127$ $\mu{\rm b}$
and $\sigma_\rho\simeq 21.5$ $\mu{\rm b}$, respectively. We
can then define the average central $\pi^-$ multiplicity
by
\begin{equation}
  \left\langle n_{\pi^-}^c\right\rangle(\kappa)=
  {\pi\over\sigma_{\rm tot}}\int dp_\bot^2\int dx
  f_c(x){\exp[-p_\bot^2/\Lambda^2]\over\Lambda^2}
  \left[x^2+{1+2\kappa\over D^2}
\left({p_\bot^2\over m_N^2}+\varepsilon^2
  \right)\right]^{-1/2}\,.\label{multpi}
\end{equation}
The integration range is determined by $p_\bot^2/(p_{\rm max}^*)^2+x^2
\le 1$.
By evaluating this formula one can see that the multiplicity
$\left\langle n_{\pi^-}^c\right\rangle$ increases
asymptotically logarithmically in $s$ for
$s\rightarrow\infty$.
Applying charge conservation to the central pion distribution,
making the above assumptions and in addition assuming $\pi^+$
and $\pi^-$ distributions to be proportional to each other
uniquely determines $d\sigma_6/dE'_{\pi^+}\equiv
d\sigma_{\gamma p\rightarrow\pi^+X}/dE'_{\pi^+}$. It is
obtained by substituting $\left[1+(2/5)(1-\sigma_\rho
/\sigma_{\rm tot})/\left\langle n_{\pi^-}^c\right\rangle
(\kappa)\right]f_c(x)+f_\rho(x)$ for $f(x)$ in
Eq.~(\ref{dsig3}).

It is now easy to compute the fractions $r_c(\kappa)$ and
$r_\rho(\kappa)$ of the incoming nucleon energy which go
into the central and leading pions, respectively
(see Appendix B). Fig.~6 shows these fractions and
the central and total $\pi^-$ multiplicities as functions of
$s$. Assuming a flat distribution for the
outgoing nucleons, we then have
\begin{eqnarray}
  {d\sigma_{p\rightarrow p}\over dE'_p}&=&
  {\sigma_\rho\over2r_\rho(\kappa)E_p}\,\Theta\left[
  {E'_p\over E_p}-1+2r_\rho(\kappa)\right]+{3\over5}\,
  {\sigma_{\rm tot}-\sigma_\rho\over2r_c(\kappa)E_p}
  \,\Theta\left[{E'_p\over E_p}-1+2r_c(\kappa)\right]
  \,,\nonumber\\
  {d\sigma_{p\rightarrow n}\over dE'_n}&=&{2\over5}\,
  {\sigma_{\rm tot}-\sigma_\rho\over2r_c(\kappa)E_p}
  \,\Theta\left[{E'_n\over E_p}-1+2r_c(\kappa)\right]
  \label{dsig4}
\end{eqnarray}
for the charge retention and charge exchange cross sections,
respectively. For the processes involving an incoming
neutron, I assume that the cross sections are also given
by the above expressions after substituting
$p\leftrightarrow n$ and $\pi^+\leftrightarrow\pi^-$
everywhere. Fig.~7 shows the differential cross sections for
production of $\pi^-$, $\pi^+$, $\pi^0$, protons, and neutrons
for an incoming proton for two different CM energies resulting
from the formalism adopted above. Fig.~8 shows the inclusive
pion production cross sections for nucleons as a
function of $s$. Fig.~9 shows all the rates at redshift $z=0$
that affect the nucleons in the energy range we consider.

Pions produced by nucleons quickly decay to EM particles and neutrinos and
feed the 
EM cascade. $\pi^0$ decays into photons ($\pi^0 \rightarrow \gamma \gamma$),
and $\pi^{\pm}$ decays to produce electrons, positrons, and neutrinos
($\pi^{\pm} \rightarrow \mu^{\pm} \nu_{\mu} (\bar{\nu}_{\mu});  \mu^{\pm}
\rightarrow e^{\pm} \nu_e (\bar{\nu}_e) \bar{\nu}_{\mu}
(\nu_{\mu})$)~\cite{HS2}. Since
the decay time of pions is very short compared to the timescale in the
problem, I assume that pions are converted into secondary particles
instantaneously. The decay spectra of the secondary particles may be
calculated 
easily \cite{Gaisser}. The expressions for the decay
spectra are given in Appendix C.

\subsubsection{Neutron $\beta$-decay}

Below $\sim10^{20}{\,{\rm eV}}$, neutron $\beta$-decay is the fastest
process among the interactions that 
affect nucleons in the problem. The neutron decay rate is $\Gamma = \Gamma_0
/\gamma_n = 1/\tau_n \gamma_n$, where $\tau_n$ is the neutron lifetime ($\tau_n
\simeq 888.6 \pm 3.5$ sec), and $\gamma_n$ is the neutron
Lorentz factor. The range $R_n$ of a neutron is given as
\begin{equation}
R_n \simeq\frac{c}{\Gamma} = c\tau_n\gamma_n \simeq 0.9 \left(
\frac{E_n}{10^{20} {\,{\rm eV}}} \right) {\,{\rm Mpc}}.
\end{equation}

In calculating the spectrum of secondary particles, I neglect 
the proton kinetic energy in the neutron rest frame, as is usually done. The
result can be found in standard textbooks such as Ref.~\cite{hm}.

\smallskip

Fig.~10 shows the energy attenuation lengths for cascade photons
and nucleons as functions of energy in the CEL approximation.

\section{Comparison with other Work}
Before I apply the propagation code to specific HECR injection
scenarios in the next section, I compare the predicted
spectra with results from other investigations for some standard
situations. For a discrete source producing a differential
injection spectrum $F_a(E)$ of particle type $a$ (in units of
number per energy per time) at redshift $z=z_i$, we obtain the
spectrum $j_a(E)$ (in units of number per area per time per
solid angle per energy) observed at $z=0$ in the following way:
I impose the boundary condition
\begin{equation}
  N_a(E,z_i)={(1+z_i)^2\over r_i^2}F_a(E)\,,\label{ssource}
\end{equation}
where $r_i$ is the comoving dimensionful source distance
corresponding to redshift $z_i$
($r_i=2H_0^{-1}[1-(1+z_i)^{-1/2}]$ in our cosmology), and
solve the propagation equations for vanishing source terms. If
we denote the resulting distribution at $z=0$ by $N_a(E)$, then
$j_a(E)=N_a(E)/(4\pi)$ and the modification factor $M_a(E,z_i)$,
defined as in Ref.~\cite{bg,Rachen}, is given by
\begin{equation}
  M_a(E,z_i)\equiv{4\pi d_L^2j_a(E)\over F_a(E)}=
  (1+z_i)^4{N_a(E)\over N_a(E,z_i)}\,,\label{modi}
\end{equation}
where I used the luminosity distance $d_L\equiv
r_i(1+z_i)$~\cite{Kolb}.

In Fig.~11 I plot the modification factors as
defined in Ref.~\cite{Rachen} for discrete sources injecting
protons with a power law at a given distance along with the
corresponding curves from Ref.~\cite{pj}. It can be seen that
our results lie somewhat between results from Refs.~\cite{pj}
and~\cite{yt}. In Fig.~12 I compare the nucleon,
$\gamma$-ray and neutrino fluxes computed
for monoenergetic proton injection at a given distance
with results from Ref.~\cite{pj}. In our prediction the
secondary $\gamma$-ray flux at the low energy side is higher than
the one given in Ref.~\cite{pj} by a factor $\simeq10$. I
attribute that to the fact that the differential multiple pion
production cross section used in our analysis (see Fig.~7) peaks
at low energies. In Fig.~13 I consider the case of power law
injection by a single source and compare
the nucleon and $\gamma$-ray fluxes with corresponding results in
Ref.~\cite{yt}. The nucleon fluxes agree well, whereas, again, our
prediction for the $\gamma$-ray flux is higher at the low energy
side and lower at the high energy side. Since Refs.~\cite{yt,pj}
do not give detailed information on their treatment of pion
production, it is hard to give an exhaustive explanation of
these differences. This, however, will not have an influence on
our considerations where secondary $\gamma$-ray production by
nucleons plays a minor role.

\section{Application to Models of HECR Origin}
We are now in a position to compute the cosmic and $\gamma$-ray fluxes
predicted by various models of HECR origin. Since there is
currently no unambiguous information on HECR composition, I
will normalize the predicted sum of
$\gamma$-ray and nucleon fluxes to the observed HECR flux. This is
done to optimally enable an explanation for the events above
$10^{20}{\,{\rm eV}}$ without overshooting the UHE flux at lower energies
(which might be explained by more conventional components) or
predicting an excessive integral flux above $10^{20}{\,{\rm eV}}$. I
estimate the uncertainty in the predicted $\gamma$-ray flux at lower
energies induced by this normalization procedure to be less than
a factor $\simeq3$.

\subsection{GUT Scale Physics Models}
As already mentioned in the introduction, it has been
suggested that HECRs may have a
nonacceleration origin~\cite{Hill,HSW,Bh0,Bh1,BR,bhs,BS,Sigl1} such as
the decay of supermassive elementary ``X'' particles
associated with Grand Unified Theories (GUTs), for example.
These particles could be radiated from topological
defects (TDs) formed in the early universe during phase transitions
caused by spontaneous breaking of symmetries implemented in
these GUTs (for a review on TDs, see~\cite{Vilenkin}). This is
because TDs, such as ordinary and
superconducting cosmic strings, domain walls and magnetic monopoles,
are topologically stable but nevertheless can release part of
their energy in the form of these X particles due to physical processes
like collapse or annihilation. The corresponding injection rate
of X particles $dn_X/dt$ as a function of cosmic time $t$ is
usually parametrized as
\begin{equation}
  {dn_X\over dt}\propto t^{-4+p}\,,\label{Xrate}
\end{equation}
where $p\geq0$ depends on the evolution of TDs. For example, X
particle release from a network of ordinary cosmic strings in
the scaling regime would correspond to $p=1$ if one assumes that
a constant fraction of the total energy in closed loops goes
into X particles~\cite{Bh0,BR}. Annihilation of magnetic monopoles
and antimonopoles~\cite{Hill,BS} predicts $p=1$ in the matter dominated and
$p=3/2$ in the radiation dominated era~\cite{SJSB} whereas the simplest
models for superconducting cosmic strings lead to
$p=0$~\cite{HSW}. A constant comoving injection rate corresponds
to $p=2$ and $p=5/2$ during the matter and radiation dominated
era, respectively.

The X particles with typical GUT scale
masses $m_X$ of the order of $10^{16}{\,{\rm GeV}}$ subsequently
decay into leptons
and quarks. The strongly interacting quarks fragment into a
jet of hadrons which results in mesons and baryons that are typically of the
order of $10^4 - 10^5$.
It is assumed that these hadrons then give rise
to a substantial fraction of the HECR flux
as well as a considerable neutrino flux.

The shapes of the nucleon and $\gamma$-ray spectra predicted
within such TD models
are thus expected to be universal (i.e., independent of the specific
process involving any specific kind of TD) at ultrahigh energies
and to be dependent only on the
physics of X particle decay. This is because at HECR energies
nucleons and $\gamma$-rays
have attenuation lengths in the cosmic microwave background (CMB)
which are small compared to the Hubble scale. Cosmological evolutionary
effects which depend on the specific TD model and are usually
parametrized by Eq.~(\ref{Xrate}) are therefore negligible.
In contrast, the predicted neutrino flux and the $\gamma$-ray
flux below the pair production threshold on the CMB [see
Eq.~(\ref{Ethr})] depend on the energy release integrated over
redshift and thus on the specific TD model.

I now discuss the particular form of the particle injection
spectra expected from
X particle release. I assume that each X-particle decays into a lepton
and a quark each of an energy approximately half of the X particle
mass $m_X$. For reasonable extragalactic field strengths, the
lepton (which I assume to be an electron in the following) will
quickly be degraded by synchrotron loss producing synchrotron
photons of a typical energy given by Eq.~(\ref{Esyn}). This energy is
typically much smaller than $10^{20}{\,{\rm eV}}$ where the resulting
contribution to the $\gamma$-ray flux is likely to be buried
below the charged CR flux. For that
reason, the GUT-scale lepton was usually omitted. However, for
high EGMF strengths the synchrotron peak can approach
$10^{20}{\,{\rm eV}}$ and thus could become relevant. For the present
analysis I will thus include the source term for the GUT-scale
lepton by writing its injection flux at energy $E$ and time $t$
as
\begin{equation}
\Phi_e(E,t)={dn_X(t)\over dt}\,\delta(E-m_X/2)\,,\label{inje}
\end{equation}
in units of particles per volume per time per energy.

The quark from X particle decay hadronizes by jet fragmentation and 
produces nucleons, $\gamma$-rays and neutrinos, the latter two from the decay 
of neutral and charged pions in the hadronic jets.
The hadronic
route is expected to produce the largest number of particles.
The resulting effective injection spectrum for particle species
$a$ from the hadronic channel can be written as  
\begin{equation}
\Phi_a(E,t)={dn_X(t)\over dt}{2\over m_X}{dN_a(x)\over dx}\,,\label{injh}
\end{equation}
where $x\equiv2E/m_X$, and 
$dN_a/dx$ is the effective fragmentation function describing the 
production of the 
particles of species $a$ from the original quark.

The spectra of the hadrons in a jet produced by the quark are, in principle, 
given by quantum chromodynamics (QCD). Suitably parametrized QCD motivated
hadronic spectra that fit well the data in collider experiments in the
GeV--TeV energies have been suggested in the literature~\cite{Hill}.
The {\it total\/} hadronic fragmentation spectrum 
$dN_h/dx$ is taken to be of the form~\cite{Hill}
\begin{equation}
{dN_h(x)\over dx}=\left\{\begin{array}{ll}
             {15\over16}x^{-1.5}(1-x)^2 & \mbox{if $x_0\leq x\leq1$}\\
             0 & \mbox{otherwise}\end{array}\right. \,,\label{frag}
\end{equation}
where the lower cutoff $x_0$ is typically taken to correspond to a cut-off 
energy $\sim1{\,{\rm GeV}}$. The spectrum Eq.~(\ref{frag}) obeys
energy conservation, $\int_{x_0}^1dxx(dN_h(x)/dx)=1$.
Assuming a nucleon content of
$\simeq3\%$ and the rest equally distributed among the three
types of pions, we can write the fragmentation spectra
as~\cite{bhs,abs}
\begin{eqnarray}
{dN_N(x)\over dx}&=&(0.03)\,{dN_h(x)\over dx}\,,
\label{frag1}\\
{dN_{\pi^+}\over dx}&=&{dN_{\pi^-}\over dx}=
{dN_{\pi^0}\over dx}=\left({0.97\over3}\right)
{dN_h(x)\over dx}\,.\nonumber 
\end{eqnarray}
From the pion injection spectra one gets the resulting
contribution to the injection spectra for $\gamma$-rays, electrons and
neutrinos by applying the formulae in Appendix C.

Independently of the spectral shapes of the predicted nucleon and $\gamma$-ray
fluxes, the question for the absolute normalization of the
injection rates $dn_X/dt$ in Eqs.~(\ref{Xrate}), (\ref{inje}) and
(\ref{injh}) arises. It has been shown,
for example for cosmic strings~\cite{Bh0,BR} and annihilation
of magnetic monopoles and antimonopoles~\cite{BS}, that at
least some TD models are capable of producing an observable HECR
flux if reasonable parameters are adopted. For the purposes of
this paper I will therefore not consider this issue and simply
adopt the normalization procedure mentioned above.

TD models of HECR origin are subject to a variety of constraints
mostly of cosmological nature. These are mainly due to the comparatively
substantial predicted energy injection at high redshift [see
Eq.~(\ref{Xrate})]. Note that more conventional CR sources like
galaxies start to inject energy only at a redshift of a few.
Using an analytical approximation for the
cascade spectrum below the pair production threshold on the CMB
resulting from X particle injection, one can derive constraints
from cascading nucleosynthesis and light element abundances,
CMB distortions, and the measured $\gamma$-ray
background~\cite{Fichtel,Digel,Osborne} in
the $100{\,{\rm MeV}}$ region~\cite{SJSB}, as well as from
observational limits on the $\gamma$-ray to charged CR flux
ratio between $10^{13}{\,{\rm eV}}$ and
$10^{14}{\,{\rm eV}}$~\cite{Karle,Sigl1}. The $100{\,{\rm MeV}}$
$\gamma$-ray background constraint was
first discussed in Refs.~\cite{Wolfendale1,hpsv}. In the context
of top-down models it was applied in Refs.~\cite{Chi1,Chi2} on
the basis of analytical approximations.

In addition, there has been a claim
recently~\cite{pj} that TD models might be ruled out
altogether due to overproduction of $\gamma$-rays in the range
between the knee and $\sim10^{19}{\,{\rm eV}}$. This
would occur for EGMFs stronger than
about $10^{-10}{\,{\rm G}}$ due to synchrotron radiation from the electronic
component of the TD induced flux which was normalized to the
observed flux at $3\times10^{20}{\,{\rm eV}}$. However, in my opinion,
the argument in Ref.~\cite{pj} suffers from several shortcomings: First,
monoenergetic injection of protons and $\gamma$-rays was used
instead of the more realistic injection spectra such as the ones
discussed above
in Eqs.~(\ref{frag}) and (\ref{frag1}). And second, only the
case of a single, discrete TD source at a fixed distance from
the observer was considered instead of more realistic source distributions and
evolution histories. Finally, electron deflection due to the EGMF which can
influence the processed spectrum from a single source was neglected.
Nevertheless, I simulated the situation of
Fig.~13 in Ref.~\cite{pj} for an EGMF of $10^{-9}{\,{\rm G}}$ on which
their claim is based.
As a result I got a spectrum whose shape is roughly similar to Fig.~13 in
Ref.~\cite{pj}, but the details of the spectrum differed somewhat, part of
which can be attributed to a different model of the radiation background. The
synchrotron peak I got was about an order of magnitude lower than in
Ref.~\cite{pj} relative to other parts of the spectrum. Most
importantly, however, we observe that if the spectrum is normalized to the
highest energy event this model would predict simply too many events above
$10^{20}\,$eV including the original injection peak. Therefore, the model
adopted in Ref.~\cite{pj} is not a realistic model for UHE CRs to start with.
I thus conclude that it is not possible to rule out
TD models on the basis of the discussion in Ref.~\cite{pj}.

Our goal here is to reexamine the constraints
based on the predicted $\gamma$-ray flux in the
regimes around $100{\,{\rm MeV}}$, between $10^{13}{\,{\rm eV}}$ and
$10^{14}{\,{\rm eV}}$, and between the knee and $10^{19}{\,{\rm
eV}}$, using our numerical techniques discussed in the previous section,
I base this on realistic injection spectra and histories as
discussed above. To
my knowledge, this has not been done yet despite its
importance for making $\gamma$-ray flux based constraints more
reliable.

The redshift range of energy injection contributing to the
$\gamma$-ray flux at energy $E_\gamma$ today is given by
$1+z\la\left(E_{\rm th}(z=0)/E_\gamma\right)^{1/2}$ where
$E_{\rm th}(z=0)$ is the PP threshold on the CMB at $z=0$ [see
Eq.~(\ref{Ethr})]. Since our interest is in the $\gamma$-ray flux at
$E_\gamma\ga100{\,{\rm MeV}}$, I maximally integrate up to
$1+z_{\rm max}=10^3$. The spectrum in this energy range converges before we
reach this redshift.
A word of caution is in order for the predicted neutrino spectra.
The UHE neutrinos interact with the universal neutrino background with $T_\nu
\sim 
1.95\,$K, and produce $l \bar{l}$ where $l=e,\mu,\tau,\nu,q,\ldots$ via $Z_0$
resonance~\cite{bhs,Yoshida}. The decay products of $\mu, \tau, q$ contain
secondary neutrinos. Here I
consider only simple absorption of UHE neutrinos, i.e.~I
integrate up to the average absorption redshift $z_{\rm abs}$ due to this
interaction~\cite{bhs}. The neutrino spectra
also converge rather fast with increasing redshift for the parameters I used
for TD models. Furthermore, the modification to
the neutrino spectra due to the cascading by the aforementioned interaction is
expected to be small for these parameters~\cite{Yoshida}. Therefore, the
neutrino spectra given in this paper are expected to be good approximations to
the real converged spectra. 
I leave a more detailed calculation of the UHE neutrino
flux to future work.

I performed simulations assuming uniform
injection rates given by Eqs.~(\ref{inje})-(\ref{frag1}) for
$m_X=10^{23}{\,{\rm eV}}$ and an
injection history given by Eq.~(\ref{Xrate}) for $p=1$
(representative of scenarios based on ordinary cosmic strings
and monopole-antimonopole annihilation) and for constant
comoving injection ($p=2$). Fig.~14 shows the results for a
negligible EGMF and assuming our IR/O background model. Note that
for a vanishing
EGMF the $\gamma$-ray flux dominates the
nucleon flux at UHEs and is higher by about an order of magnitude
compared to predictions within the CEL
approximation [compare Figs.~14(a) and~15]. This is due
to the influence of non-leading particles on the
development of the EM cascade.
Fig.~16 shows the dependence of the results on the EGMF
and the IR/O background. For an EGMF strength
$\ga10^{-11}{\,{\rm G}}$, the $\gamma$-ray flux is determined by
photon absorption and is thus harder. It is suppressed below a few
$10^{20}{\,{\rm eV}}$ and dominates at higher energies which is
in contrast to the case of a negligible EGMF [compare
Figs.~14(a) and~16(a),(b)]. This scenario has
the potential of explaining a possible gap in the HECR
spectrum~\cite{slsb}.
On the other hand, the neutrino flux is typically at least one order of
magnitude larger than the other components. However, we note that the
probability that these UHE neutrinos generate a shower in the atmosphere is 
smaller than $10^{-5}$~\cite{Elbert}.
We also find that the predicted
integral neutrino flux above $\simeq 10^{20}\,$eV is about $10^4$ times
smaller than current limits from the Fr\'ejus experiment~\cite{Frejus}. 

In the case of a negligible EGMF and absence of an IR/O background
[see Fig.~16(c)], our normalization procedure leads to a
$\gamma$-ray background below $\sim10^{14}{\,{\rm eV}}$ which
is about 20 times lower than analytical estimates
adopting a normalization based on the CEL
approximation for the $\gamma$-ray component
alone~\cite{SJSB}. This is caused by the aforementioned
influence of the non-leading EM particles on the UHE flux on
which the normalization depends. For given backgrounds,
EGMF strength, and flux normalization at HECR energies, the
$\gamma$-ray background flux below $\sim10^{14}{\,{\rm eV}}$
is proportional to the total energy injection which
increases monotonically with decreasing $p$. Comparison with the
$\gamma$-ray background observed around $100{\,{\rm MeV}}$~\cite{Fichtel}
and recently up to $\simeq 10{\,{\rm GeV}}$~\cite{Digel,Osborne} clearly
rules out the cases with $p\la1$ within our IR/O background
model and negligible EGMF [see Fig.~14(a)]. For an EGMF near its
currently believed upper limit $\simeq10^{-9}{\,{\rm G}}$~\cite{magnet},
proper normalization of the different predicted spectral shape
at UHEs leads to an increase of the predicted low energy
$\gamma$-ray background by about a factor 5, thus tightening the
constraint somewhat (Fig.~16).
The $\gamma$-ray flux level between $\sim10^{11}{\,{\rm eV}}$ and
$\sim10^{14}{\,{\rm eV}}$ is very sensitive to the IR/O background, and in
the extreme case of absence of any IR/O flux it
increases by about a factor $10^2$ relative to the level
predicted by our IR/O background model. At the same time,
the flux below $\simeq10{\,{\rm GeV}}$ goes down by about a
factor of 10 for vanishing IR/O flux [see Fig.~16(c)]. 
I stress that in any case, the scenarios considered here are
currently neither
constrained by the limit on the $\gamma$ to charged CR flux
ratio below $100{\,{\rm TeV}}$~\cite{Karle}, nor by the synchrotron peak
between the knee and $\sim10^{20}{\,{\rm eV}}$. 

Analytical arguments~\cite{SJSB} suggests that for a
given normalization of the spectra at the highest energies and a
given injection history, the
total injected energy and thus the $\gamma$-ray flux below the
PP threshold is roughly proportional to $m_X^{2-q}$. Here, a
fragmentation function $N_a(x)$ was assumed which is roughly proportional
to $x^{-q}$ for $x\ga E_{\rm obs}/m_X$, i.e. $q=1.5$ in the case
of Eq.~(\ref{frag}).
This allows one to rescale the above constraints to different values of
$m_X$. For example, for $p=1$ the constraint on $q$ is roughly
\begin{equation}
  q\ga2-\frac{3/2}{3+\log\left(m_X/10^{23}{\,{\rm eV}}\right)}
  \,.\label{qconstr}
\end{equation}
However, the effect of the cascading and the EGMF complicate the problem
considerably because the UHE spectrum depends sensitively on those effects.
More accurate estimates can be achieved only by a separate
numerical simulation of the case
$m_X\gg10^{23}{\,{\rm eV}}$. I leave that to a forthcoming
letter which will summarize the results~\cite{LSC}.

In order to mitigate or avoid overproduction of the $\gamma$-ray
background, based on analytical considerations of the cascade
spectrum, Chi {\it et al.}~\cite{Chi2} have recently suggested
somewhat different injection spectra. In this case injection of
nucleons, $\gamma$-rays, and neutrinos is again given by
Eq.~(\ref{injh}) where the following fragmentation functions are
adopted:
\begin{equation}
{dN_N(x)\over dx}=A_N\,x^{-1.5}\,,\quad
{dN_\nu(x)\over dx}={dN_\gamma(x)\over dx}=A_\gamma\,x^{-2.4}
\,,\label{wolf1}
\end{equation}
with
\begin{equation}
{A_\gamma\over A_N}\simeq0.028\,
\left({10^{15}\,{\rm GeV}\over m_X}\right)^{0.9}\,.\label{wolf2}
\end{equation}
The condition~(\ref{wolf2}) comes from the
requirement~\cite{Chi2}
that the photon-to-nucleon $(\gamma/N)$ ratio at injection
at energy $E=10^{20}{\,{\rm eV}}$, 
i.e., at $x=2\times10^{20}\,{\rm eV}/m_X$ be $\simeq60$. The spectra 
Eq.~(\ref{wolf1}) are absolutely normalized such that the total
quark energy $m_X/2$ is injected between $E=5\times10^{19}{\,{\rm eV}}$ and
and $E=m_X/2$.

Fig.~17 shows results obtained by assuming the
fragmentation functions given by
Eqs.~(\ref{wolf1}), (\ref{wolf2}). The $100{\,{\rm MeV}}$
$-10{\,{\rm GeV}}$
$\gamma$-ray background constraint is basically unchanged
from the case of the QCD motivated injection spectra for
vanishing EGMF on which it depends more weakly. Note that
this scenario has the potential to explain a HECR spectrum
continuing beyond $10^{20}{\,{\rm eV}}$ without any break or
gap~\cite{slsb}.

\subsection{Gamma Ray Burst Models}
Recently, it has been suggested that UHE CR could be associated
with cosmological GRBs~\cite{Wax1,Vietri,Wax2}. This was mainly
motivated by an
apparent numerical coincidence: Assuming that each (cosmological) GRB
releases an amount of energy in the form of UHE CRs which is
comparable to the total $\gamma$-ray output normalized
to the observed GRB rate (about $10^{51}{\,{\rm erg}}$ per burst), the
predicted and the observed UHE CR flux at the Earth are
comparable. It should be mentioned, however, that it is not
clear whether constraints on cosmological GRB distributions
are consistent with HECR observations~\cite{Quashnock}.

In these models protons are accelerated to UHE via first order
Fermi acceleration. Since there are no firm predictions for the
injection spectrum, I assume the hardest possible spectrum
proportional to $E^{-2}$ up to a maximal energy of $10^{23}{\,{\rm eV}}$.
Furthermore, assuming a constant
comoving injection rate up to some maximal redshift $z_{\rm
max}$, we can write
\begin{equation}
  \Phi_p(E,t)\propto t^{-2}E^{-2}\,\Theta(z_{\rm max}-z)
  \Theta(10^{23}{\,{\rm eV}}-E)\,.\label{grbinj}
\end{equation}
The authors of Ref.~\cite{Wax2} pointed out that bursting
sources in combination with deflection of protons in the EGMF
could lead to UHE CR spectra with a time variability on a scale
of $\sim50{\,{\rm yr}}$. This might allow reasonable fits to the
observed HECR spectrum. However, I only consider the continuous injection of
CRs in this paper for illustrative purposes.
Since the $\gamma$-ray background
depends only on the average flux, the only uncertainty in
its flux level comes from the fit to the HECR events. I
estimate the uncertainty introduced by normalizing the average flux
to be less than a factor 1.5.
Fig.~18 shows the results for various values of $z_{\rm
max}$. I conclude that these models are currently unconstrained
by the $\gamma$-ray background, although they still have difficulty
explaining the highest energy CR events.

\section{Conclusions}

I have performed detailed numerical simulations for the
propagation of extragalactic nucleons, $\gamma$-rays, and
electrons in the energy range between $10^8{\,{\rm eV}}$ and
$10^{23}{\,{\rm eV}}$. My goal was to explore constraints on various
models of HECR origin from a comparison of predicted and
observed $\gamma$-ray fluxes at lower energies. The main focus
thereby is on models which associate HECRs with GUT scale
physics or with cosmological GRBs. 

I find that at present
the TD scenarios are primarily constrained by the observed
$\gamma$-ray background between $\simeq 100{\,{\rm MeV}}$ and
$\simeq 10{\,{\rm GeV}}$ but not by the limit on the $\gamma$ to
charged CR flux ratio below $100\,{\rm TeV}$. 
The CEL approximation usually does not take the IR/O background
into account, and thus may not be directly compared to the numerical
calculation because the presence of the IR/O background may affect the
$\gamma$-ray flux level at $100\,{\rm MeV}$ by an order of magnitude. There is
also a significant difference for the UHE spectrum between predictions by the
CEL approximation and
my numerical simulation.
For an EGMF strength $\la 10^{-11}\,{\rm G}$ the TD models yield the
$\gamma$-ray flux which is at about the same level as or below the
current observed flux, depending on the adopted parameters.
On the other hand, an EGMF stronger than $\sim 10^{-11}\,{\rm G}$ stops
the cascade at UHEs and the UHE end of the spectrum is suppressed
significantly. Thus, the level of the $\gamma$-ray flux at about $100\,{\rm
MeV}$ is higher relatively, tightening the constraints.
However, these results are rather insensitive to different models of the IR/O
background \cite{coaha}, although
they are somewhat dependent on the poorly known universal radio background
flux.

I conclude that TD scenarios with QCD
motivated injection spectra up to energies $\la10^{23}{\,{\rm eV}}$
are still viable if injection occurs uniformly or from a
discrete source. This is in contrast to a recent claim in the
literature~\cite{pj}. In case of uniform injection this assumes an
injection history motivated by energy release from a network of
cosmic strings in the scaling regime or from
monopole-antimonopole annihilation ($p=1$). Higher injection energy
cutoffs are allowed for either a weaker source evolution or for
injection spectra somewhat steeper than the QCD motivated
spectra. For EGMF strengths larger than $\simeq10^{-10}{\,{\rm G}}$,
some of the predicted TD spectra have the potential to explain a
possible gap in the HECR spectrum.
The cosmological GRB scenarios recently suggested in the
literature~\cite{Wax1,Vietri,Wax2} are currently unconstrained
by these limits.

With the arrival of the anticipated Pierre Auger Cosmic Ray
Observatories~\cite{Cronin2}, it
is expected that the UHE end of the CR spectrum will be known with much better
accuracy. Constraints derived from the influence of CR
propagation on the observed spectrum will then be one of the
most powerful tools in discriminating between models of HECR
origin.

\section*{Acknowledgments}

This paper is for fulfillment of part of the Doctoral degree requirement in
the Department of Physics at the University of Chicago for the author. The
author would like to thank G.~Sigl and P.~S.~Coppi for their guidance
throughout this ongoing project and for invaluable discussions and
collaboration. This paper would have been impossible without their effort. The
author also thanks D.~N.~Schramm for his advice and many insightful comments,
and F.~A.~Aharonian for many valuable discussions
on cosmic and $\gamma$-ray propagation and for providing the impetus to get
this project started. This work was supported by the DOE, NSF and NASA at the
University of Chicago, by the DOE and by NASA through grant NAG5-2788 at
Fermilab. The financial support by the POSCO Scholarship Foundation is also
gratefully acknowledged.


\newpage
\section*{Appendix A: Triplet pair production}
The expressions for the differential spectra of produced pairs and the
recoiling electron (positron) for TPP by a very energetic
electron on a soft photon can be found in many papers~\cite{mast1,haug,bors}.
I adopt the analytic
approach used in \cite{haug}. For an interaction of an electron of
energy $E$ 
with a photon of energy $\epsilon$ ($E \gg \epsilon$), the double
differential cross section with respect to the positron produced
with energy $E_+$ at solid angle $\Omega_+$ can be expressed as
\begin{equation}
\frac{d^2\sigma}{dE_+ d\Omega_{p_+}} = \sigma_T \cdot \frac{3\alpha}{16\pi^3}
\frac{p_+}{p \cdot k} \frac{(\rho_t^2-4)^{1/2}}{\rho_t} \int A_t
d\Omega_{p'},\label{ddcs}
\end{equation}
where $p \cdot k$ is the scalar product of the initial electron and
photon four-momenta, $p_+$ is the magnitude of the produced positron
three-momentum, $\Omega_{p'}$ is the solid angle of the
recoiling electron, and $\rho_t$ and $A_t$ are given in Ref.~\cite{haug}. The
single differential cross section with respect to the positron energy may be
obtained by integrating Eq.~(\ref{ddcs}) over the positron solid angle
$\Omega_+$ numerically. The differential cross section for the produced
electron is identical to that for the positron due to symmetry. In doing the
integral, it is useful to use the approximation where the dependence of the
cross section on the azimuthal angle of the outgoing particles in the cosmic
ray frame is neglected, as was mentioned before.

Finally, we can obtain the total TPP cross section by
integrating Eq.~(\ref{ddcs}) numerically over the kinematic
range of electron and positron energies given by
\begin{equation}
E_{\rm max,min} = \frac{E_{\rm tot} (s_0-m_e^2) \pm P_{\rm tot}
[s_0 (s_0- 4m_e^2)]^{1/2}} {2s_0+m_e^2},
\end{equation}
where $s_0 \equiv \epsilon (E+p)$, and $E_{\rm tot}$ and $P_{\rm tot}$ are
the total incident energy and momentum, respectively.

I also give here the kinematic energy range for the outgoing
electron and positron and the recoiling proton in case of pair
production by protons (Section 3.2.1):
\begin{equation}
E_{\rm max,min} = \frac{E_{\rm tot}(s_0-m_e \lambda) \pm P_{\rm
tot} [s_0^2-2 m_e s_0(m_e+\lambda)+m_e^2 (m_e^2-m_N^2)/2]^{1/2}}
{2s_0+m_N^2},
\end{equation}
where $s_0$ is as defined previously, $\lambda
\equiv[(m_e^2+m_N^2)/2]^{1/2}$.

\section*{Appendix B: Photopion production}
First, I express the differential cross sections for
single pion production in terms of the CRF energies
$E'_a$, where $a=N,\pi$:
\begin{eqnarray}
  {d\sigma_i\over dE'_N}\,,\,{d\sigma_i\over dE'_\pi}&=&
  {2\pi\over E_N}\,{1+2\kappa\over
D}
  \,\sum_{j=0}^Ka_{ij}(\kappa)\left(x^*\right)^j\nonumber\\
  &&\times\left\{\begin{array}{ll}
           1 & \mbox{for $i=1$ (charge retention)}\\
           \left(1-\beta^*x^*\right)^{-1/2} & 
           \mbox{for $i=2,3$ (charge exchange)}
         \end{array}\right.\label{dsigCRF}
\end{eqnarray}
where $\beta^*=D
/(\kappa+\varepsilon^2/2)$ and $x^*$ can be
expressed as a function of $E'_N$ or $E'_\pi$ and $E_N$:
\begin{equation}
  x^*=
  \frac{(1+2\kappa)(E'_N/E_N)-1-\kappa-\varepsilon^2/2}{D}=
  -\frac{(1+2\kappa)(E'_\pi/E_N)-\kappa-\varepsilon^2/2}{D}
  \,.\label{xprime}
\end{equation}
Finally, I compute the fractions $r_c(\kappa)$ and
$r_\rho(\kappa)$ of the incoming nucleon energy going into
the central and leading pions. These fractions are given
by integrating the differential cross section
$d\sigma/dE_\pi$ for the respective process, weighted
by the pion energy $E_\pi$, in the CRF, and dividing by
the corresponding total cross section. Using Eq.~(\ref{dsig1}),
\begin{equation}
  E_\pi={E_N\over m_N}\,
  {E_\pi^*-p_\|^*\over(1+2\kappa)^{1/2}}=
  {E_N\over m_N}{p_{\rm max}^*\over(1+2\kappa)^{1/2}}
  \left[\left(x^2+{p_\bot^2+m_\pi^2\over(p_{\rm max}^*)^2}
  \right)^{1/2}-x\right]\,,\label{Epi}
\end{equation}
and Eq.~(\ref{pmax}), we end up with
\begin{eqnarray}
  r_\rho(\kappa)&=&{\pi\over\sigma_\rho}\,
{D
  \over1+2\kappa}\label{rrho}\\
  &&\times\int dp_\bot^2\int
dx\,2f_\rho(x)\frac{\exp[-p_\bot^2/\Lambda^2]}{\Lambda^2}
  \left\{1-x\left[x^2+\frac{1+2\kappa}{D^2}
\left({p_\bot^2\over m_N^2}+\varepsilon^2
  \right)\right]^{-1/2}\right\}\,.\nonumber
\end{eqnarray}
Again, the integration ranges are obtained by the requirement $p_\bot^2/(p_{\rm
max}^*)^2 + x^2 \le 1$.
The formula for $r_c(\kappa)$ can be obtained from this by
substituting $\sigma_\rho\rightarrow\sigma_{\rm tot}-\sigma_\rho$
and $2f_\rho(x)\rightarrow\left[3+(2/5)(1-\sigma_\rho
/\sigma_{\rm tot})/\left\langle n_{\pi^-}^c\right\rangle
(\kappa)\right]f_c(x)$, where
$\left\langle n_{\pi^-}^c\right\rangle(\kappa)$ was given
in Eq.~(\ref{multpi}).

\section*{Appendix C: Pion decay spectra}

First, we define the decay spectrum $N_a(E)$ as the differential number
of the secondary particle $a$ at energy $E$. Then the spectrum is normalized
as
\begin{equation}
\int dEN_a(E)= n_a,
\end{equation}
where $n_a$ is the number of particles $a$ produced by decay of a single
pion. For example, $n_{\gamma} = 2$ for $\pi^0$ decay.

First, the photon spectrum from $\pi^0$ decay is
\begin{equation}
N_\gamma(E)=\frac{2}{E_{\pi}}\quad\mbox{for $E\le E_\pi$}\,.
\end{equation}

Before we calculate the charged pion decay spectra, we note the fact that the
pions produced from photopion production are always relativistic. Thus, we
may make the relativistic approximation for both pions and the resulting muons.
For example, the decay spectra of $\pi^+$ read:
\begin{eqnarray}
N_{e^+}(E)&=&N_{\bar{\nu}_\mu}(E)\simeq \left\{ \begin{array}{ll}
\frac{1}{(1-r) E_\pi}
(A_0+A_2 z^2 + A_3 z^3) & \mbox{for $E \le rE_\pi$} \\
\frac{1}{(1-r) E_\pi} (B_0 + B_0'\ln z+B_2 z^2+B_3 z^3) & \mbox{for $rE_\pi
\la E \le E_\pi$},
\end{array}\right.\nonumber\\
N_{\nu_\mu}(E)&=&{1\over(1-r)E_\pi}\quad
\mbox{for $E\leq(1-r)E_\pi$}\\
N_{\nu_e}(E)&\simeq&\left\{ \begin{array}{ll}
\frac{1}{(1-r)E_\pi} (C_0 +C_2 z^2 +C_3 z^3) & \mbox{for $E \le r E_\pi$}\\
\frac{1}{(1-r) E_\pi} (D_0 +D_0' \ln z +D_1 z+D_2 z^2 +D_3 z^3) & \mbox{for $r
E_\pi \le E \le E_\pi$},
\end{array}\right.\,,\nonumber
\end{eqnarray}
where $r \equiv m_\mu^2/m_\pi^2$, $z \equiv E/E_\pi$, and coefficients are
given as 
\[
(A_0,A_2,A_3)=(0.94486,-2.7892,1.2397),
\]
\[
(B_0,B_0',B_2,B_3)=(-2.4126,-2.8951,4.3426,-1.9300),
\]
\[
(C_0,C_2,C_3)=(1.1053,-4.46883,3.71887),
\]
and
\[
(D_0,D_0',D_1,D_2,D_3)=(13.846,5.37053,-28.1116,20.0558,-5.7902).
\]
The average energies of the secondary particles are $\langle E_{e^+}
\rangle=\langle E_{\bar{\nu}_\mu} \rangle = 0.265 E_\pi$, $\langle E_{\nu_e}
\rangle = 0.257 E_\pi$, and $\langle E_{\nu_\mu} \rangle = 0.213 E_\pi$
respectively. The decay spectra from $\pi^-$ are obtained by substituting
particles accordingly. 



\begin{figure}
\centering\leavevmode
\epsfxsize=5.5in
\epsfbox{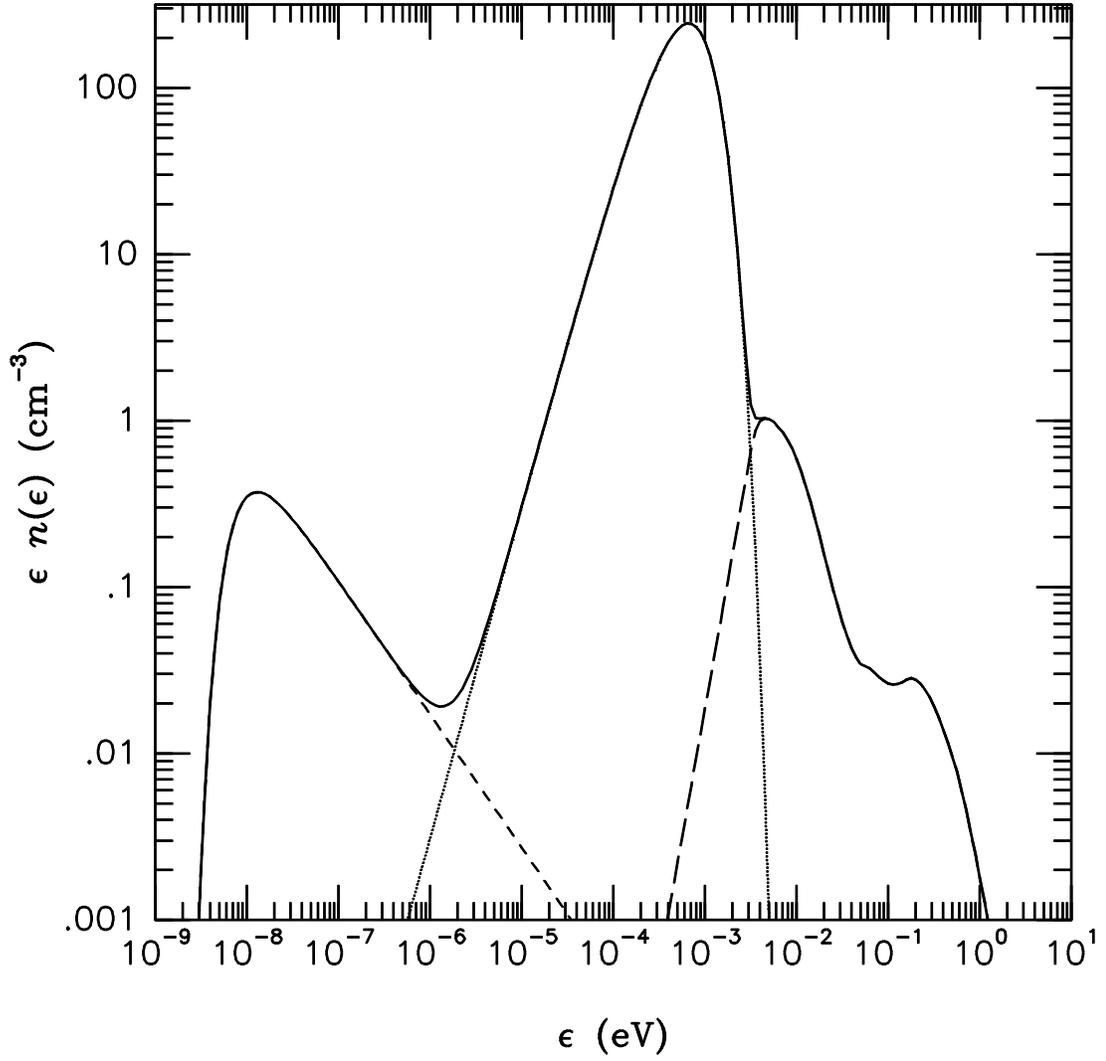}
\smallskip
\caption[...]{The universal background radiation
intensity spectrum at $z=0$ (solid line) used in our model. The
separate contributions from the radio (short dashed line), the
IR/O (long dashed line) background, and the CMB (dotted
line) are also shown.}
\end{figure}

\begin{figure}
\centering\leavevmode
\epsfxsize=5.5in
\epsfbox{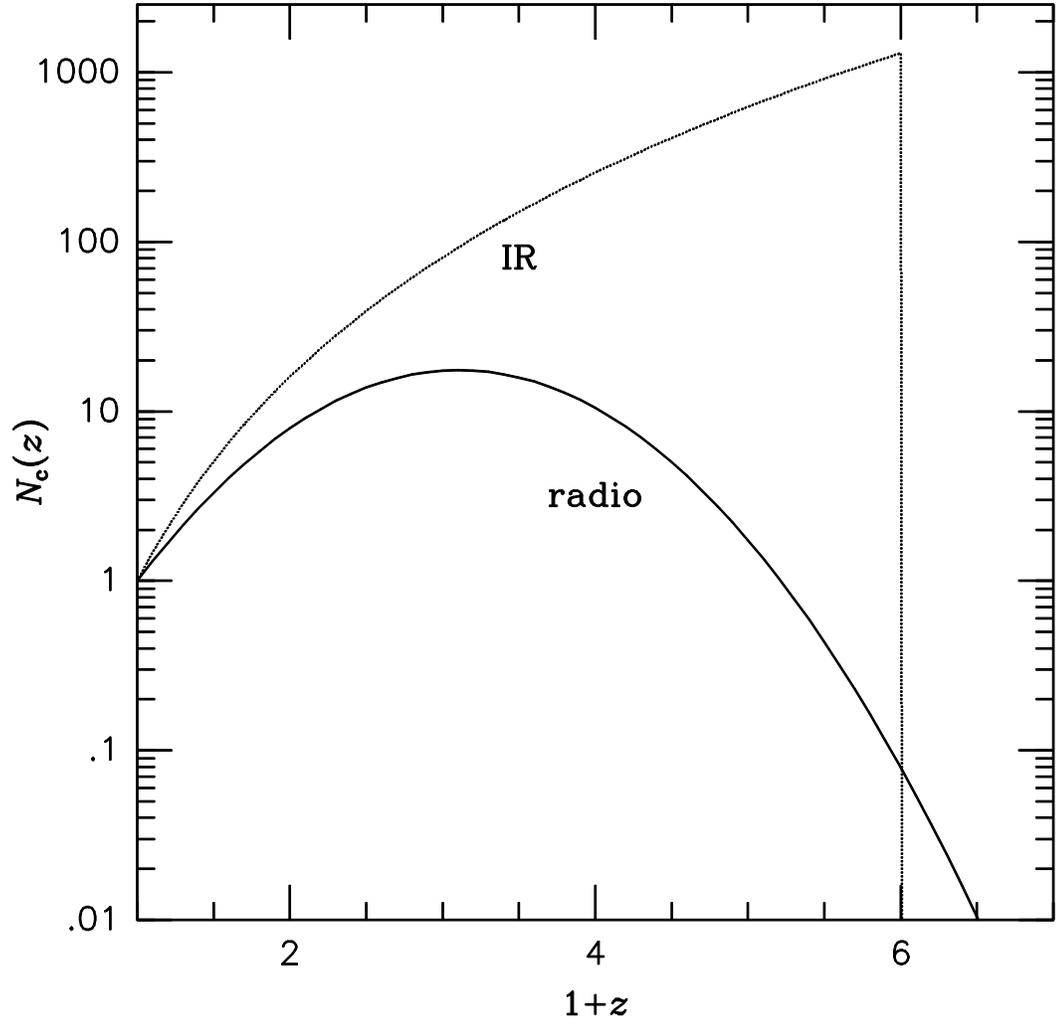}
\smallskip
\caption[...]{The effective comoving density of radio
and IR/O sources whose luminosities are normalized at $z=0$, as a
function of redshift. This corresponds to $N_c(z)$ in
Eq.~(\ref{backinj}). The IR/O source density is assumed to cut off
at $z=5$.}
\end{figure}

\begin{figure}
\centering\leavevmode
\epsfxsize=5.5in
\epsfbox{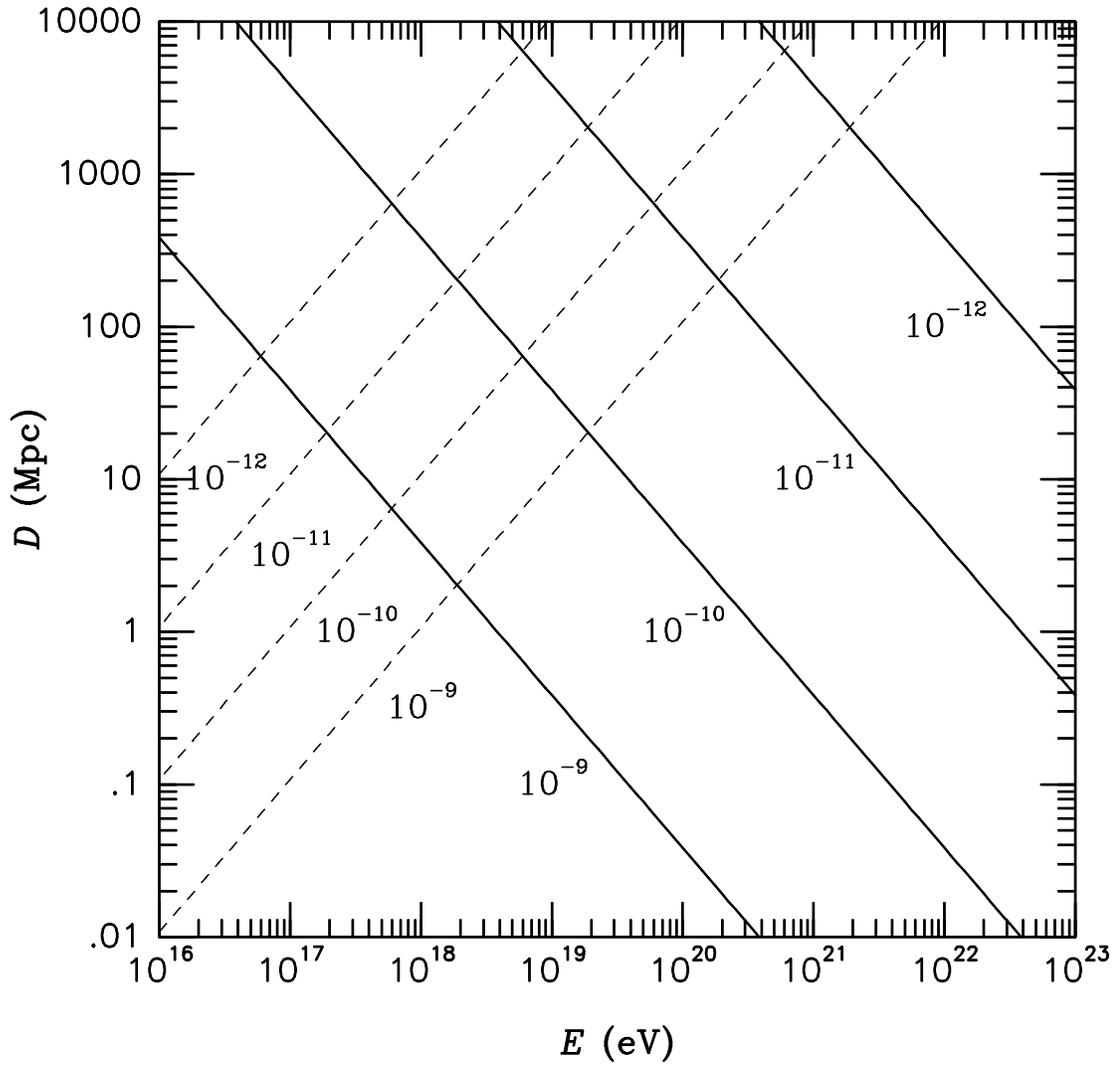}
\smallskip
\caption[...]{Gyroradii (dashed lines) and
synchrotron loss lengths (solid lines) of electrons for various
strengths of the EGMF in units of gauss (G) as indicated.}
\end{figure}

\begin{figure}
\centering\leavevmode
\epsfxsize=5.5in
\epsfbox{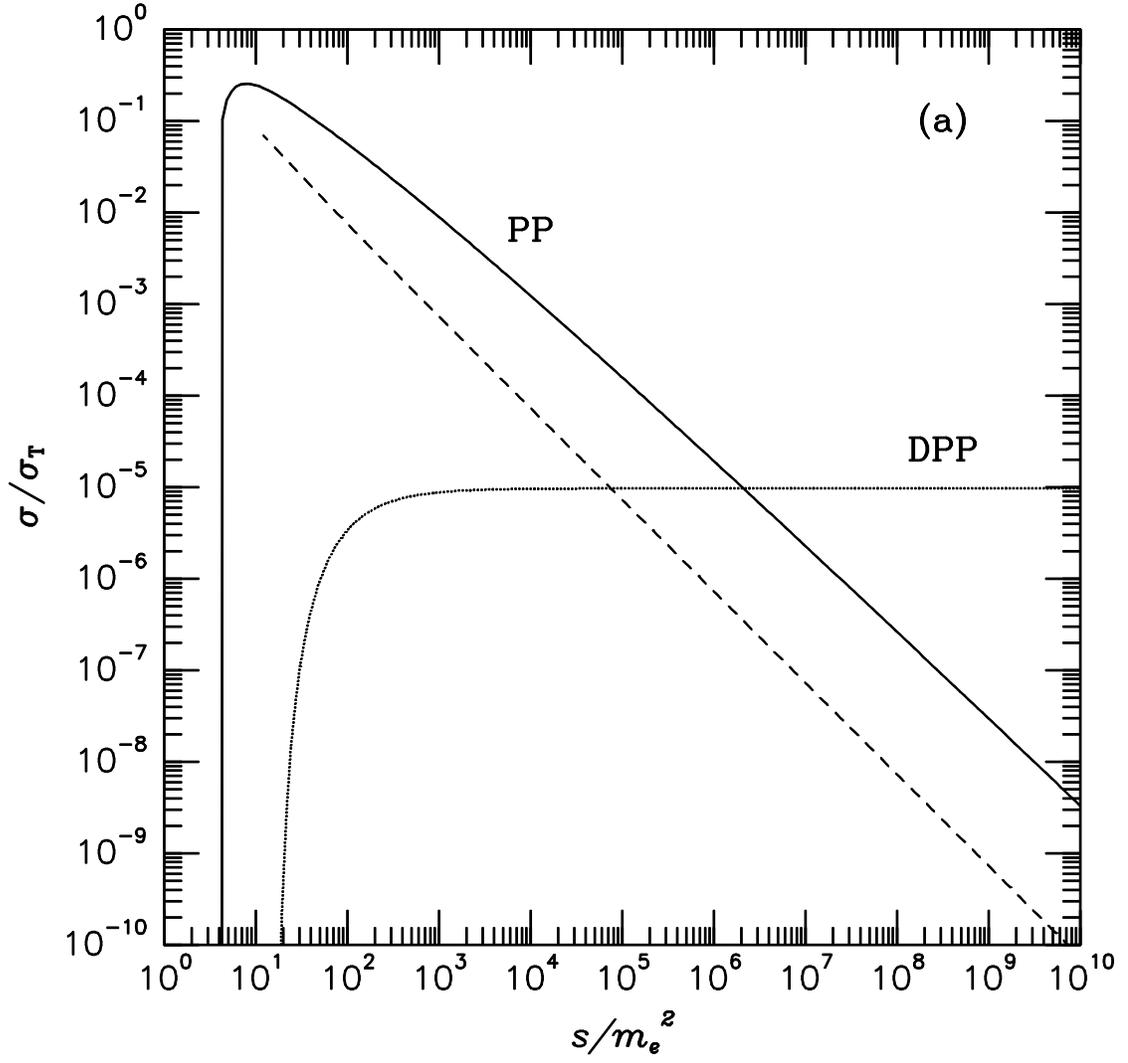}
\smallskip
\caption[...]{The total cross sections, $\sigma(s)$,
and the cross sections times the average inelasiticity,
$\sigma(s)\eta(s)$, which is proportional to the fractional
energy loss rate of the leading particle: (a) For PP (solid line
and short dashed line, respectively) and DPP (dotted line).}
\end{figure}
\setcounter{figure}{3}
\newpage
\begin{figure}
\centering\leavevmode
\epsfxsize=5.5in
\epsfbox{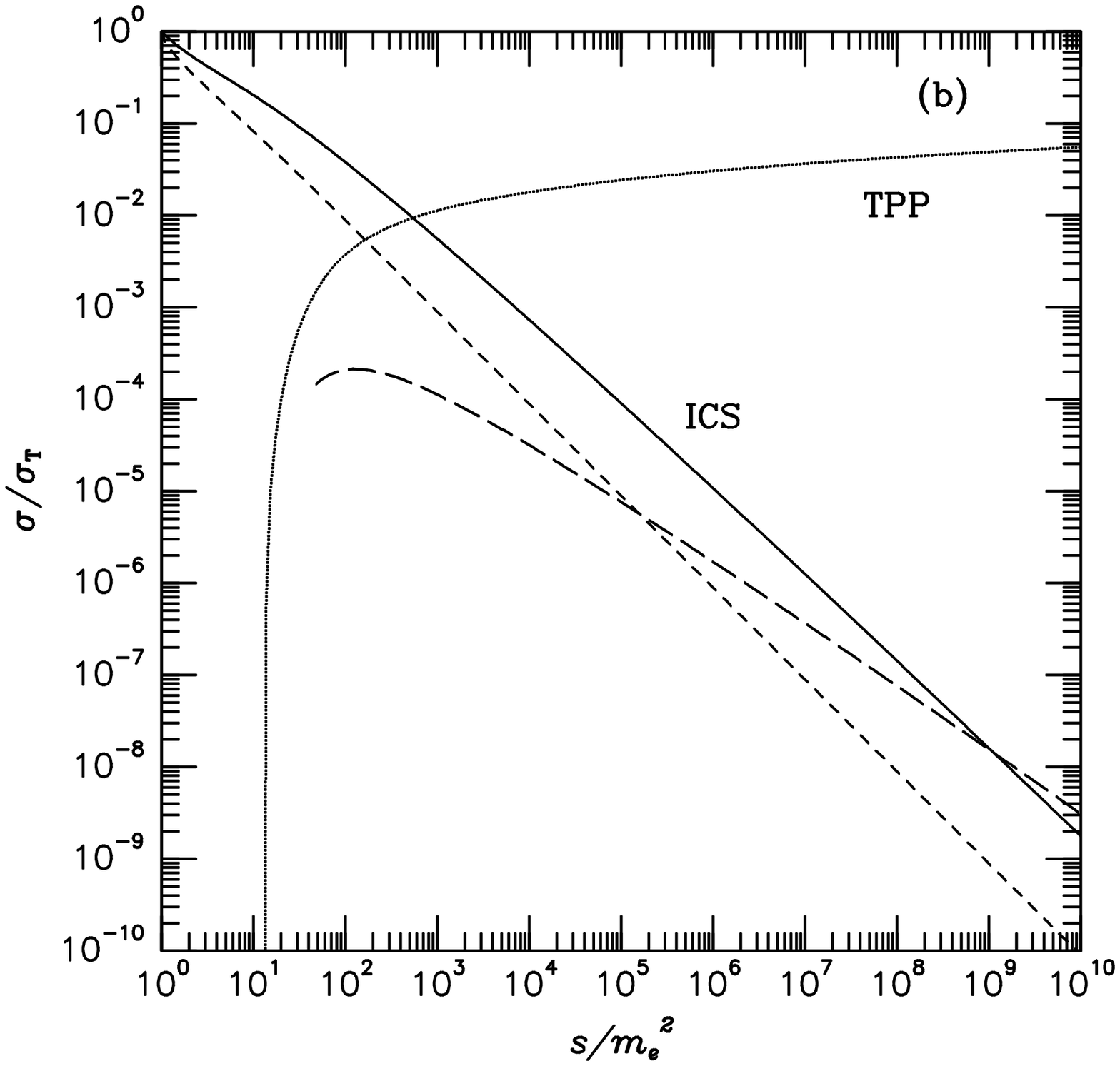}
\smallskip
\caption[...]{(b) For ICS (solid line and short dashed line,
respectively) and TPP (dotted line and long dashed line, respectively).}
\end{figure}

\begin{figure}
\centering\leavevmode
\epsfxsize=5.5in
\epsfbox{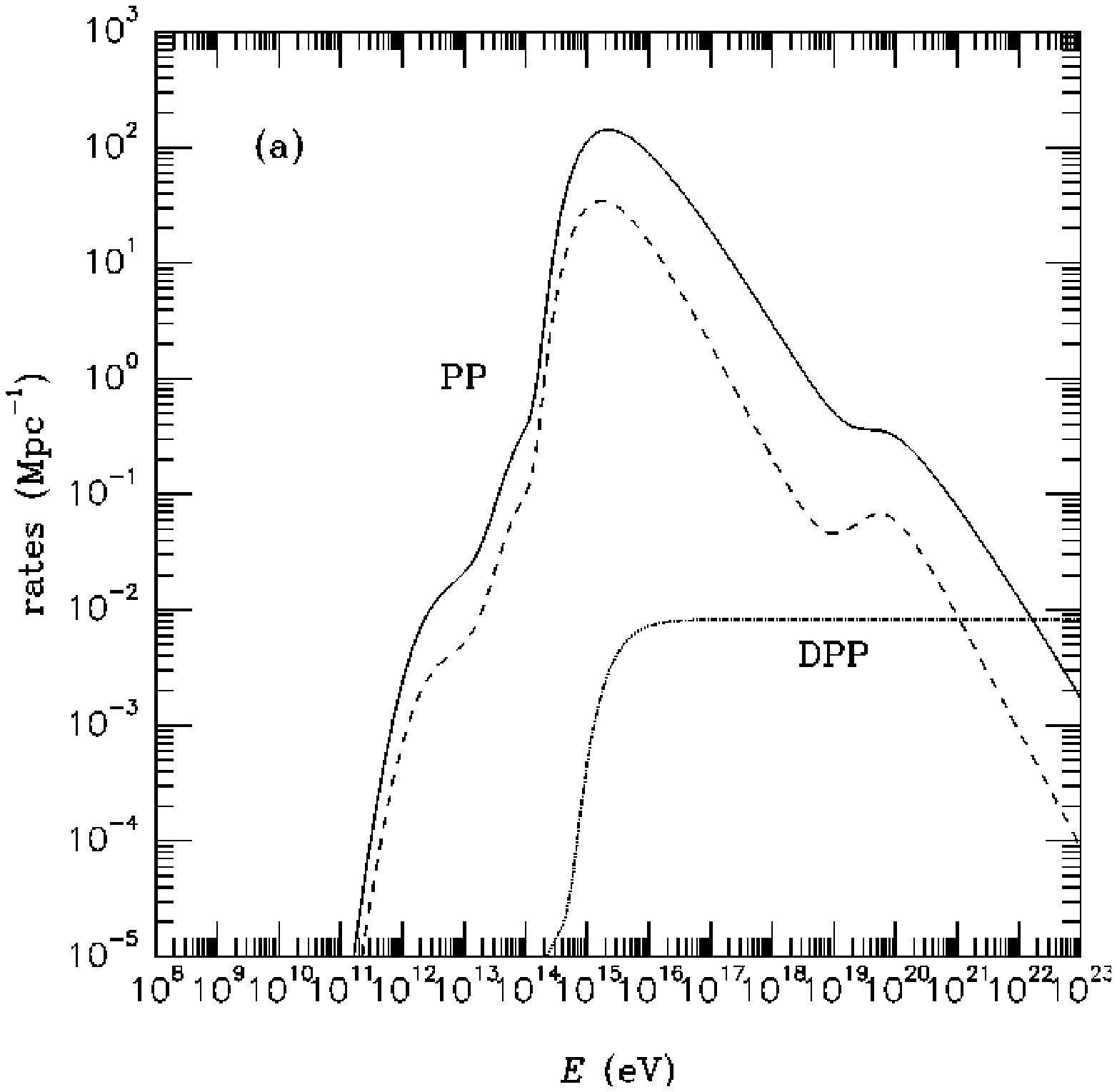}
\vspace{-0.3in}
\caption[...]{The relevant interaction rates at $z=0$
that affect the photons and electrons in the
energy range we consider. The key is identical to the key for
Fig.~4. The rates are calculated by folding the total
cross sections and inelasticity weighted cross sections with
the present background photon spectrum shown in Fig~2.}
\end{figure}
\setcounter{figure}{4}
\newpage
\begin{figure}
\centering\leavevmode
\epsfxsize=5.5in
\epsfbox{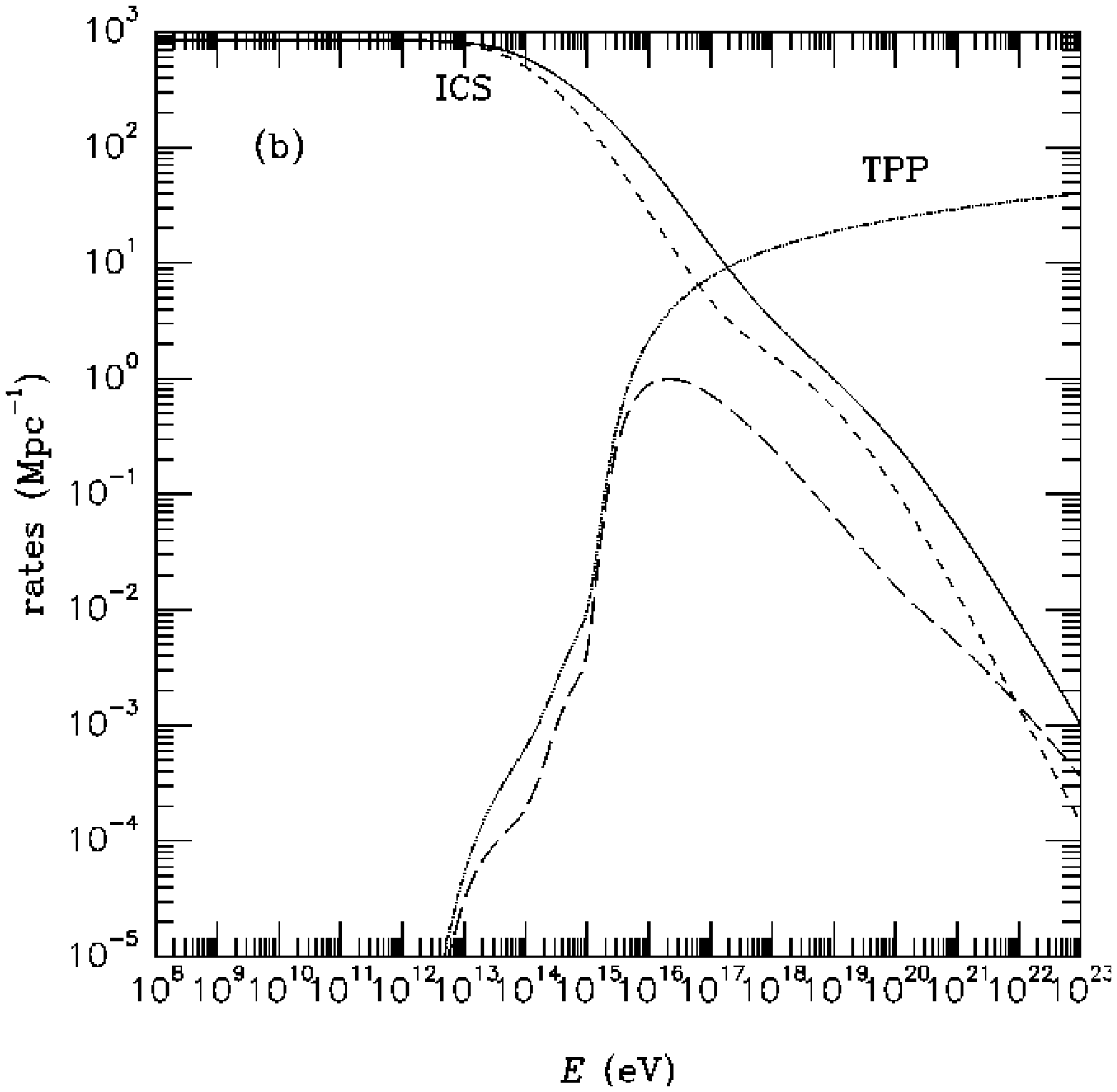}
\smallskip
\caption[...]{(b).}
\end{figure}

\begin{figure}
\centering\leavevmode
\epsfxsize=5.5in
\epsfbox{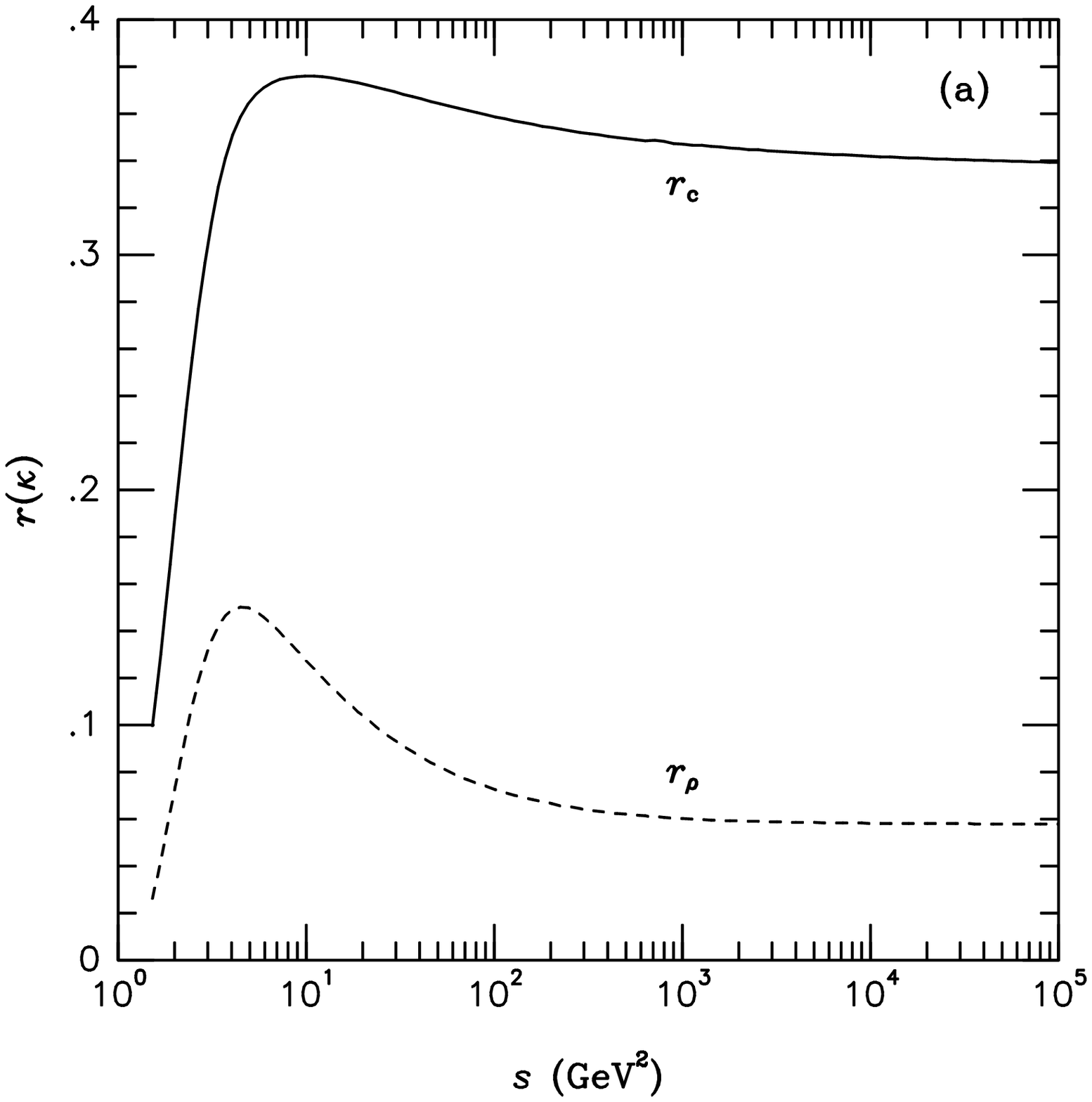}
\smallskip
\caption[...]{(a) The fractions $r_c$ and
$r_\rho$ of the incoming nucleon energy which goes
into the central and leading pions, respectively, as functions
of $s$.}
\end{figure}
\setcounter{figure}{5}
\newpage
\begin{figure}
\centering\leavevmode
\epsfxsize=5.5in
\epsfbox{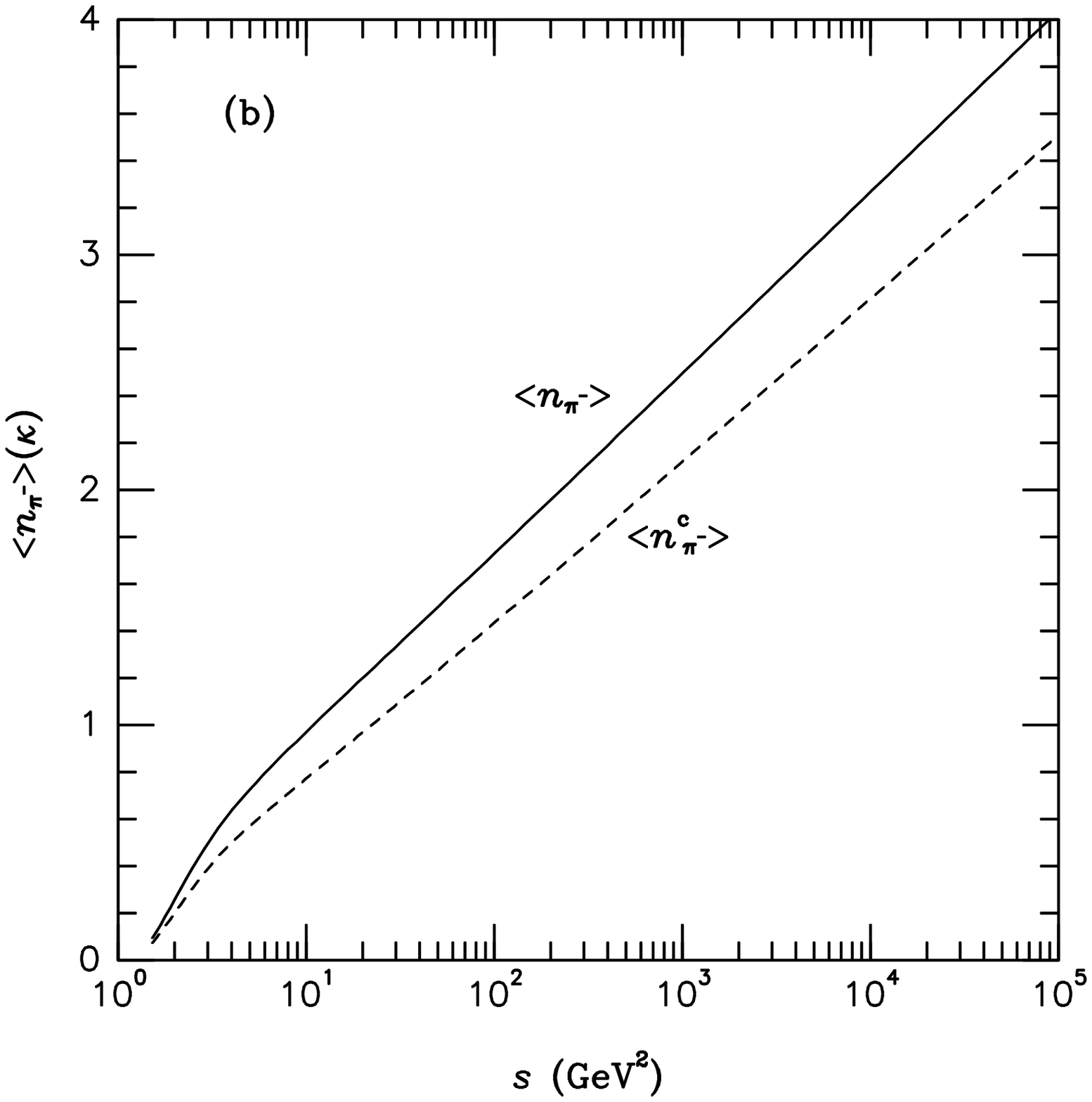}
\smallskip
\caption[...]{(b) The average central and total $\pi^-$
multiplicities, $\left\langle n_{\pi^-}^c\right\rangle$ [see
Eq.~(\ref{multpi})] and $\left\langle n_{\pi^-}\right\rangle$
[resulting by substituting $f_c(x)\rightarrow f(x)$ in
Eq.~(\ref{multpi})] as functions of $s$.}
\end{figure}

\begin{figure}
\centering\leavevmode
\epsfxsize=5.5in
\epsfbox{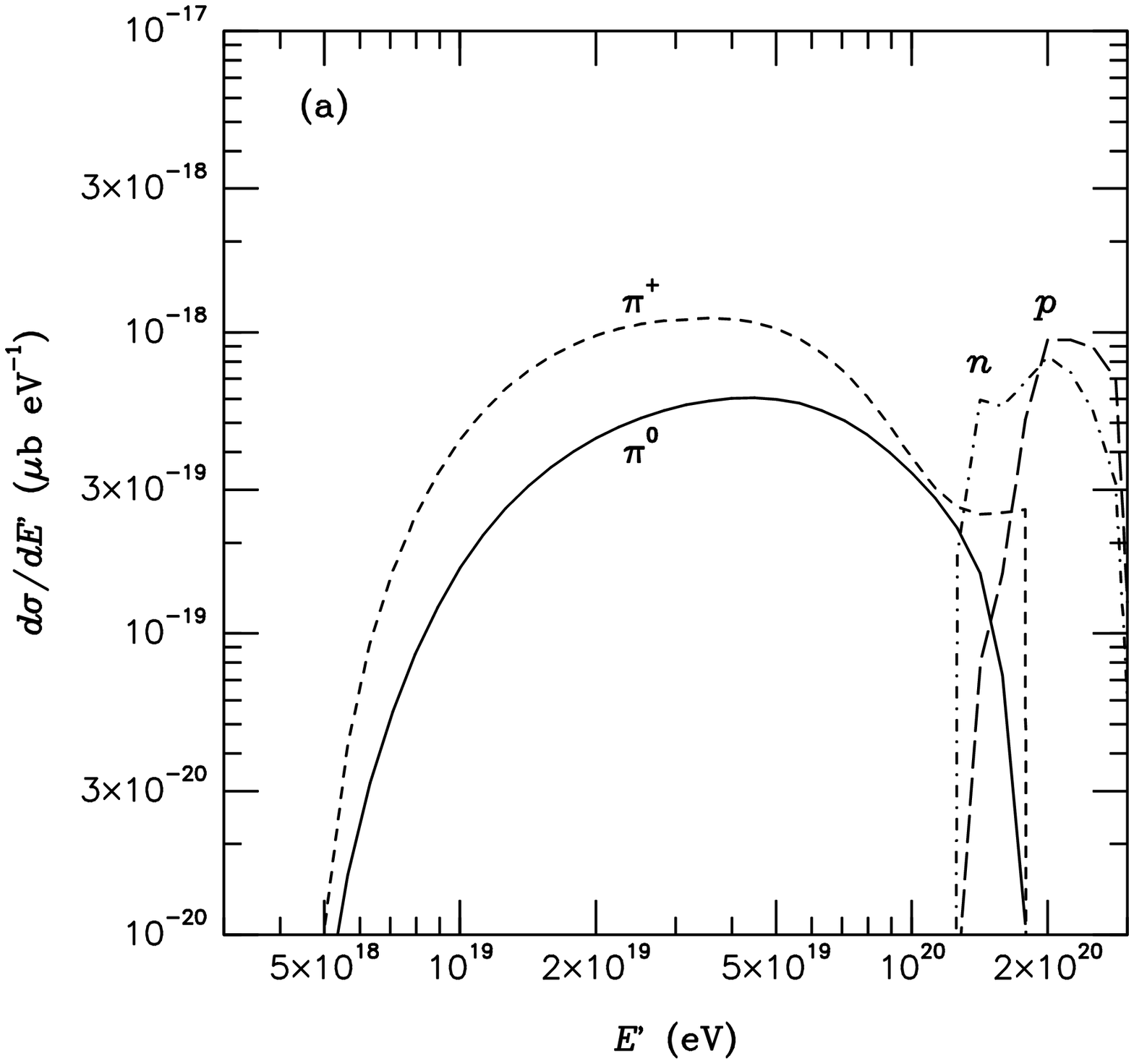}
\vspace{-0.3in}
\caption[...]{The differential cross sections for
production of $\pi^-$ (dotted lines), $\pi^+$ (short dashed
lines), $\pi^0$ (solid lines), protons (long dashed lines), and
neutrons (dash-dotted lines) for the collision of a proton of
energy $E$ with a background photon at squared CM energy $s$,
from the formalism adopted in Section 3.2.2: (a) For
$E=3\times10^{20}{\,{\rm eV}}$, $s=2.1{\,{\rm GeV}}^2$.}
\end{figure}
\setcounter{figure}{6}
\newpage
\begin{figure}
\centering\leavevmode
\epsfxsize=5.5in
\epsfbox{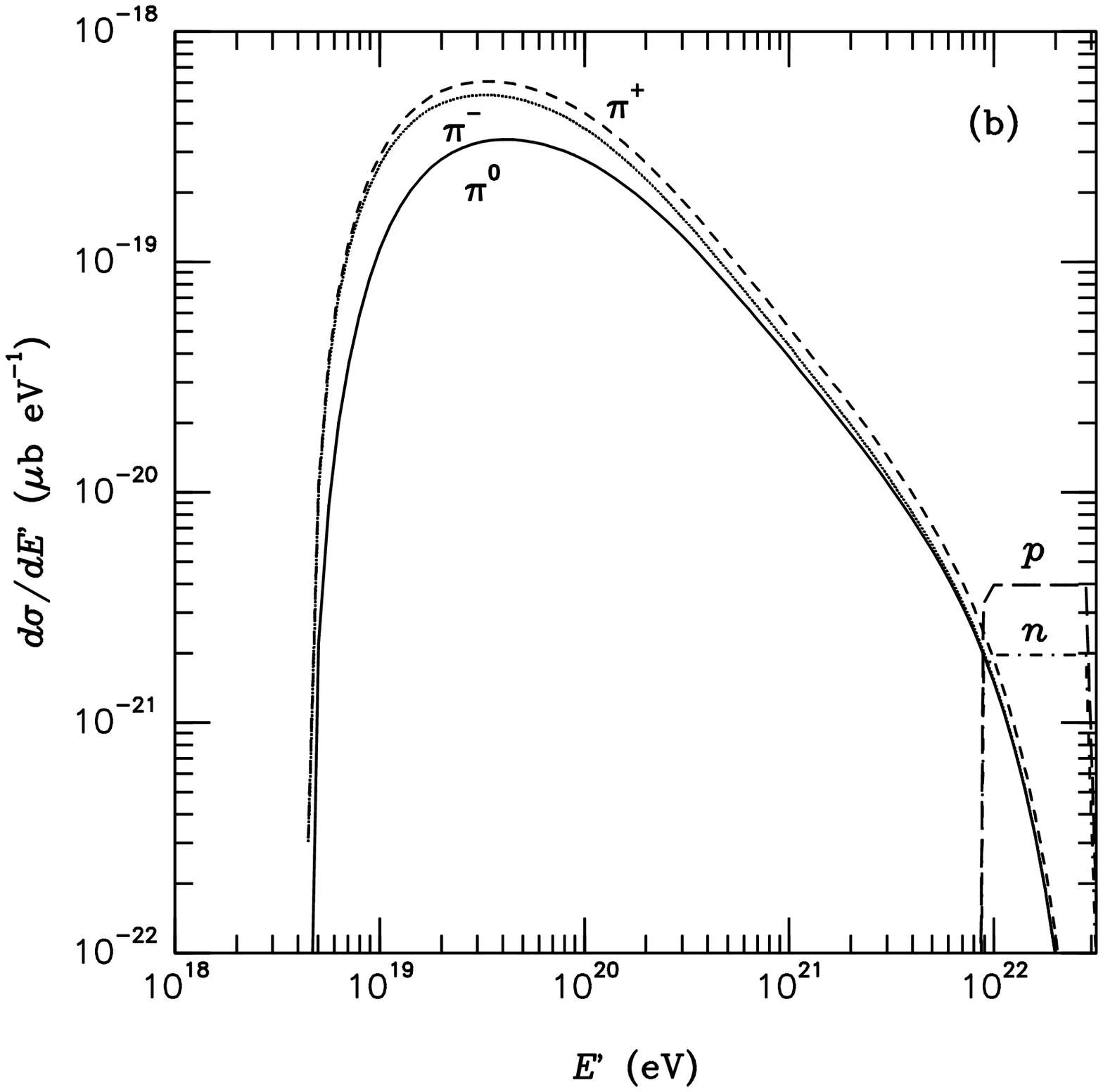}
\smallskip
\caption[...]{(b) For $E=3\times10^{22}{\,{\rm eV}}$, $s=120{\,{\rm GeV}}^2$.}
\end{figure}

\begin{figure}
\centering\leavevmode
\epsfxsize=5.5in
\epsfbox{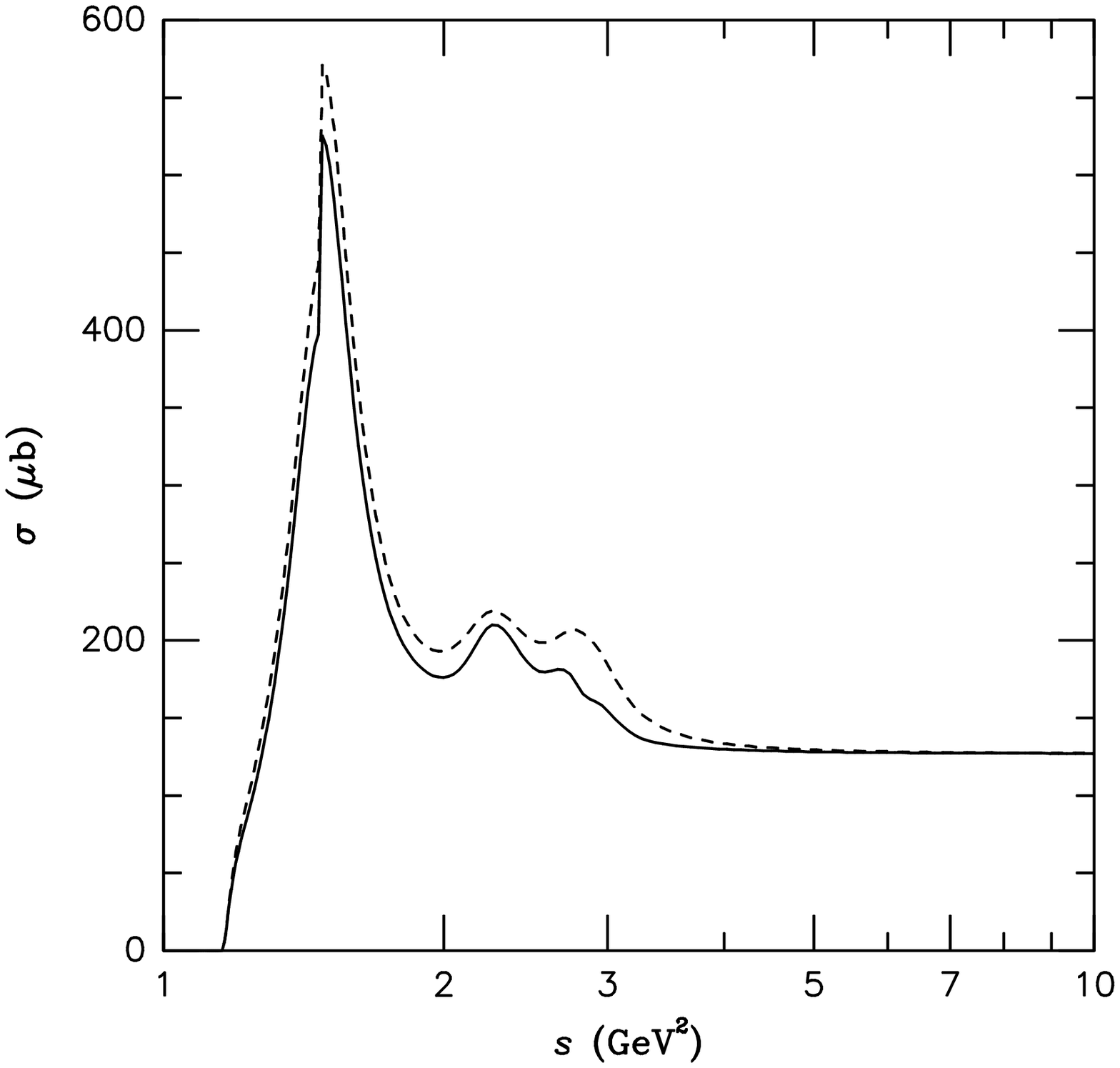}
\smallskip
\caption[...]{The inclusive multiple pion production
cross section for protons (solid line) and neutrons (dashed
line) as a function of $s$.}
\end{figure}

\begin{figure}
\centering\leavevmode
\epsfxsize=5.5in
\epsfbox{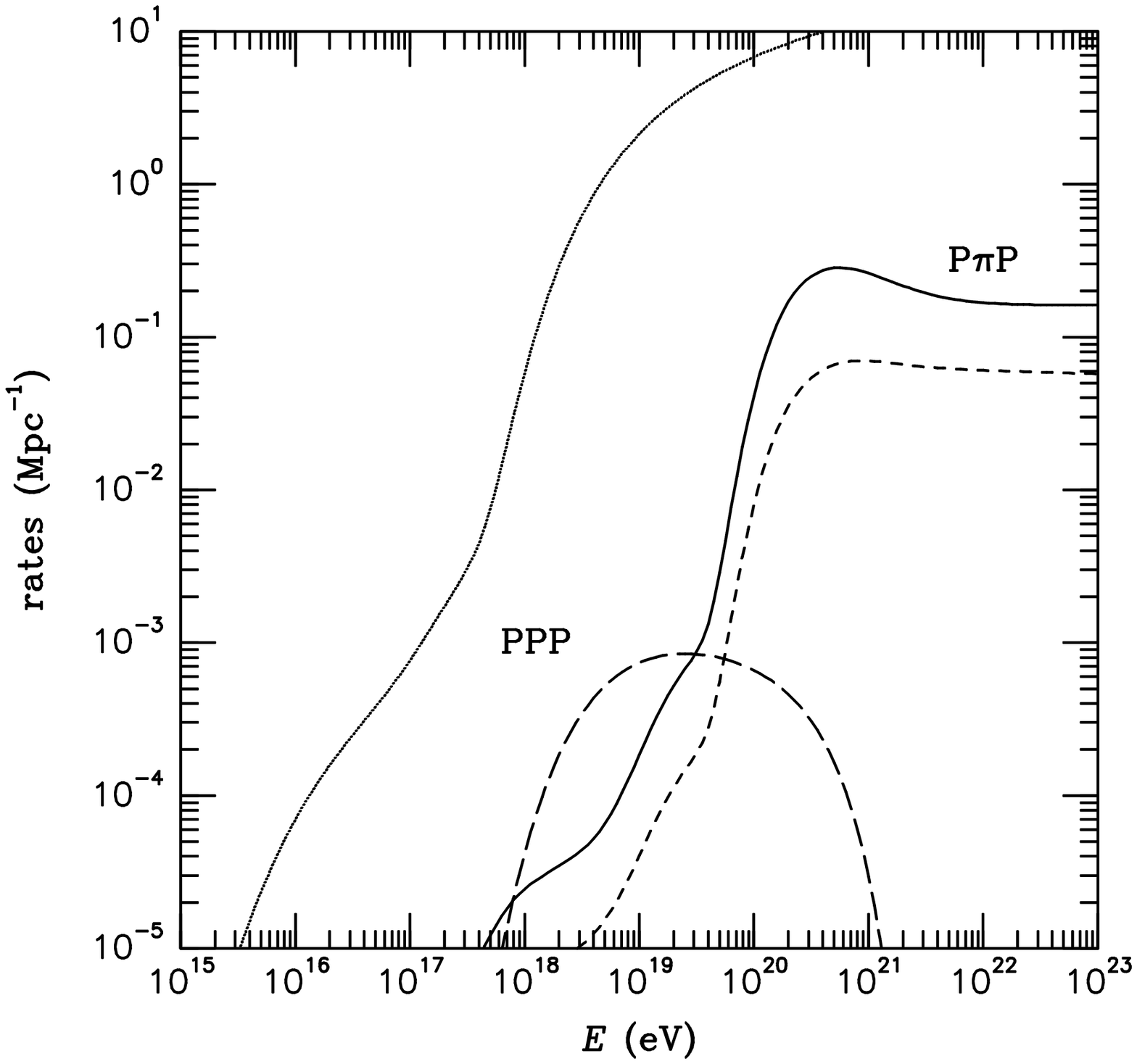}
\vspace{-0.3in}
\caption[...]{The interaction rates and energy
attenuation rates for multiple pion production (solid line and
short dashed line, respectively) and PPP (dotted line and long
dashed line, respectively). The rates were obtained by folding
the cross sections and inelasticity weighted cross sections with
the present background photon spectrum shown in Fig~2.}
\end{figure}

\begin{figure}
\centering\leavevmode
\epsfxsize=5in
\epsfbox{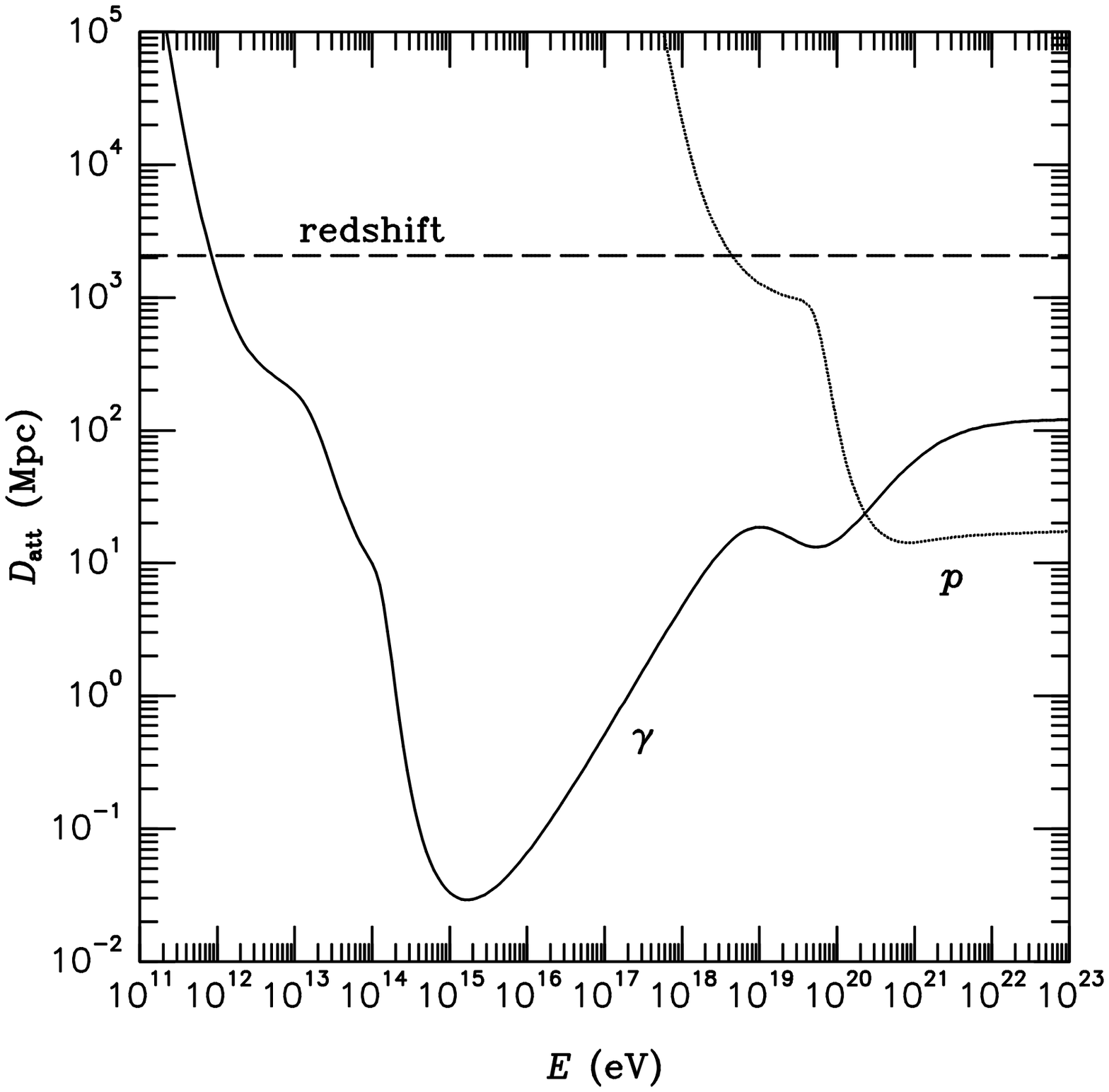}
\smallskip
\caption[...]{The energy attenuation lengths for cascade photons
and for protons as a function of energy assuming the radiation
background photon spectrum shown in Fig.~2.
These curves were obtained by running the code over
small distances and ignoring
the production of non-leading particles, which corresponds to the
CEL approximation.}
\end{figure}
\clearpage

\begin{figure}
\centering\leavevmode
\epsfxsize=5.5in
\epsfbox{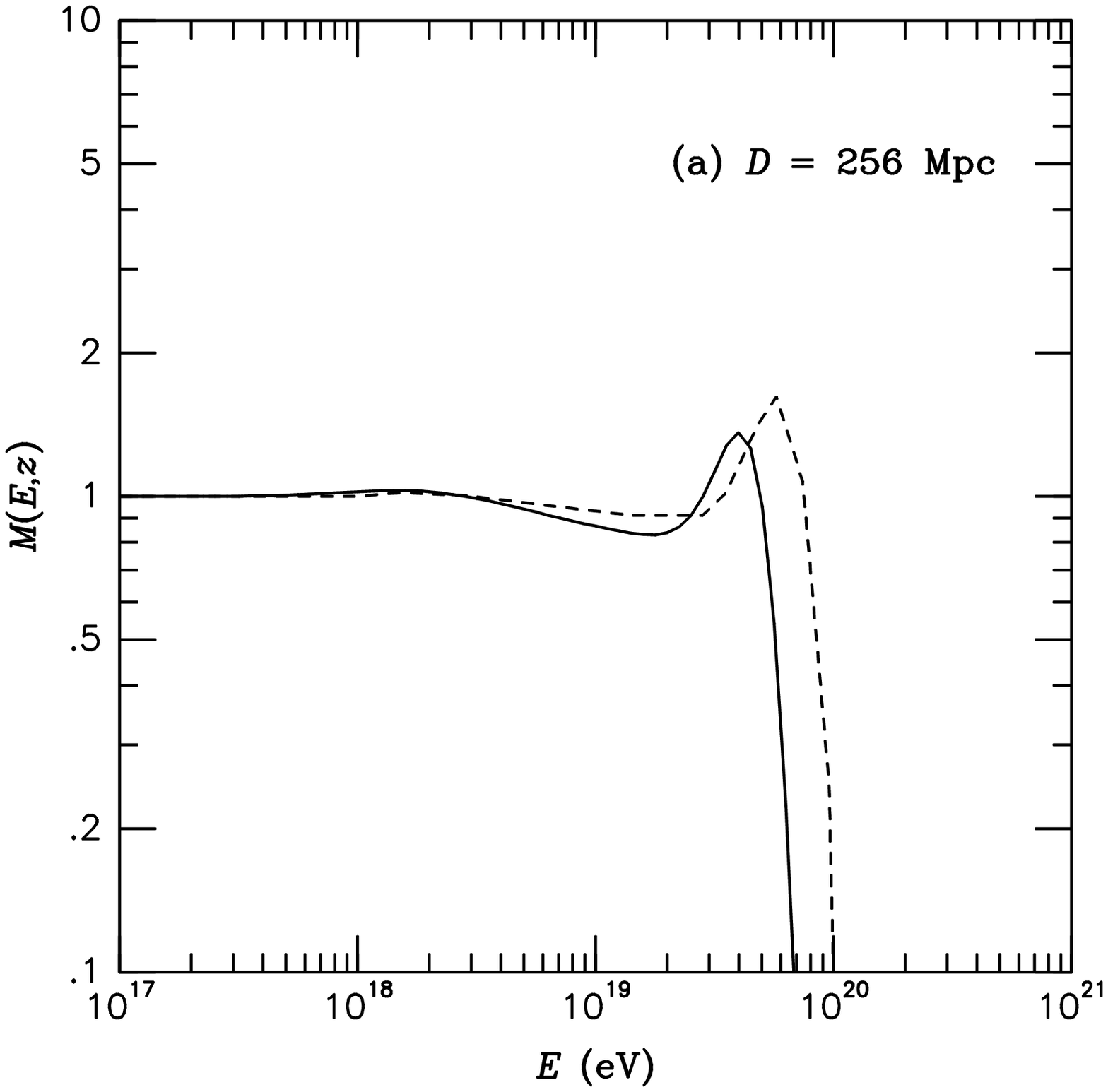}
\vspace{-1in}
\caption[...]{The modification factors as
defined in Ref.~\cite{Rachen} for discrete sources injecting a
$E^{-2}$ proton spectrum extending up to $3\times10^{20}{\,{\rm eV}}$
at a given distance $d$ or redshift $z$ resulting from our
analysis (solid lines). Also shown are the corresponding curves from
Protheroe and Johnson~\cite{pj} (dashed lines): (a) For
$d=256{\,{\rm Mpc}}$. For further
comparison with results from Refs.~\cite{Rachen,yt} see the
discussion in Ref.~\cite{pj}.}
\end{figure}
\setcounter{figure}{10}
\newpage
\begin{figure}
\centering\leavevmode
\epsfxsize=5.5in
\epsfbox{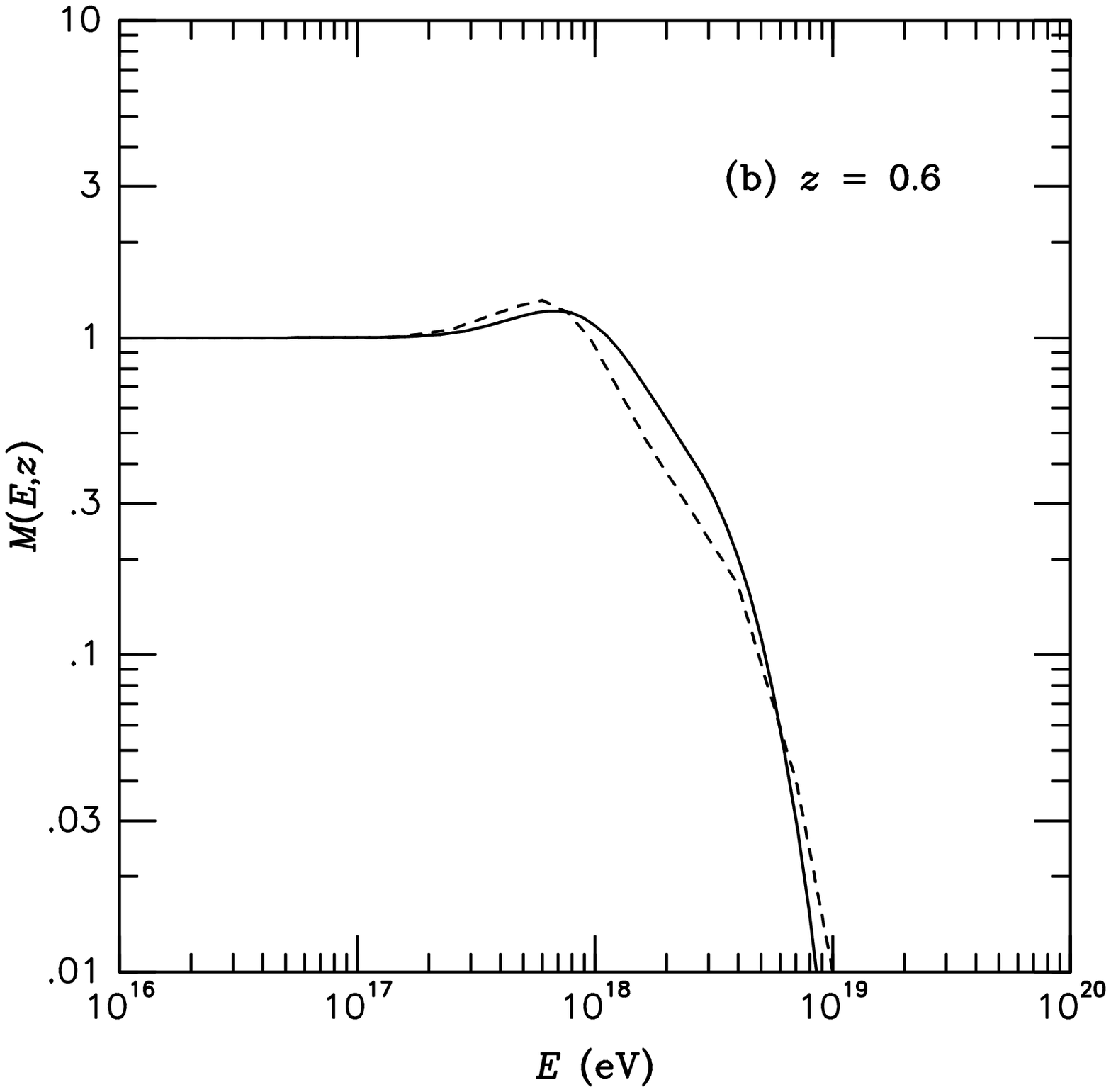}
\smallskip
\caption[...]{(b) For $z=0.6$.}
\end{figure}
\setcounter{figure}{10}
\newpage
\begin{figure}
\centering\leavevmode
\epsfxsize=5.5in
\epsfbox{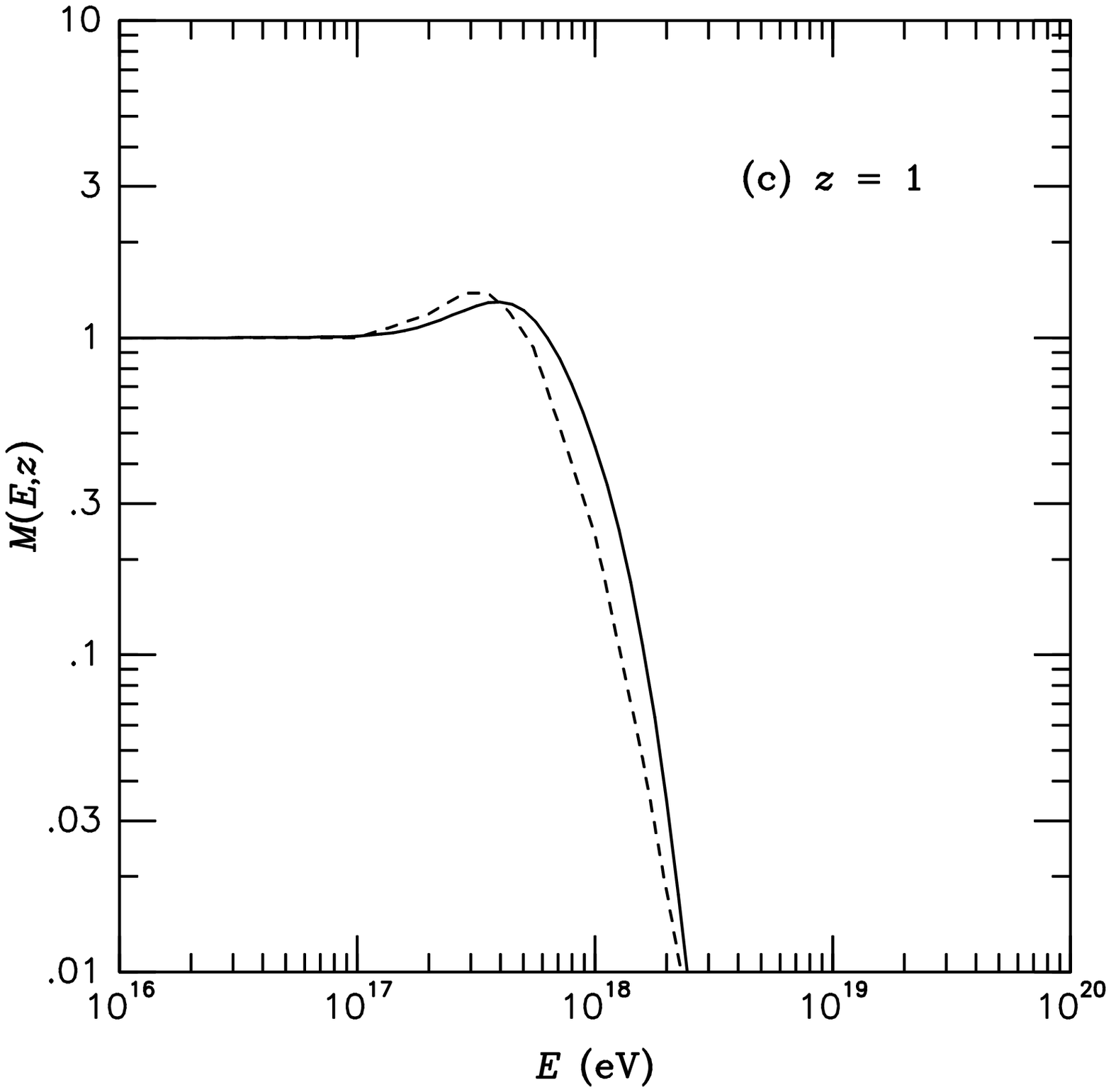}
\smallskip
\caption[...]{(c) For $z=1$.}
\end{figure}

\begin{figure}
\centering\leavevmode
\epsfxsize=5in
\epsfbox{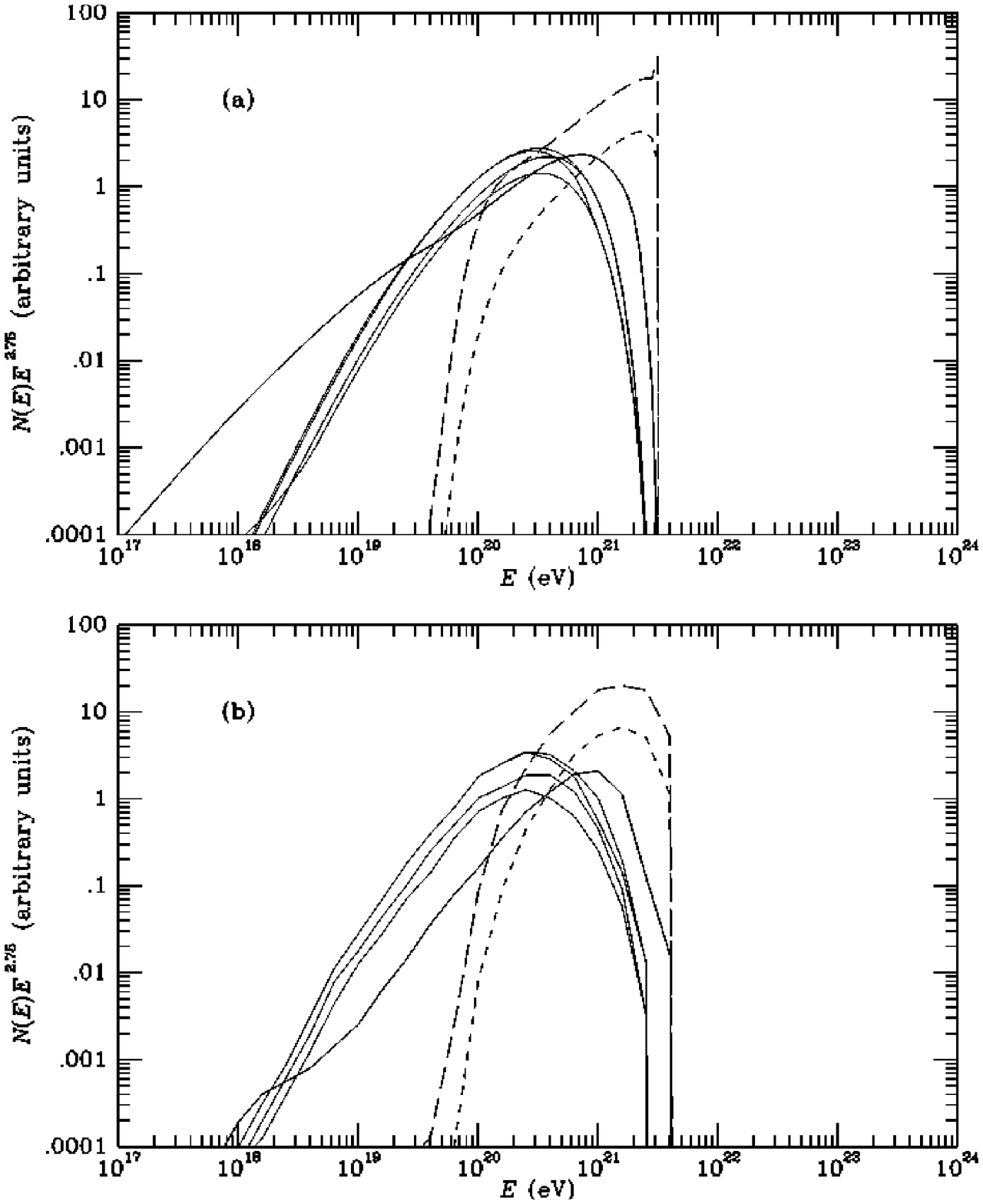}
\smallskip
\caption[...]{The differential fluxes of $\gamma$-rays
(solid line), nucleons (long dashed line), neutrons (short
dashed line) and $\nu_\mu$, $\bar\nu_\mu$, $\nu_e$, $\bar\nu_e$
(thin solid lines in decreasing order) for monoenergetic proton
injection at an energy $E=10^{21.5}{\,{\rm eV}}$ and a distance
$d=32{\,{\rm Mpc}}$: (a) Result from our analysis in arbitrary units;
(b) Corresponding results from Ref.~\cite{pj}.}
\end{figure}

\begin{figure}
\centering\leavevmode
\epsfxsize=5in
\epsfbox{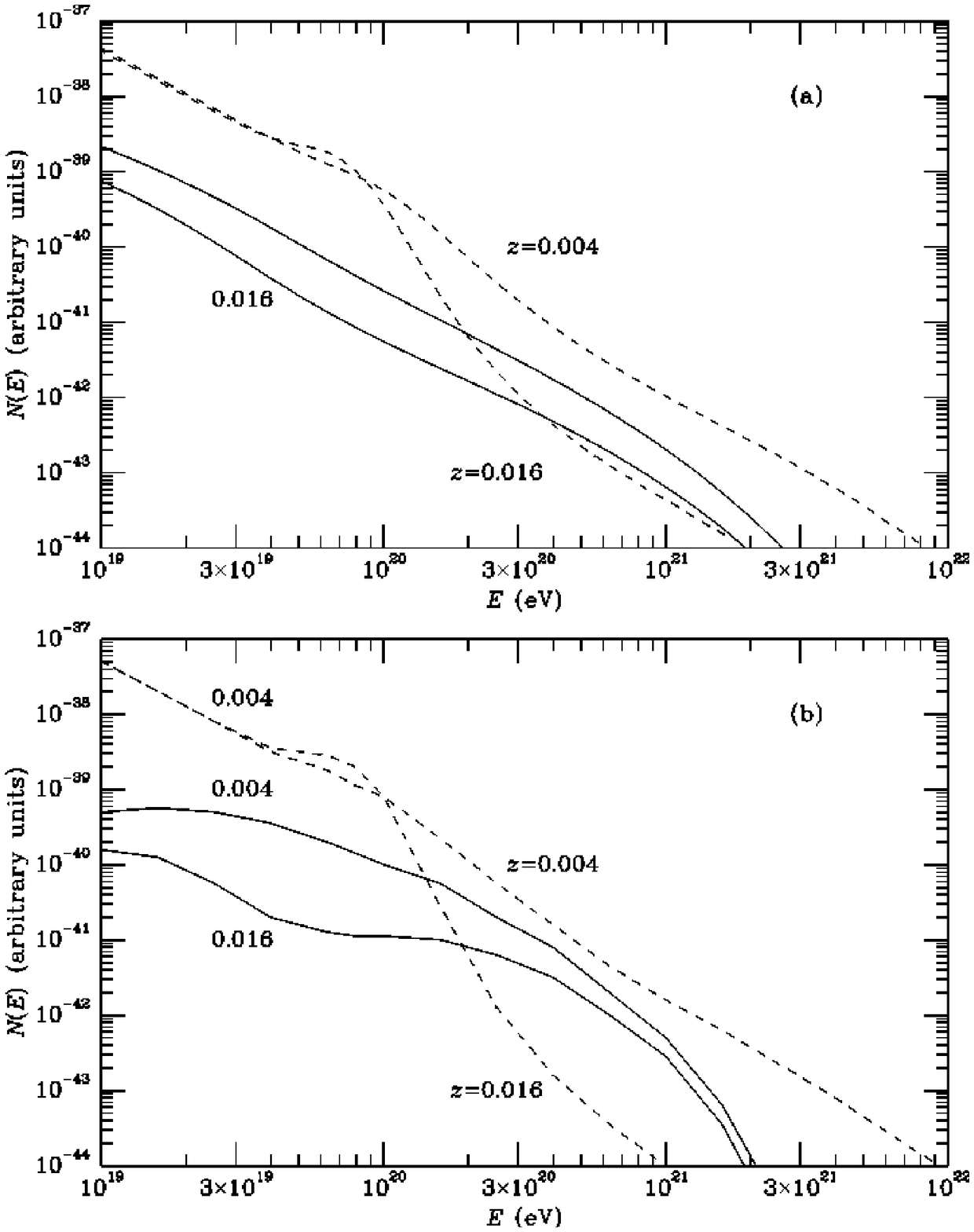}
\smallskip
\caption[...]{The differential fluxes of $\gamma$-rays
(solid lines) and nucleons (long dashed lines) from a discrete
source injecting a $E^{-2}$ proton spectrum extending up to
$10^{22}{\,{\rm eV}}$, located
at the redshift indicated: (a) Result from our analysis in
arbitrary units; (b) Corresponding results from Ref.~\cite{yt}.}
\end{figure}

\begin{figure}
\centering\leavevmode
\epsfxsize=5.5in
\epsfbox{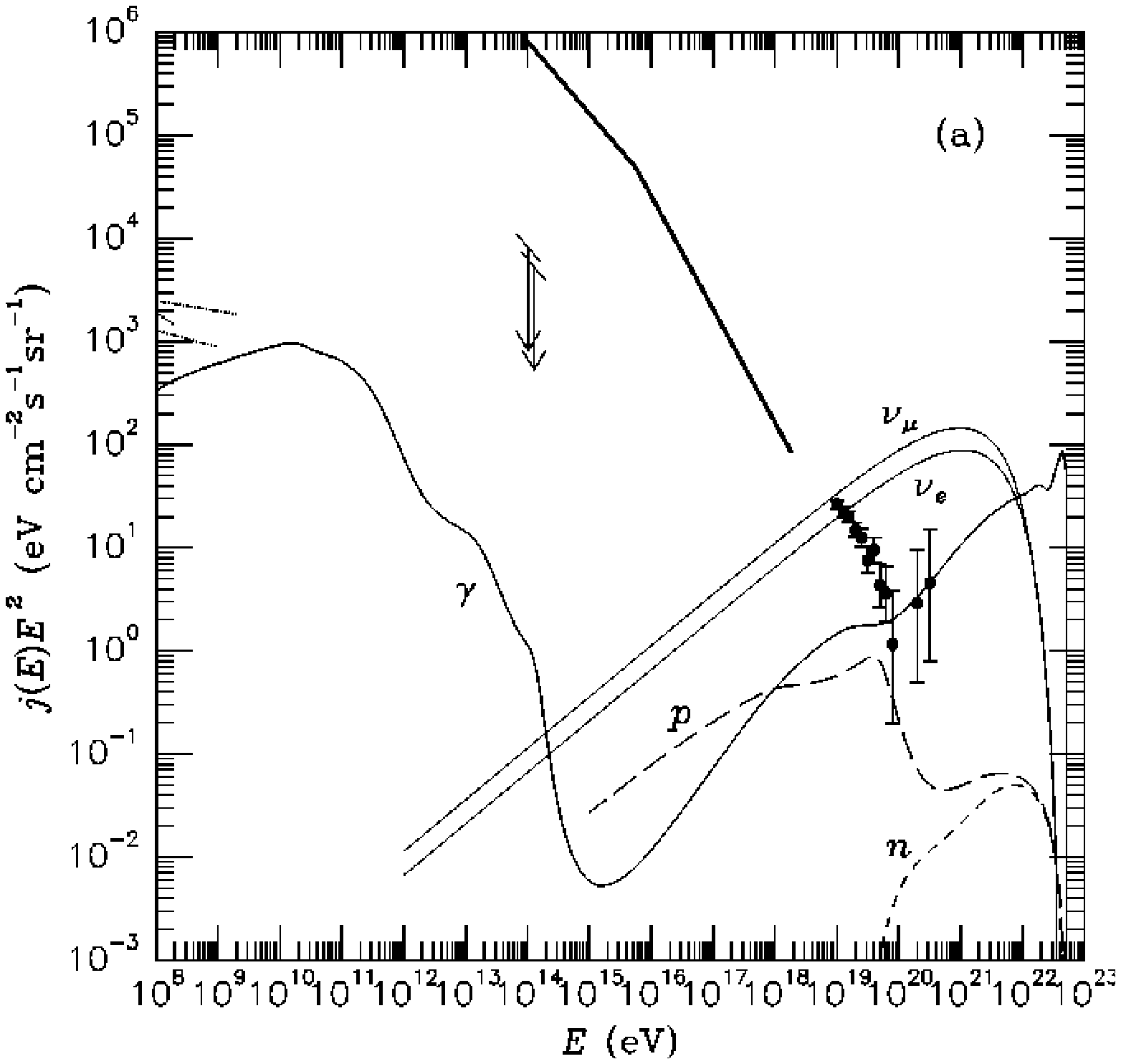}
\vspace{-1.5in}
\caption[...]{Predictions for the differential
fluxes of $\gamma$-rays (solid line), protons (long dashed
line), neutrons (short dashed line), and $\nu_\mu (\bar\nu_\mu),\nu_e
(\bar\nu_e)$ (thin solid lines in decreasing order) by a typical
topological defect scenario for a vanishing EGMF. This model
assumes uniform
injection rates with spectra given by Eq.~(\ref{inje}) and
(\ref{injh}) for the QCD motivated fragmentation functions
Eq.~(\ref{frag}) and (\ref{frag1}) for
$m_X=10^{23}{\,{\rm eV}}$. The injection history is given by
Eq.~(\ref{Xrate}) for:
(a) $p=1$. Also shown are the combined data from the Fly's
Eye~\cite{FE1,FE2} and the 
AGASA~\cite{AG1,AG2} experiments above $10^{19}{\,{\rm eV}}$ (dots with
error bars), piecewise power law fits to the charged CR flux
(thick solid line) and observational upper limits on the
$\gamma$-ray flux around $100{\,{\rm MeV}}$ from
Refs.~\cite{Digel,Fichtel,Osborne} (dotted lines in decreasing
order). The arrows indicate the limits on the $\gamma$-ray to
charged CR flux ratio from Ref.~\cite{Karle}.}
\end{figure}
\setcounter{figure}{13}
\newpage
\begin{figure}
\centering\leavevmode
\epsfxsize=5.5in
\epsfbox{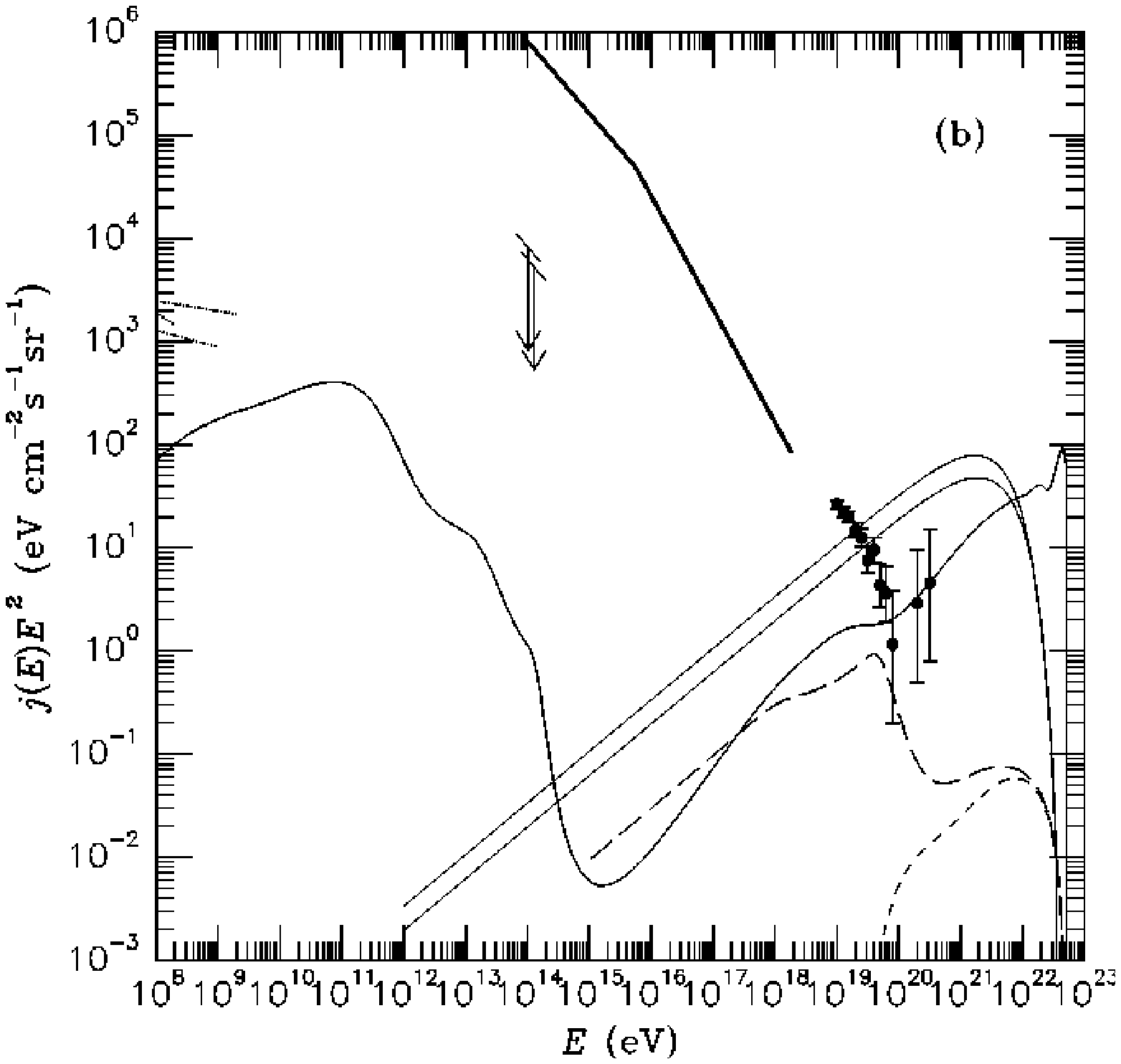}
\smallskip
\caption[...]{(b) $p=2$.}
\end{figure}

\begin{figure}
\centering\leavevmode
\epsfxsize=5.5in
\epsfbox{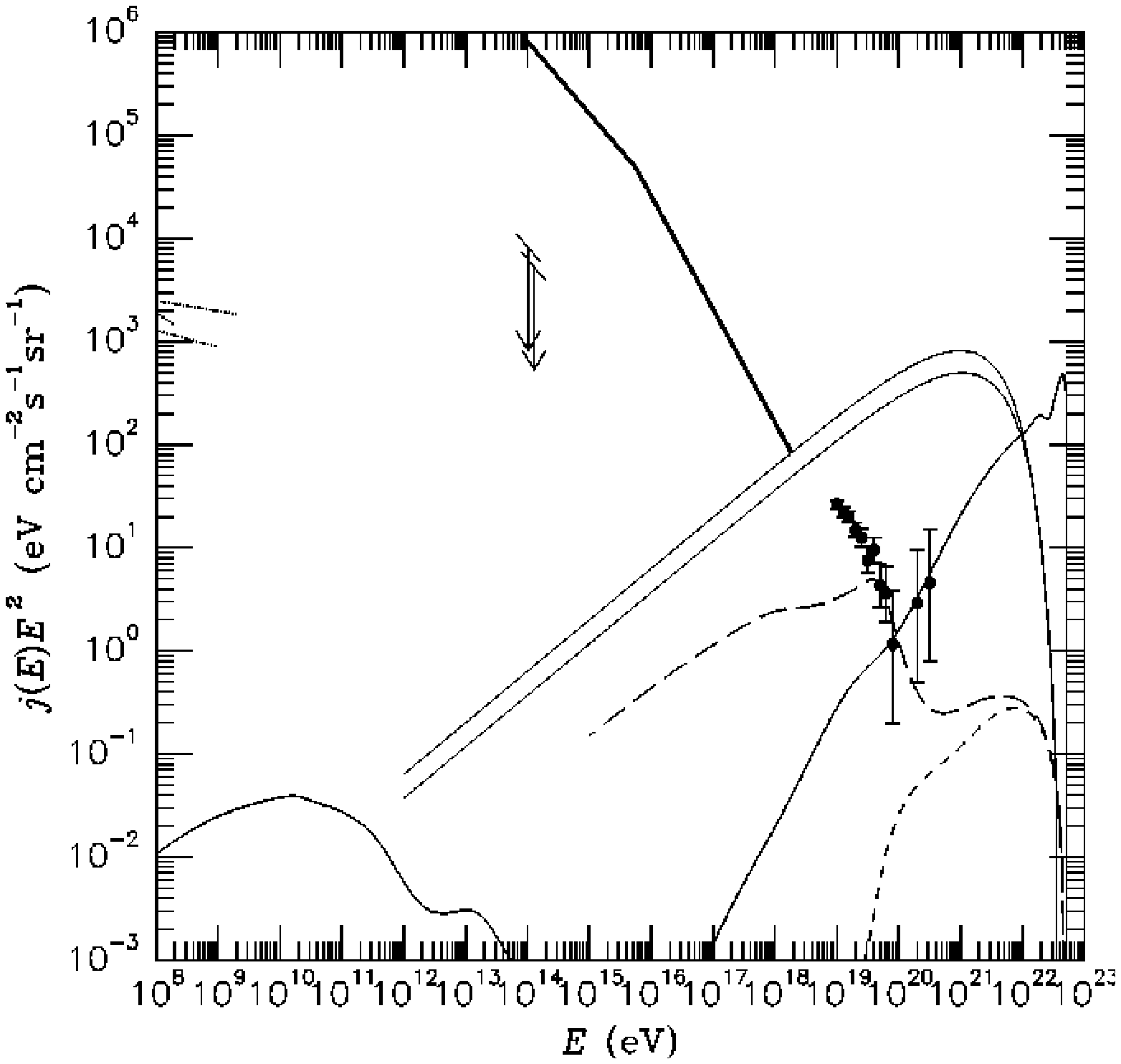}
\smallskip
\caption[...]{Same as Fig.~14(a), but neglecting
non-leading particles in the EM cascade development. Only the UHE part is
important in comparison with Fig.~14(a).}
\end{figure}

\begin{figure}
\centering\leavevmode
\epsfxsize=5.5in
\epsfbox{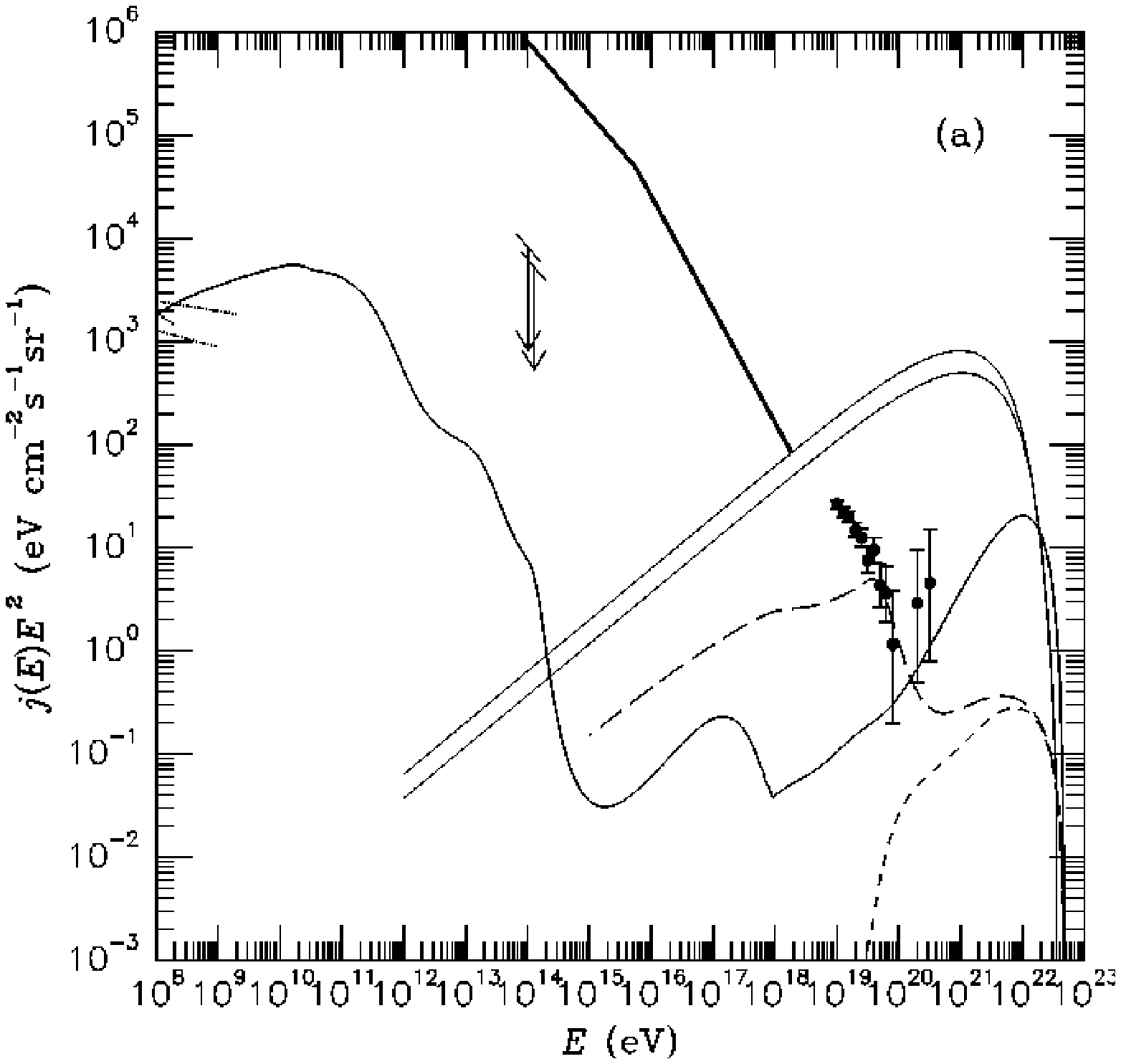}
\smallskip
\caption[...]{(a) Same as Fig.~14(a), but for an EGMF of
$10^{-9}\,$G.}
\end{figure}
\setcounter{figure}{15}
\newpage
\begin{figure}
\centering\leavevmode
\epsfxsize=5.5in
\epsfbox{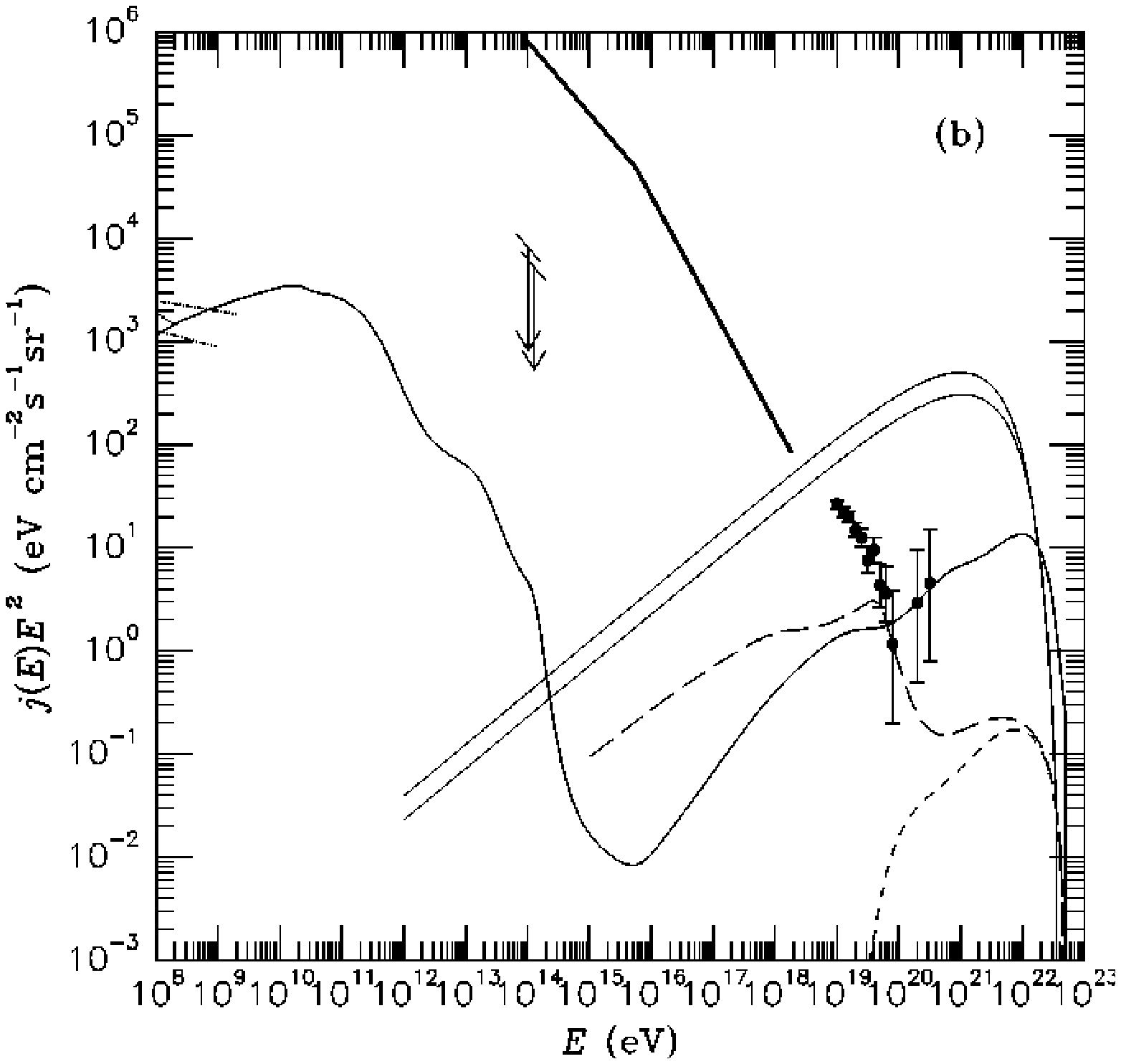}
\smallskip
\caption[...]{(b) Same as Fig.~14(a), but for an
EGMF of $10^{-11}\,$G.}
\end{figure}
\setcounter{figure}{15}
\newpage
\begin{figure}
\centering\leavevmode
\epsfxsize=5.5in
\epsfbox{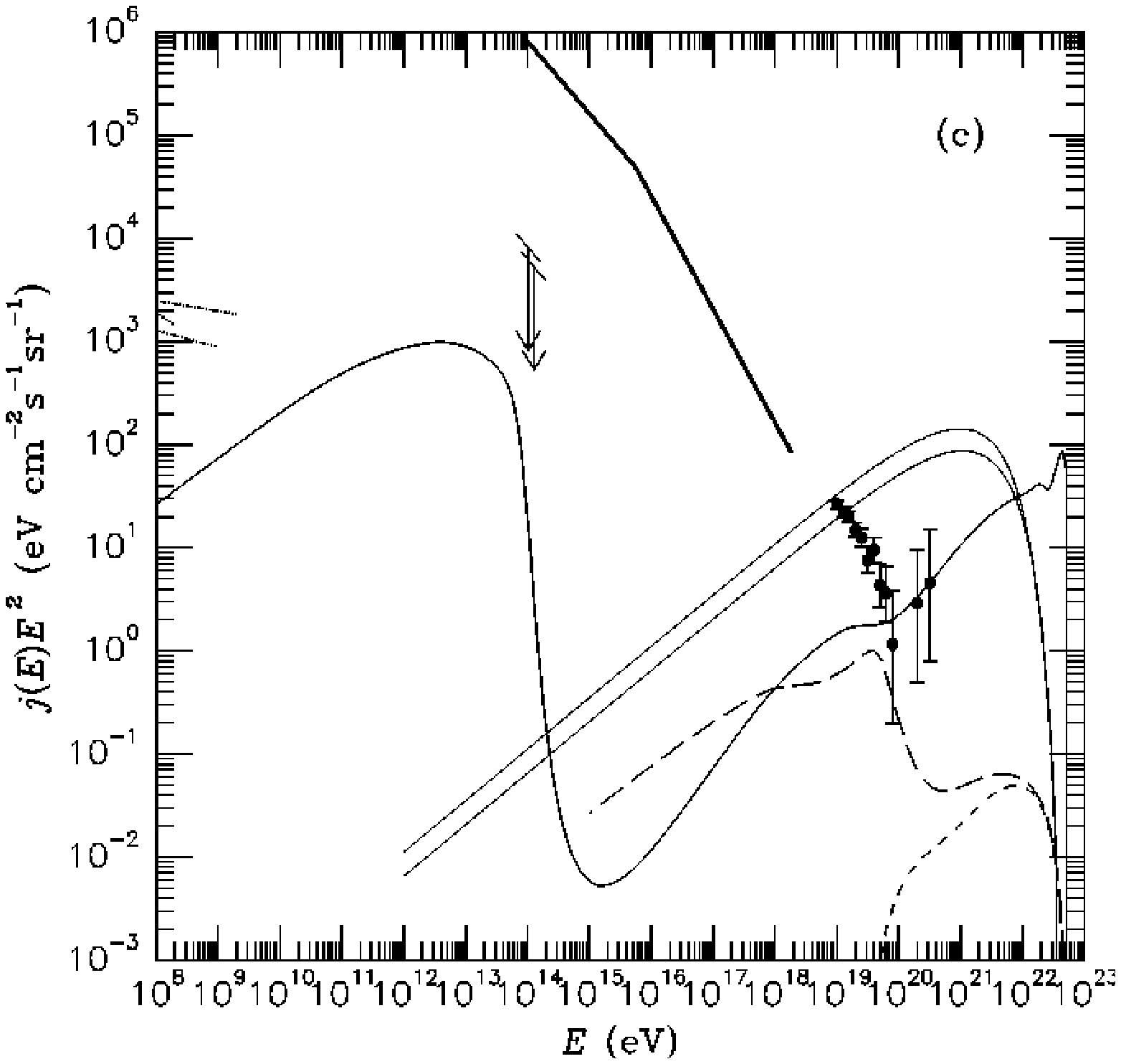}
\vspace{-0.2in}
\caption[...]{(c) Same as Fig.~14(a),
but assuming absence of any IR/O background. These panels
demonstrate the EGMF and IR/O background dependence of
the predicted HECR flux shape and composition and its influence
on the prediction for the $\gamma$-ray background at lower
energies.}
\end{figure}

\begin{figure}
\centering\leavevmode
\epsfxsize=5.5in
\epsfbox{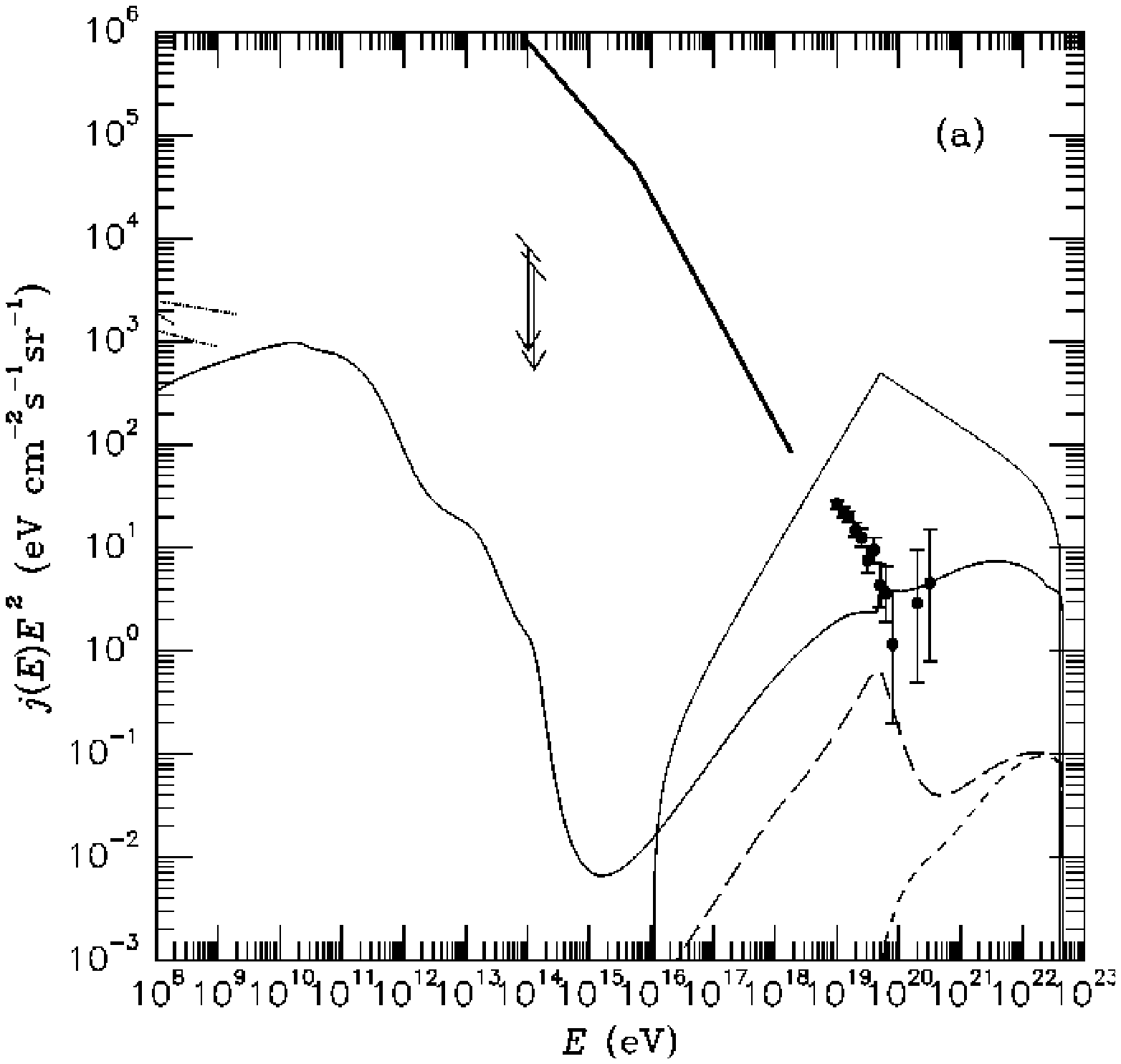}
\smallskip
\caption[...]{(a) Same as Fig.~14(a), but assuming the
fragmentation functions given by
Eqs.~(\ref{wolf1}), (\ref{wolf2}) with $m_X=10^{23}{\,{\rm eV}}$.}
\end{figure}
\setcounter{figure}{16}
\newpage
\begin{figure}
\centering\leavevmode
\epsfxsize=5.5in
\epsfbox{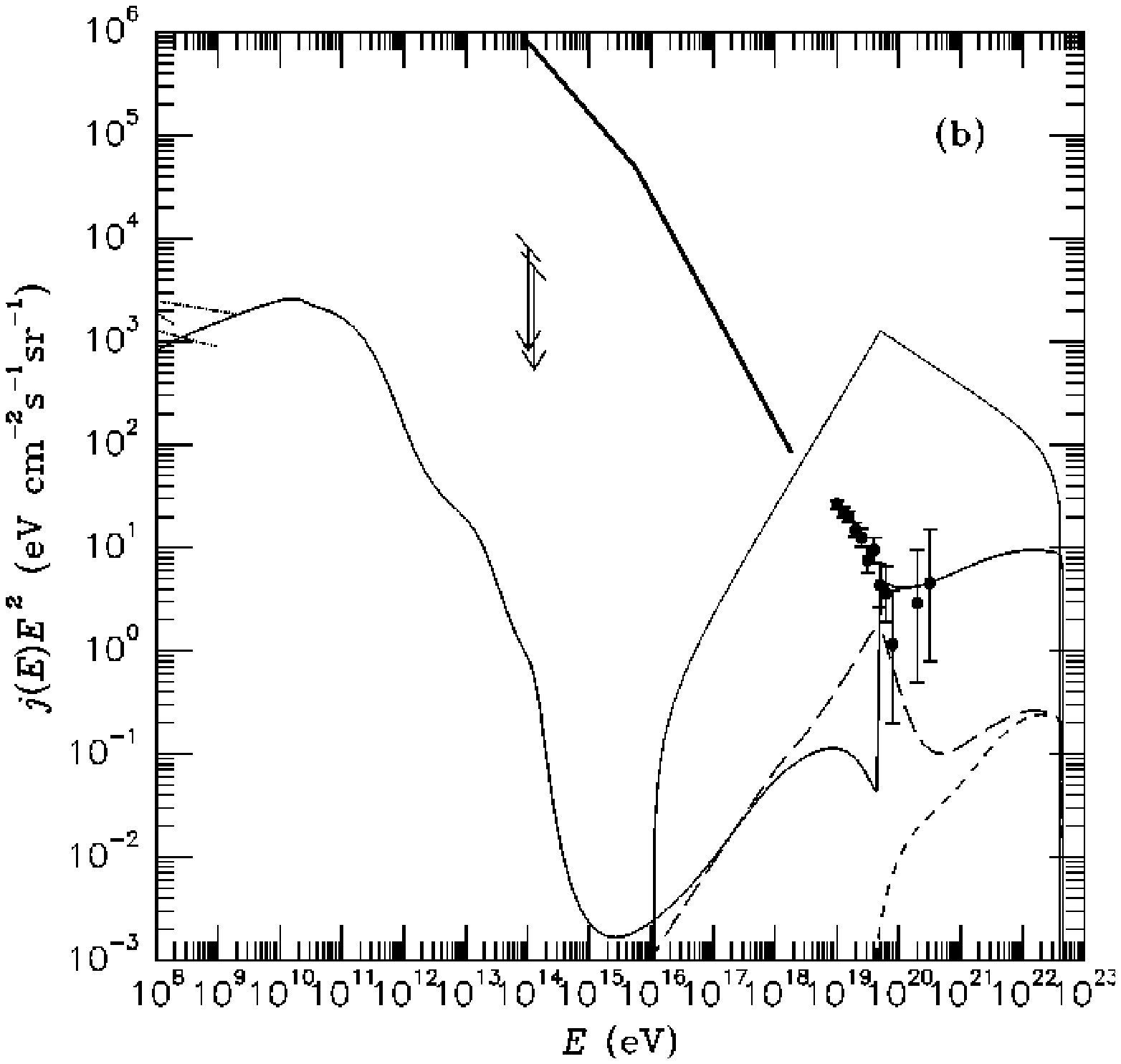}
\smallskip
\caption[...]{(b) Same as (a), but for an EGMF of $10^{-9}{\,{\rm
G}}$.}
\end{figure}

\begin{figure}
\centering\leavevmode
\epsfxsize=5.5in
\epsfbox{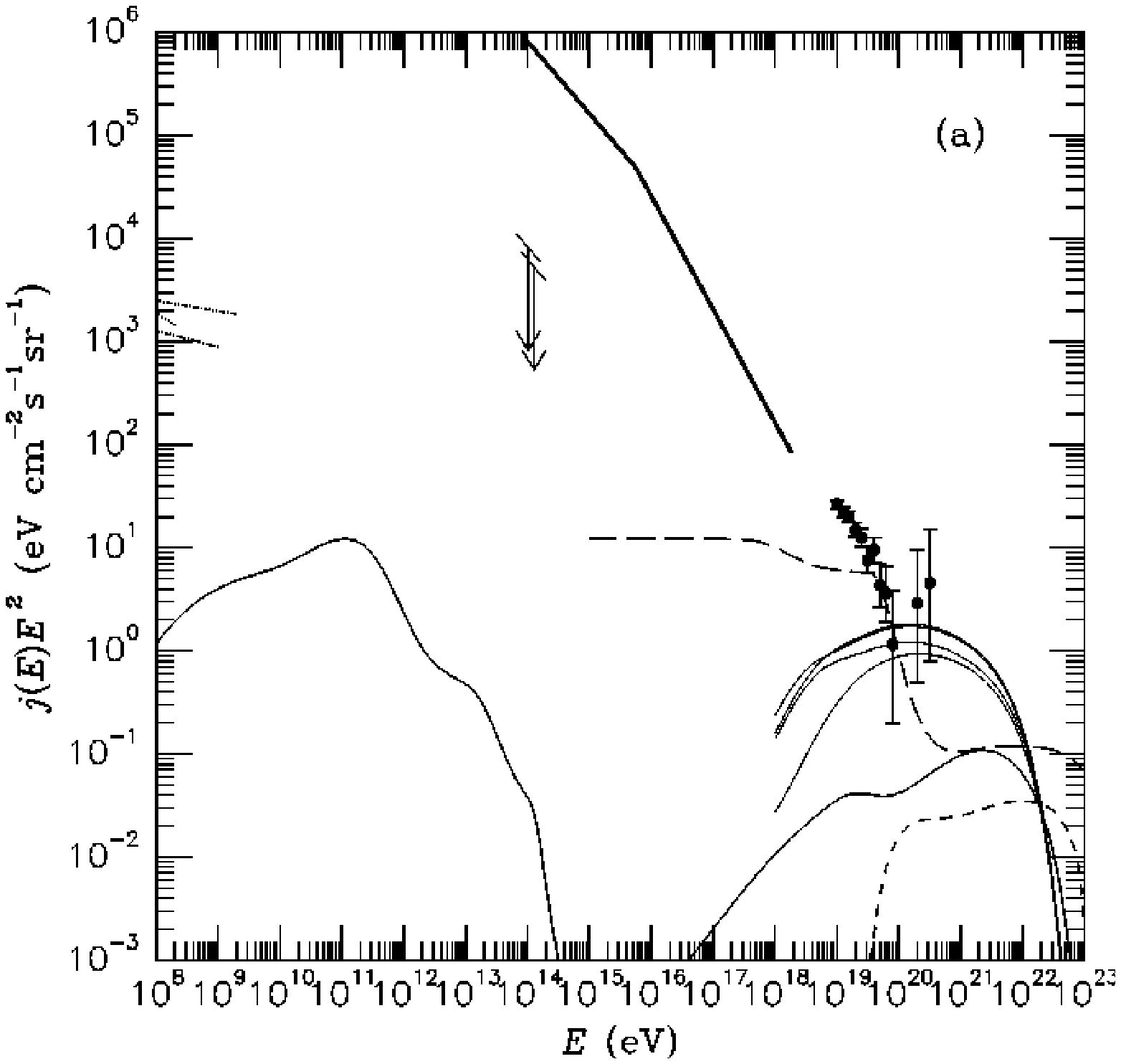}
\vspace{-0.3in}
\caption[...]{Predictions for the differential
fluxes of $\gamma$-rays (solid line), protons (long dashed
line), neutrons (short dashed line), and $\nu_\mu$,
$\bar\nu_\mu$, $\nu_e$, $\bar\nu_e$ (thin solid lines from the top)
by the GRB injection scenario given by
Eq.~(\ref{grbinj}) for vanishing EGMF for: (a) $z_{\rm max}=1$. Observational
data and constraints are
presented as in Fig.~14.}
\end{figure}
\setcounter{figure}{17}
\newpage
\begin{figure}
\centering\leavevmode
\epsfxsize=5.5in
\epsfbox{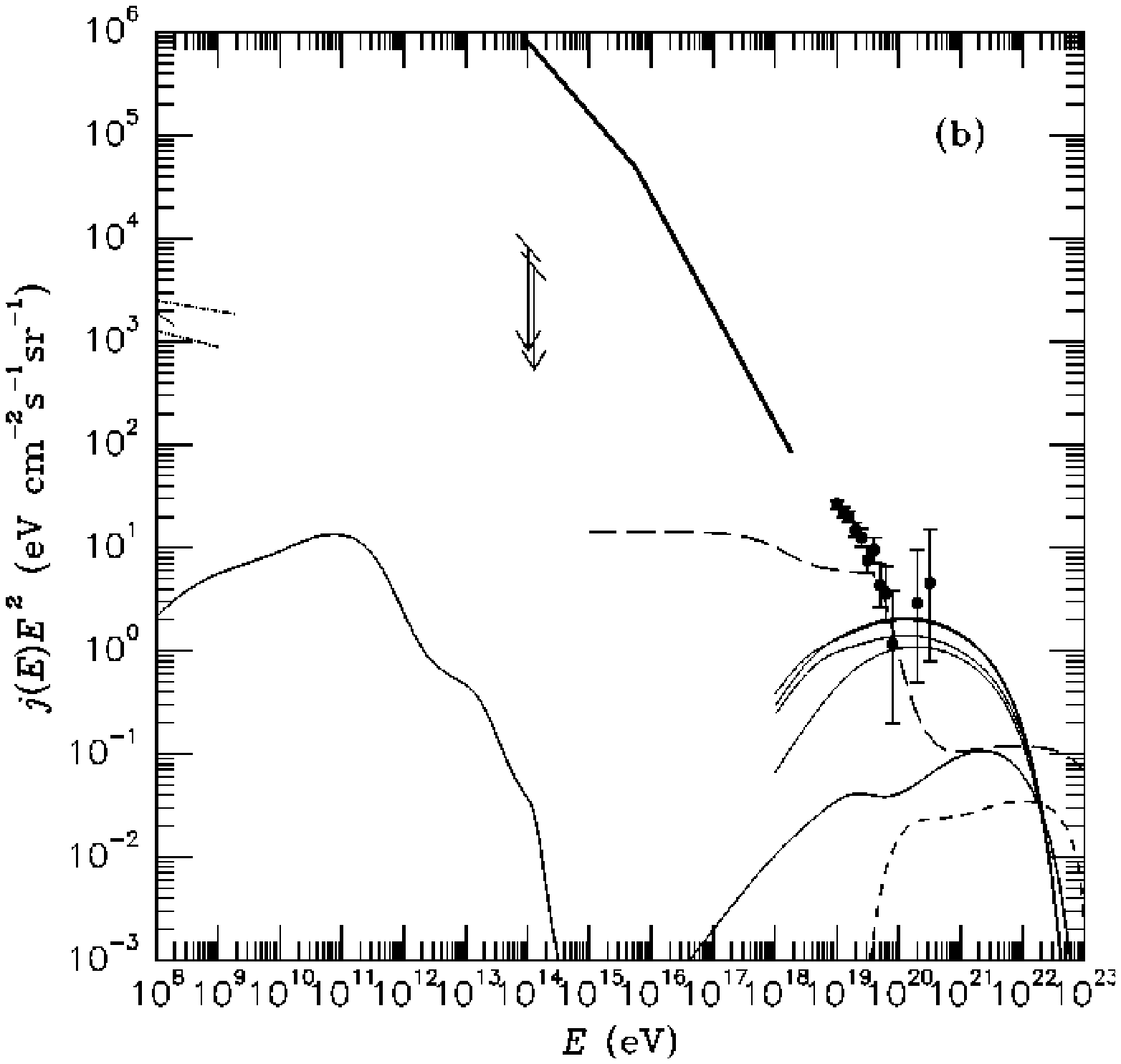}
\smallskip
\caption[...]{(b) $z_{\rm max}=4$.}
\end{figure}

\end{document}